%% file: bounded_arxivstyle.tex
\documentclass[10pt]{article}

\usepackage{amsmath}
\usepackage{amssymb}

\usepackage{graphicx}

\usepackage{cite}

\usepackage{color} 

\usepackage[labelfont=bf,labelsep=period,justification=raggedright]{caption}

\makeatletter
\renewcommand{\@biblabel}[1]{\quad#1.}
\makeatother

\date{}

\usepackage{graphicx}
\usepackage{latexsym}
\usepackage{algorithmic}
\usepackage{algorithm}
\usepackage{extpfeil}
\usepackage{morefloats}
\usepackage{hyperref}

\newcommand{\integral}[2]{\int_{#1}^{#2} }
\newcommand{\exponential}[1]{\exp \left( {#1} \right) }

\newcommand{\XX}{\mathbf{X}}
\newcommand{\xx}{{{\mathbf{x}}}}

\newcommand{\zz}{\mathbf{z}}
\newcommand{\hh}{\mathbf{h}}
\newcommand{\xiv}{\mathbf{\xi}}
\newcommand{\psiv}{\mathbf{\psi}}

\newcommand{\srewrites}[1]{\xmapsto{#1}}

\newcommand{\SSAL}{\textsc{SSAn}}

\newcommand{\deriv}[1]{{#1}^\prime}

\newcommand{\Probab}{\mathcal{P}}
\newcommand{\Exp}[1]{\mathcal{E}xp({#1})}

\newcommand{\N}{\mathbb{N}}
\newcommand{\R}{\mathbb{R}}

\begin{document}

\begin{flushleft}
{\Large
\textbf{The interplay of intrinsic and extrinsic bounded  noises  in genetic networks
} } \\ \bigskip


Giulio Caravagna$^{1,\star}$, 
Giancarlo Mauri$^{1}$, 
Alberto d'Onofrio$^{2,\star,\ast}$
\\ \bigskip

{\bf1} Dipartimento di Informatica, Sistemistica e Comunicazione, Universit\`a degli Studi Milano-Bicocca, Viale Sarca 336, I-20126 Milan, Italy.
\\
{\bf 2} Department of Experimental Oncology, European Institute of Oncology, Via Ripamonti 435, 20141 Milan, Italy.
\\  \medskip

$\star$ Equal contributors. \\
$\ast$ Corresponding author: \texttt{alberto.donofrio@ifom-ieo-campus.it}
\end{flushleft}

\section*{Abstract}

After being considered  as a nuisance to be  filtered out,  it became recently clear that biochemical noise plays a complex role, often  fully functional, for a genetic network.
The influence of intrinsic and extrinsic noises on genetic networks has intensively been investigated in last ten years, though  contributions on the co-presence of both are sparse.
Extrinsic noise  is usually modeled as an unbounded  white or colored gaussian stochastic process, even though realistic stochastic perturbations are  clearly bounded. In this paper we consider  Gillespie-like stochastic models of nonlinear networks, i.e. the intrinsic noise, where  the
model jump rates are affected by colored bounded extrinsic noises synthesized by a suitable biochemical state-dependent Langevin system.  These systems are described by a master equation, and  a  simulation algorithm to analyze them is derived. This new modeling paradigm should enlarge the class of systems amenable at modeling.
 
We investigated the influence of  both amplitude and autocorrelation time of  a extrinsic Sine-Wiener  noise on: $(i)$ the Michaelis-Menten approximation of noisy enzymatic reactions, which  we show to be applicable also in co-presence of both intrinsic and  extrinsic noise,  $(ii)$ a model of enzymatic futile cycle and $(iii)$ a genetic toggle switch. In $(ii)$ and $(iii)$  we show that the presence of a bounded extrinsic noise induces qualitative modifications in the probability densities of the involved chemicals, where new modes emerge, thus suggesting the possibile functional role of bounded noises.

\section*{Author Summary}
Realistic modeling  the dynamics of genetic networks is a non-trivial task that requires choosing a suitable level of abstraction. 
For example, within cells the molecules interacting in a network can be present in either small or large numbers. 
In the former case their time course is typically characterized by wild random oscillations closely mimicking the randomness of chemical reactions. In the latter, instead,  these fluctuations are invisible, due to the ``law of large numbers''.  So in this case  the molecular concentrations should theoretically stabilize to nice smooth steady states. 

However, the presence of perturbing external factors may induce noisy fluctuations more or less disrupting the theoretical ``nice behavior''.  Surprisingly, this disruption may be constructive: a single-function network may gain additional biological functionalities in presence of perturbations. In real world, these two kinds of randomness are not separated: proteins of a specific genetic network can be present in small number and be perturbed by external noises.  This is actually a topic only sparsely explored in systems biology. 

A factor that makes complex the study of these phenomena is that external disturbances are classically represented through unbounded Gaussian noises, which are actually unrealistic and may induce non-biological artifacts.  Thus we focus on studying systems with both small number of molecules and external bounded noises.

After defining an algorithmic framework including the co-presence of intrinsic and extrinsic stochastic effects, we apply the developed algorithm to investigate simple motifs of relevance in Systems Biology.  First, we investigate the possibility of applying the ``fast reactions elimination''  approximation to enzymatic networks affected by bounded noises. Then, we study a simple model of futile cycle and a toggle switch. In both these two cases we show that the addition of extrinsic bounded noises induces the emergence of new possible stochastic states.
\section*{Introduction}

\input{sections/introduction}

\input{sections/simulation}

\input{sections/noises}
\input{sections/examples}

\input{sections/conclusions}

\end{document}

%% file: sections/introduction.tex

Multistability is a key concept of Systems Biology since the first pioneering investigations on the dynamics of genetic networks \cite{GlassKauffman,Griffith,Simon,Thomas}. Indeed, it was quite soon understood - both experimentally and theoretically - that multiple locally stable equilibria allows for the presence of multiple functionalities, even in small groups of interplaying proteins \cite{Alon,ZH1,Fussenegger,Iannaccone,BK,SiegalGaskins,Zhdanov2,Angeli,Xiong,Zhdanov3}. 

A second key concept is that the dynamic behavior of a network is never totally deterministic, but it exhibits more or less strong stochastic fluctuations due to its interplay with many, and mainly unknown, other networks, as well as with various random signals coming from the extracellular world.  For long time the stochastic effects due these two classes of interactions were interpreted as a disturbance inducing undesired jumps between states or, with marginally functional role, as an external initial input directing towards one of the possible final states of the network in study. In any case, in the important scenario of deterministically monostable networks the stochastic behavior under the action of extrinsic noises was seen as monomodal. In other words, external stochastic effects were seen similarly as in radiophysics, namely as a disturbance more or less obfuscating the real signal, to be controlled by those pathways working as a low-pass analog filter \cite{detw,RaoWolfArkin}. For these reasons, a number of theoretical and experimental investigations focused on the existence of noise-reducing sub-networks \cite{detw,Thattai,Becskei1}. However,  it has been recently shown the existence of fundamental limits on filtering noise \cite{lestas}.


Moreover, if noises were only pure nuisances, there would be an interesting consequence. Indeed, in such a case a monostable network in presence of noise should exhibit more or less large fluctuations around the unique deterministic equilibrium. In probabilistic languages this means that the probability distribution of the total signal (noise plus deterministic signal) should be a sort of ``bell'' centered more or less at the deterministic equilibrium, i.e. the probability distribution should be ``unimodal''. However, at the end of seventies it became clear in statistical physics that the real stochastic scenario is far more complex, and the above-outlined correspondence between deterministic monostability and  stochastic monomodality in presence of external noise was seriously challenged\cite{hl}.
Indeed, it was shown that many systems that are monostable in absence of external stochastic noises have, in presence of random Gaussian disturbances, multimodal equilibrium probability densities. 
This counter-intuitive phenomenon was termed noise-induced transition \cite{hl}, and it has been shown relevant also in genetic networks \cite{hasty,Arkin}.

We above focused mainly on external random perturbations acting on genetic networks. In the meantime, experimental studies revealed the other and equally important role of stochastic effects in biochemical networks by showing that many important transcription factors, as well as other proteins and mRNA, are present in cells with very low concentrations, i.e. with a small number of molecules \cite{ghae,siggia,Becskei2}. Moreover, it was shown that RNA production is not continuous, but instead it has the characteristics of stochastic bursts \cite{cfx}. Thus, a number of investigations has focused on this internal stochastic effect, the ``intrinsic noise'' as some authors term it \cite{G77,ThattaiPNAS}. In particular, it was shown - both theoretically and experimentally - that also the intrinsic noise may induce multimodality in the discrete probability distribution of proteins \cite{Arkin,To}. However, the fact that intrinsically stochastic systems may exhibit behaviors similar to systems affected by extrinsic gaussian noises was very well known in statistical and chemical phsyics, where this was theoretically demonstrated  by approximating the exact Chemical Master Equations with an appropriate Fokker-Planck equation \cite{Gardiner,Hanggi,G80}, an approach leading to the  Chemical Langevin Equation \cite{G00}.

Thus, after that for some time noise was mostly seen as a nuisance, more recently it has finally been appreciated that the above-mentioned and other noise-related phenomena may in many cases have a constructive, functional role (e.g. see \cite{eldar,losick} and references therein). Indeed, for example, noise-induced multimodality allows a transcription network for reaching states that would not be accessible if the noise was absent \cite{Arkin,eldar,losick}. Phenotype variability in cellular populations is probably the most important macroscopic effect of intracellular noise-induced multimodality \cite{eldar}.

In Systems Biology, from the modeling point of view Swain and coworkers \cite{siggia} were among the first to study the co-presence of both intrinsic and extrinsic randomness, by stressing the synergic role in modifying the velocity and average in the context of the basic network for the production and consumption of a single protein, in absence of feedbacks. These and other important effects were shown, although nonlinear phenomena such as multimodality were absent.
The above study is also remarkable since: $(i)$ it has stressed the role of the autocorrelation time of the external noise and,  differently from other investigations, $(ii)$ it has stressed that modeling the external noise by means of a Gaussian noise, either  white or colored, may induce artifacts. In fact, since the perturbed parameters may become negative, the authors employed a lognormal positive noise to model the extrinsic perturbations. In particular, in \cite{siggia} a noise obtained by exponentiating the classical Orenstin-Uhlenbeck noise was used \cite{hl}.

From the data analysis point of view, You and collaborators \cite{You} and Hilfinger and Paulsson \cite{hilf} recently proposed interesting methodologies to infer by convolution the contributions of extrinsic noise also in some nonlinear networks, including a synthetic toggle switch \cite{You}. 

Our general aim here is manifold. 
Indeed, we want to start by investigating the co-presence of both extrinsic and intrinsic randomness in nonlinear genetic networks in the important case where the external perturbations are not only non-Gaussian, but also bounded. Indeed, by imposing a bounded extrinsic noise we increase the degree of realism of a model, since the external perturbations must not only preserve the positiveness of reaction rates, but must also be bounded. Moreover, it has also been shown in other contexts such as oncology and statistical physics that: $(i)$ bounded noises deeply impact on the transitions from unimodal to multimodal probability distribution of state variables \cite{wio,SWiener,onofr1,onofr2,onofr3} and $(ii)$ the dynamics of a system under bounded noise may be substantially different from the one of systems perturbed by other kinds of noises, for example there is dependence of the behavior on the initial conditions \cite{onofr1,onofr3}.

This approach opens a number of questions, two of which we shall try to assess here. The first question is to identify a suitable mathematical framework to represent mass-action biochemical networks perturbed by bounded noises (or simply left-bounded), which in turn can depend on the state of the system. To this extent, in the first part of this work we derive a master equation for these kinds of systems in terms of the {differential Chapman-Kolgomorov equation} (DCKE) \cite{Gardiner,Olaf} and propose a combination of the Gillespie's Stochastic Simulation Algorithm (SSA) \cite{G76,G77} with a state-dependent Langevin system, affecting the model jump rates, to simulate these systems.

The second question relates to the possibility of extending, in this ``doubly stochastic'' context, the Michaelis-Menten Quasi Steady State approximation (QSSA) for enzymatic reactions \cite{Murray}. We face the validity of the QSSA in copresence of both types of noise in the second part of this work, where we numerically investigate the classical Enzyme-Substrate-Product network. The application of QSSA in this network has been recently investigated by Gillespie and coworkers when extrinsic noise is absent \cite{GillMM}. Based on our results, we propose the extension of the above structure also to more general networks than those ruled by the rigourous mass-action law via a stochastic QSSA.

Finally, in the third part we investigate the interplay between intrinsic randomness and extrinsic bounded noises in two cases of interest in biology: $(i)$ a futile cycle \cite{Arkin} and $(ii)$ a genetic toggle switch \cite{ZH1}, which is a fundamental motif for cellular differentiation and for other switching functions. As expected,  the co-presence of both intrinsic stochasticity and  bounded extrinsic random perturbations suggests the presence of possibly unknown functional roles for noise in both networks. 
The described noise-induced phenomena are shown to be strongly related to physical characteristics of the extrinsic noise such as the noise amplitude and its autocorrelation time.

%% file: sections/simulation.tex
\section*{Methods}

\subsection*{Noise-free stochastic chemically reacting systems} \label{sec:noise-free}

We start by recalling the Chemical Master Equation and the Stochastic Simulation Algorithm (SSA) by Doob and Gillespie \cite{G76,G77}. Systems where the jump rates are time-constant are hereby referred to  as stochastic noise-free systems. 
We consider a well stirred  system of molecules  belonging to $N$ chemical species $\{S_1, \ldots,S_N\}$ interacting through $M$ chemical reactions ${R_1, \ldots, R_M}$.  We represent the (discrete) state  of the target system with a $N$-dimensional integer-valued  vector  $\XX(t)=(X_1(t), \ldots, X_N(t))$ where $X_i(t)$ is the  number of molecules of species $S_i$ at time $t$.  To each reaction $R_j$ is  associated its  stoichiometric vector  $\nu_j=(\nu_{1,j},\ldots,\nu_{N,j})$, where $\nu_{i,j}$ is the change in the $X_i$ due to one $R_j$ reaction. The stoichiometric vectors form the $N \times M$ stoichiometry matrix
$
D = 
\left[ 
\begin{matrix}
\nu_1 & \nu_2 & \ldots & \nu_M 
\end{matrix}
\right]$. Thus, given $\XX(t)=\xx$ the firing of  reaction $R_j$ yields the new state $\xx + \nu_j$.
A propensity function $a_j(\xx)$  \cite{G76,G77} is associated  to each $R_j$ so that $a_j(\xx)dt$, given $\XX(t)=\xx$, is the probability of reaction $R_j$ to fire in state $\xx$ in the infinitesimal interval $[t, t+dt)$. Table \ref{table:propensity} summarizes the analytical form of such functions \cite{G76}. For more generic form of the propensity functions (e.g. Michaelis-Menten, Hill kinetics) we refer to \cite{CHAPTERGILL}.

\begin{table}[t] 
\begin{center}
    \begin{tabular}{ lc|cl}
      Order & Reaction  & Propensity \\ \hline \hline
  $0$-{th}  & $\emptyset \srewrites{k} S_w$  & $k$ &  \\ \hline
  $1$-{st}  & $S_i \srewrites{k} S_w$  & $k X_i(t)$ \\ \hline
  $2$-{nd}  & $2S_i \srewrites{k}S_w$  & $k X_i(t) (X_i(t) - 1)/2$ \\
               & $S_i + S_{i'} \srewrites{k} S_w$& $k X_i(t) X_{i'}(t)$
    \end{tabular}
\end{center}
\caption{Analytical form of the propensity functions \cite{G76}.}
\label{table:propensity}
\end{table}

We recall the definition of the {\em Chemical Master Equation} (CME) \cite{Doob1,Doob2,G76,G77} describing the time-evolution of the probability of a system to occupy each one of a set of states. We study the time-evolution of $\XX(t)$, assuming that the system was initially in some state $\xx_0$ at time $t_0$, i.e. $\XX(t_0)=\xx_0$.  We denote with $\Probab(\xx,t\mid\xx_0,t_0)\equiv\Probab(\xx,t\mid\omega)$ the probability that, given $\XX(t_0)=\xx_0$, at time $t$ it is $\XX(t)=\xx$. From the usual hypothesis that at most one reaction fires in the infinitesimal interval $[t, t+dt)$, it follows that the time-evolution of $\Probab(\xx,t\mid\omega)$ is given by the following partial differential equation termed ``master equation'' 
\begin{equation} \label{eq:cme}
{\partial_t \Probab(\xx,t\mid\omega)}  =\sum_{j=1}^M  \Probab(\xx -\nu_j,t\mid\omega) a_j(\xx-\nu_j) 
 -\Probab(\xx,t\mid\omega) a_j(\xx) \,.
\end{equation}
The CME is a special case of the more general Kolmogorov Equations  \cite{Kolm}, i.e. the differential equations corresponding to the time-evolution of stochastic Markov   jump processes.   As it is well known, the CME can be solved analytcally only for a very few simple systems, and normalization techniques are sometimes adopted to provide approximate solutions \cite{CMENormalization}. However, exact algorithmic realization of the process associated to the CME are possible by using the Doob-Gillespie Stochastic Simulation Algorithm (SSA) \cite{Doob1,Doob2,G76,G77}, summarized as Algorithm \ref{alg:SSA}. Although equivalent formulations exist \cite{GIBSONBRUCK,G76,G77}, as well as some approximations \cite{SLOWSCALE,TAULEAP,CHAPTERGILL},  here we consider its  {\em Direct Method} formulation without loss of generality.

\begin{algorithm}[!h]
\caption{SSA ($t_0$, $\xx_0$, $T$)}
\label{alg:SSA}
\begin{algorithmic}[1]
\STATE set $\xx \gets \xx_0$ and $t\gets t_0$;
\WHILE{$t < T$} 
\STATE $a_0(\xx) \gets \sum_{j=1}^M a_j(\xx);$
\STATE let $r_1$, $r_2 \sim U[0,1];$
\STATE $\tau \gets   a_0(\xx)^{-1} \ln({r_1}^{-1})$;
\STATE let $j$ such that $\sum_{i=1}^{j-1} a_i(\xx) < r_2 \cdot a_0(\xx) \leq \sum_{i=1}^j  a_i(\xx)
$;
\STATE set $\xx \gets \xx + \nu_j$ and $t\gets t+\tau$;
 \ENDWHILE
\end{algorithmic}
\end{algorithm}

The SSA is an {exact} dynamic Monte-Carlo method describing a statistically correct trajectory of a discrete non-linear Markov process, whose probability density function is  the solution of equation (\ref{eq:cme}) \cite{Feller}.  The SSA computes a {single} realization of the process $\XX(t)$, starting from state $\xx_0$ at time $t_0$ and up to time $T$. Given $\XX(t)=\xx$ the putative time $\tau$ for the next reaction to fire is chosen by sampling an {exponentially} distributed random variable, i.e. $\tau \sim \Exp{a_0(\xx)}$ 
where $a_0(\xx)=\sum_{j=1}^M a_j(\xx)$ and $\sim$ denotes the equality in law between random variables.  The  reaction to fire $R_j$ is chosen with weighted probability $a_j(\xx)/a_0(\xx)$, and the system state is updated accordingly.


The correctness of the SSA comes from the relation between the jump process and the CME \cite{Feller,G76}. In fact,  the probability, given $\XX(t)=\xx$, that the next reaction in the system occurs in the infinitesimal  time interval $[t + \tau, t + \tau + d\tau)$, denoted $\Probab(\tau \mid \xx,t)$,  follows 
 \begin{equation} \label{eq:SSA-tau} \small
\Probab(\tau \mid \xx,t) = \sum_j \Probab(\tau, j \mid \xx,t) = a_0(\xx) \exponential{-\integral{0}{\tau}a_0(t')dt'} 
= a_0(\xx)  e^{-a_0(\xx)\tau} 
\end{equation}
since $\Probab(\tau, j \mid \xx,t)=a_j(\xx) \exponential{-a_0(\xx)\tau}$ is the probability distribution of the putative time for the next firing of $R_j$, and the formula follows by the independency of the reaction firings. Notice that in equation (\ref{eq:SSA-tau}) $a_0(t')$ represents the propensity functions evaluated in the system state at time $t' > t$, i.e. as if they were time-dependent functions. In the case of noise-free systems that term evaluates as $a_0(\xx)$ for any $t \in[t, t+\tau]$, i.e. it is indeed time-homogenous whereas in more general cases it may not, as we shall discuss later.  Finally, the probability of the reaction to fire at $t+\tau$ to be $R_j$ follows by conditioning on $j$, that is
\begin{equation}\label{eq:SSA-j}  
\Probab(j \mid \tau ; \xx,t) = \dfrac{\Probab(\tau, j \mid \xx,t)}{\Probab(\tau \mid \xx, t)} = \dfrac{a_j(\xx) \exponential{-a_0(\xx)\tau}}{a_0(\xx) \exponential{-a_0(\xx)\tau}} = \dfrac{a_j(\xx)}{a_0(\xx)} \, .
\end{equation}

\subsection*{Noisy stochastic chemically reacting systems} \label{sec:noisy-sys}
We now introduce a theory of stochastic chemically reacting systems with (un)bounded noise in the jump rates by combining Stochastic Differential Equations and the SSA. Here we consider  a system where each propensity function may be affected by a {\em extrinsic}  noise term. In general, such a term can be either a time or state-dependent function,  and  the propensity function  for reaction $R_j$ reads now as
\begin{align} \label{eq:prop-noise0}
a_j(\xx,t) =  a_j(\xx) L^\ast_j(t) \, ,
\end{align}
where $a_j(\xx)$ is a propensity function of a type listed in Table \ref{table:propensity}. The noisy perturbation term  $L^*_j(t)$ is positive and bounded by some $C_j  \le +\infty $, i.e.
\begin{equation} \label{eq:mlb}
0  \le  \,L^\ast_j(t) \le C_j  
\end{equation}
so we are actually considering both bounded and right-unbounded noises, i.e.  $C_j=+\infty$. In the former case we say that the $j$-th extrinsic noise is bounded, in the latter that it is left-bounded.
%
Note that in  applications we shall mainly consider  unitary mean perturbations, that is
\[
 \langle L^\ast_j(t) \rangle =1 \,. 
 \]
We consider here that the extrinsic noisy disturbance $L^*_j(t)$ is a function of a more generic $\Sigma$-dimensional noise $\xiv(t)$ with $1 \le \Sigma \le M$ so we write $L^*_j(t)=L_j(\xiv(t))$ and equation (\ref{eq:prop-noise0}) reads as
\begin{align} \label{eq:prop-noise}
a_j(\xx,t) =  a_j(\xx) L_j(\xiv(t))\, .
\end{align} 
Notice that the use of a vector in  equation (\ref{eq:prop-noise}) provides the important case of multiple reactions sharing the same noise term, i.e. the reactions may be affected in the same way by a unique noise source.
In equation (\ref{eq:prop-noise})   $L_j$  is a continuous functions $L_j:  \R^{\Sigma} \to \R$ and $\xiv(t) \in \R^{\Sigma}$ is a colored and, in general, non-gaussian noise that may depend on the state $\XX(t)$ of the chemical system. The  dynamics of $\xiv(t)$ is described  by a  $\Sigma$-dimensional Langevin system 
\begin{equation} \label{eq:noise}
\xiv^{\prime}(t) = f(\xiv,\XX(t)) + g(\xiv,\XX(t))\eta(t) \,.
\end{equation}
Here,  $\eta$ is a $\Sigma$-dimensional vector of uncorrelated white noises of unitary intensities, $g$ is a $\Sigma \times \Sigma$ matrix which we shall mainly consider the be  diagonal  and $f, g_{h,k}: \R^{\Sigma}\times \R^{N} \to \R^{\Sigma}$.
When $\xiv(t)$  does not directly depend on $\XX(t)$, i.e. the extrinsic noise depends on an external source, which is the kind of noise we mainly consider, equation (\ref{eq:noise}) reduces to
\begin{equation} \label{eq:noiseIMP}
\xiv^{\prime}(t) = f(\xiv) + g(\xiv)\eta(t)\, .
\end{equation}
We stress that  the ``complete" Langevin system in equation (\ref{eq:noise}) is not a mere analytical exercise, but it has the aim of phenomenologically modeling extrinsic noises that are not totally independent of the process  in study.


\subsubsection*{The  Chapman-Kolmogorov Forward Equation} 

When a discrete-state jump process as one of those described in previous section is linked with a continuous noise  the state of the stochastic process is  the vector
\begin{align} \label{eq:vectorzz}
 \zz = (\xx,\xiv) & \quad \text{where}\quad \xx \in \N^N, \xiv \in \R^\Sigma \, ,
\end{align}
and  the state space of the process is now  $ \N^N\times \R^\Sigma$.  Our total process can be considered as a particular case of the general Markov process where diffusion, drift and discrete finite jumps are all co-present for all state variables \cite{Gardiner,Olaf}. For this very general family of stochastic processes the dynamics of the  probability of being in some state $\zz$ at time $t$, given an initial state $\zz_0$ at time $t_0$ shortly denoted as $\omega$,  is described by the \textit{differential Chapman-Kolgomorov equation}  (DCKE) \cite{Gardiner, Olaf}, whose generic form is
 \begin{align} \label{eq:dgkeq0} 
\partial_t \Probab\Big(\zz,t\mid \omega\Big)= &-\sum_{j} \partial_{\zz_j}A_j(\zz,t)\Probab\Big(\zz, t\mid\omega\Big)  
+ \frac{1}{2} \sum_{i,j}\partial_{\zz_i,\zz_j}B_{i,j}(\zz,t)\Probab\Big(\zz,t\mid\omega\Big)  \\
&+ \int \Big[W(\zz\mid \hh,t )\Probab\Big(\zz,t\mid\omega\Big)-W(\hh \mid \zz,t )\Probab\Big(\hh,t\mid\omega\Big) \Big] d\hh \, .  \nonumber
\end{align}
Here $A_j$ forms the drift vector for $\zz$, $B_{i,j}(\zz,t)$  the diffusion matrix and $W$ the jump probability.  For an elegant derivation of the DCKE  from the integral  Chapman-Kolgomorov equation \cite{Kolm} we refer to \cite{Olaf}.
This equation  describes various  systems, in fact we  remind that $(i)$ the Fokker-Planck equation is a particular case of the DCKE without jumps (i.e. $W(\zz \mid \hh,t )=0$), $(ii)$ the CME in equation (\ref{eq:cme}) is the DCKE without  brownian motion and drift  (i.e. $A(\zz,t) =0$ and $B(\zz,t)=0$), $(iii)$ the Liouville equation is the DCKE without brownian motion and jumps (i.e. $A(\zz,t) =0$ and $W(\zz \mid \hh,t )=0$) and  $(iv)$ the ODE with jumps correspond to the case where only diffusion is absent (i.e. $B(\zz,t)=0$). 

We stress that, at the best of our knowledge, this is the first time where a master equation for stochastic chemically reacting systems combined with bounded noises is considered.  Let \begin{align} \label{eq:genprocess}
\Probab\Big( (\xx,\xiv),t\mid(\xx_0,\xiv_0),t_0\Big)\equiv\Probab(\zz,t\mid\omega)
\end{align}
be the   probability that at time $t$ it is $\XX(t)=\xx$ and $\xiv(t) = \xiv$, given   $\XX(t_0)=\xx_0$ and $\xiv(t_0) = \xiv_0$. The time-evolution of $\Probab(\zz,t\mid\omega)$ is  equation (\ref{eq:dgkeq0})  where drift and diffusion are given by the Langevin  equation (\ref{eq:noise}), that is
\begin{align} 
A = f(\xiv,\xx) && B = g^T \times g
 \end{align}
with $\times$ the standard vector multiplication and $g^T$ the transpose of $g$. Moreover, since  only finite  jumps are possible, then the jump functions and diffusion satisfy 
\begin{align}
 \label{eq:decoup}
\partial_{\zz_i\zz_j}B_{i,j}(\zz,t)=0 &&  W\Big((\xx, \xiv) \mid  (\xx,\xiv^\ast),t \Big)=0 
\end{align}
for any $i,j=1,\ldots,N$, and noise $\xiv^\ast \in  \R^\Sigma$. Summarizing, for the systems we consider 
the  DCKE in equation (\ref{eq:dgkeq0}) reads as 
\begin{align} \label{eq:DCKE} \small
\partial_t \Probab\Big((\xx,\xiv),t\mid\omega\Big)= &- \sum_{j=1}^M \partial_{\zz_j} f_j(\xiv,\xx)\Probab\Big((\xx,\xiv),t\mid\omega\Big)  +\frac{1}{2} \sum_{i=1}^M \sum_{j=1}^M\partial_{\zz_i\zz_j}B_{i,j}(\xiv,\xx)\Probab\Big((\xx,\xiv),t\mid\omega\Big)\\
&+ \sum_{j=1}^M  \Probab\Big((\xx -\nu_j,\xiv),t\mid\omega\Big) a_j(\xx-\nu_j) L_j(\xiv(t))   - \Probab\Big((\xx,\xiv),t\mid\omega\Big) \sum_{j=1}^M  a_j(\xx)L_j(\xiv(t)) \nonumber \,.
\end{align}
This equation is the natural generalization of the CME in equation (\ref{eq:cme}), and completely characterize noisy systems. As such, however, its realization can be prohibitively difficult and is hence convenient to define  algorithms to perform the simulation of  noisy systems.

\subsubsection*{The  SSA with Bounded Noise}

We now define the {\em Stochastic Simulation Algorithm with Langevin Noise} (\SSAL{}).  The algorithm performs a realization of the stochastic process underlying the system where a (generic) realization of the noise is assumed. As for the CME and the SSA,  this corresponds to computing a realization of a process satisfying equation (\ref{eq:DCKE}). The \SSAL{} takes inspiration from the (generic) SSA with time-dependent propensity functions \cite{Anderson} as well as the SSA for hybrid deterministic/stochastic systems \cite{AlfonsiHybrid,Alfonsi,TIS1,TIS2},  thus  generalizing the jump equation (\ref{eq:SSA-tau}) to a time inhomogeneous distribution, which we discuss in the following.


For a system with $M$ reactions the time evolution equation for $\XX(t$) is
\begin{align}
d\XX(t)=\sum_{j=1}^M\nu_jN_j(t)
\end{align}
where $\{N_j(t) \mid t \geq t_0\}$ is the stochastic process counting the number of times that $R_j$ occurs in $[t_0 , t]$ with initial condition $N_j(t_0) = 0$. For Markov processes $N_j(t)$ is an inhomogeneous Poisson process satisfying
\begin{align} \label{eq:Nj}
\Probab\Big( N_j(t + dt) - N_j(t) = 1 \mid \xx \Big)  = a_j(\xx, t)dt =   a_j(\xx) L_j(\xiv(t))dt 
\end{align}
when $\XX(t)=\xx$. In hybrid systems this is is a doubly stochastic Poisson process with time-dependent intensity, in our case 
this is a Cox process \cite{Cox,Cox2} since the intensity itself is a stochastic process, i.e. it depends on the stochastic noise. More simply,  in noise-free systems, this equation evaluates as $ a_j(\xx)dt$, thus denoting a time homogeneous Poisson process. As in \cite{AlfonsiHybrid,TIS1,TIS2,ExpTrans1,ExpTrans2} such a process ca be transformed in a time homogenous Poisson process with parameter $1$, and  a simulation algorithm can be exploited. Let us denote with $T_j(t)$ the time at next occurrence of  reaction $R_j$  after time $t$, then
\begin{align} 
\Probab\Big( T_j(t)  \in [t, t+dt] \mid \xx \Big)  = a_j(\xx) L_j(\xiv(t))dt  + o(dt)
\end{align}
follows by equation (\ref{eq:Nj}) and higher order terms vanish by the usual hypothesis  that the reaction firings are locally independent, as in the derivation of equation (\ref{eq:cme}). Given the system to be in state $\xx$ at time $t$, the transformation
\begin{align} 
A_j(t, t + \tau)  = a_j(\xx) \int_{t}^{t+\tau}  L_j(\xiv(t'))dt' 
\end{align}
which is a monotonic (increasing)  function of $\tau$ is used to determine the putative time for $R_j$ to fire. Given a sequence $r_{j,k}$ of independent exponential random variables with mean $1$ for $j=1, \ldots, M$ and $k \in \N$, equation (\ref{eq:Nj}) implies that
\begin{align} 
N_j(t)  = \sum_{n=1}^\infty \mathbf{1}_{\{ \sum_{k=1}^n \eta_{j,k} \leq A_j(t, t_0)\}} \, .
 \end{align}
 This provides that, if the systems is in state $\XX(t)=\xx$, then the next  time for the next reaction  firing of $R_j$ is the smallest time $\tau>0$ such that 
 \begin{align} 
A_j(t+\tau, t)  = r
 \end{align}
 with $r \sim \Exp{1}$, and thus the next jump of the overall system is taken as the minimum among all possible times, that is by solving equality
 \begin{align}  \label{eq:SSAL-tau}
\sum_{j=1}^M A_j(t+\tau, t)  = \sum_{j=1}^M a_j(\xx)\int_{t}^{t+\tau}  L_j(\xiv(t'))dt' = r
 \end{align}
 with $r \sim \Exp{1}$. This holds because $\min\{T_j \mid j=1, \ldots, M\}$ is still exponential with parameter  $\sum_{j=1}^M A_j(t+\tau, t)$ and the jumps are independent. We remark that for a noise-free reaction $A_j(t+\tau, t) = \tau a_j(\xx)$, thus suggesting that the combination of noisy and noise-free reactions is straightforward. The index of the reaction to fire is instead a random variable following
 \begin{equation}\label{eq:SSAL-j}  
\Probab(j \mid \tau ; \xx,t) = \dfrac{a_j(\xx) L_j(\xiv(t+\tau))}{\sum_{i=1}^Ma_i(\xx) L_i(\xiv(t+\tau))} \, .
\end{equation}

\begin{algorithm}[!t]
\caption{\SSAL{} ($t_0$, $\xx_0$, $T$)}
\label{alg:SSAL}
\begin{algorithmic}[1]
\STATE set $\xx \gets \xx_0$ and $t\gets t_0$;
\WHILE{$t < T$} 
\STATE let $r_1$, $r_2 \sim U[0,1]$;
\STATE find $\tau$ by solving  $\sum_{j=1}^M A_j(t+\tau, t)  = \ln(r_1^{-1})$ and   $\xiv(t')$ for $t' \in [t, t+\tau]$;
\STATE let $j$ such that $\sum_{i=1}^{j-1} a_i(\xx, t+\tau) < r_2 \sum_{k=1}^M a_k(\xx,  t+\tau) \leq \sum_{i=1}^j  a_i(\xx, t+\tau)
$;
\STATE set $\xx \gets \xx + \nu_j$ and $t\gets t+\tau$;
 \ENDWHILE
\end{algorithmic}
\end{algorithm}

The \SSAL{} is  Algorithm \ref{alg:SSAL}.  Step $4$ is the (parallel) solution of
both equation (\ref{eq:SSAL-tau}) and  Langevin system (\ref{eq:noise}), step 
 $5$ samples values for $j$ according to equation  (\ref{eq:SSAL-j}). As far as step $4$ is concerned, it is worth nothing that  given $\XX(t)=\xx$ for any $\tau$ the Langevin equation (\ref{eq:noise}) depends only on $\xiv(t)$ and the constant $\xx$. 
To this extent,  a single trajectory  of the vectorial noise process in  $[t,t+\tau]$ is
\begin{align}\label{eq:noise-realization} 
\overline{\xi}_{t,\tau} = \{ (t, \xiv(t) )\}  
\cup  \{ (t^s, \xiv(t^s)) \mid t < t^s<t+\tau\} 
\cup \{ (t+\tau, \xiv(t+\tau) )\} \, .
\end{align}
This is a discretization of a continuous noise, thus inducing an approximation, but is in general the only possible approach.
To reduce approximation errors  the maximum size of the jump in the noise realization, i.e. the {\em noise granularity} $\Delta_s=t^{s+1} - t^s$, should be much smaller than the minimum autocorrelation time of the perturbing stochastic processes $L_j(\xiv(t))$.

 Finally,  the integral  in equation (\ref{eq:SSAL-tau}), evaluated in step $4$,  is a conventional Lebesgue integral since the perturbation $L_j(\xiv(t))$ is a colored stochastic process  \cite{stratonovich}. As an example, given $\overline{\xi}_{t,\tau}$  by  a linear interpolation scheme it holds
 \begin{equation}\label{integrapprox}
\integral{t}{t+\tau}L_j(\xiv(s)) ds \approx \sum_{t^s\in\overline{\xi}_{t,\tau}}\Delta_s \left( \min\{L_j^s,L_j^{s+1}\}+ \frac{1}{2}|L_j^s-L_j^{s+1}| \right)  
\end{equation}
where  $L_j^s = L_j(\xiv(t^s))$  for  $t^s\in\overline{\xi}_{t,\tau}$ and $\Delta_s$ the noise granularity.

\subsubsection*{Extension to non mass-action nonlinear kinetic laws} \label{sec:SQSSA}

Large networks with large chemical concentrations, i.e. characterized by deterministic behaviors, are amenable to significant simplifications by means of the well known {\em Quasi Steady State Approximation} (QSSA) \cite{Alon,Murray,GillMM,GillMM9}. 
The validity conditions underlying these assumptions are very well-known in the context of deterministic models \cite{Murray} , despite not much being known for the corresponding stochastic models. Recently,  Gillespie and coworkers \cite{GillMM}  showed that, in the classical  Michaelis-Menten Enzyme-Substrate-Product  network, a kind of {\em Stochastic QSSA} (SQSSA) may be applied as well, and that in such its limitations are identical to the deterministic QSSA.
Thus, it is of interest to consider SQSSAs also in our ``doubly stochastic" setting, even though possible  pitfalls may arise due to the presence of the extrinsic noises. As an example, in Section \ref{sec:enzyme} we will present the results of the numerical experiments similar to those of \cite{GillMM}, with the purpose of validating the SQSSA for noisy Michaelis-Menten enzymatic reactions.

Of course, in a SQSSA not only the propensities may be nonlinear function of state variables, but they may  depend nonlinearly also on the perturbations, so that instead of the elementary perturbed propensities we shall have generalized perturbed propensities of the form
$$ \alpha_j(\xx,\psiv(t)) $$
where $\psiv$ is a vector with elements $\psi_j = L_j(\xiv)$ for $j=1, \ldots, M$. This makes possible,  within the above outlined limitation for the applicability of the SQSSA,  to write a DCKE for these systems as
\begin{align} \label{eq:DCKE-QSSA}
\partial_t \Probab\Big((\xx,\xiv),t\mid\sigma\Big)&\approx - \sum_{j=1}^M \partial_{\zz_j} f_j(\xiv,\xx)\Probab\Big((\xx,\xiv),t\mid\sigma\Big) \\
&+\frac{1}{2} \sum_{i=1}^M \sum_{j=1}^M\partial_{\zz_i\zz_j}B_{i,j}(\xiv,\xx)\Probab\Big((\xx,\xi),t\mid\sigma\Big)\nonumber\\
&+ \sum_{j=1}^M  \Probab\Big((\xx -\nu_j,\xi),t\mid\sigma\Big) \alpha_j(\xx-\nu_j, \psiv(t))  \nonumber \\
& - \Probab\Big((\xx,\xi),t\mid\sigma\Big) \sum_{j=1}^M  \alpha_j(\xx,\psiv(t)) \nonumber \,.
\end{align}
As far as the simulation algorithm is concerned, it remains quite close to Algorithm \ref{alg:SSAL} provided that the jump times are sampled according to the following distribution
\begin{equation} \label{eq:QSSAnSSA-tau}
p(\tau \mid \xx,t) = \alpha_j(\xx,\psiv(t+\tau)) \exponential{-\integral{t_n}{t_n+\tau}\alpha_j(\xx,\psiv(k)) dk}   \, .
\end{equation}

%% file: sections/examples.tex
\section*{Results}

We performed \SSAL{}-based analysis of some simple biological networks, actually present in most complex realistic networks. We start by studying  the legitimacy of the stochastic Michaelis-Menten approximation  of when noise affects enzyme kinetics \cite{GillMM}. Then we study the role of the copresence of intrinsic and etrinsic bounded noises in a  in a model of enzymatic futile cycle \cite{Arkin} and, finally, in a bistable "toggle switch" model of gene expression \cite{Zhdanov, Zhdanov2}.  All the simulations have been performed by a \textsc{Java} implementation of the \SSAL{} (freely downloadable at \url {http://sites.google.com/site/giuliocaravagna/}) running on a  cluster of $15$ dual-core nodes with $2.0\, Ghz$ processor and $1\, GB$ of memory.

\paragraph{The Sine-Wiener noise \cite{SWiener}.} The bounded noise $\mu(t)$ that we use in our simulations is obtained by applying a bounded continuous function $h: \mathbb{R} \to  \mathbb{R}$ to a random walk $W$, i.e. $W^{\prime} (t) = \eta(t)$ with $\eta(t)$ a white noise\footnote{The underlying white-noise process $W(t)$ is generated at times $\{t_i \mid i \geq 0\}$ according to  the  recursive schema $W(t_{i+1}) = W(t_{i}) + r_i \sqrt{\Delta_W}$ with initial condition $W(t_0) = r_0$. Here $r_i \in {\cal N}(0,1)$  and $\Delta_W=t_{i+1} - t_i$ for $i \geq 0$ is its {\em discretization step}; it has to satisfy $\Delta_W \ll \tau$ and thus we chose to be $\Delta_W \approx \tau/100$. Notice that, as intuitive  the noise autocorrelation is expected to deeply impact on the simulation times.}. We have
\begin{align*}
\mu(t) & = h(W) 
\end{align*}
 so that for some $\beta \in \mathbb{R}$ it holds $ -\beta \leq h(W) \leq   \beta$.
The effect of the truncation of the tails induced by the approach here illustrated is that, due to this ``compression", the stationary probability densities of this class of processes satisfy 
$$\Probab(\mu= |\beta|)= +\infty .$$
Probably the best studied bounded stochastic process obtained by using this approach is the so-called Sine-Wiener noise \cite{SWiener}, that is 
\begin{equation}
\mu(t) = \beta \sin\left( \sqrt{ \dfrac{2}{\tau}}W(t) \right) 
\end{equation}
where $\beta$ is the {\em noise intensity} and $\tau$ is the {\em autocorrelation time}. The average and the variance of this noise are:
\begin{align*}\label{tbw}
\langle \mu(t) \rangle = 0 &&
\langle \mu(t)^2 \rangle = \beta^2/2
\end{align*}
and its autocorrelation is such that \cite{SWiener}
\[
\langle \mu(t) \mu(t+z) \rangle = \dfrac{\beta^2}{2} \exp\left(\dfrac{-z}{\tau}\right) \left[ 1-\exp\left(\dfrac{-4t}{\tau}\right) \right]\, .
\]
Note that, since we mean to use noises of the form $1 + \mu(t)$, i.e. the unitary-mean perturbations in equation (\ref{eq:prop-noise}), then the noise amplitude must be such that $ 0 \le \beta \le 1$. 

For this noise, the probability density is the following \cite{CaiLin}:
$$\Probab(\mu) = \frac{1}{\pi \sqrt{\beta^2 - \mu^2}} . $$
By these properties, this noise can be considered a realistic extension of the well-known symmetric dichotomous Markov noise \cite{Bena} $a(t)$,  whose stationary density is $\frac{1}{2}\left( \delta(a-\beta) + \delta(a+\beta) \right)$, for $a \in \mathbb{R}$ and  $\delta$ the Dirac delta function.

\subsection*{Enzyme kinetics} \label{sec:enzyme}

Enzyme-catalyzed reactions are fundamental for life, and in deterministic chemical kinetics theories are often conveniently represented in an approximated non mass-action form, the well-known Michaelis-Menten kinetics \cite{Alon,Murray,GillMM}. Such approximation of the exact mass-action model is based on a Quasi Steady-State Assumption (QSSA) \cite{Murray,GillMM9}, valid under some well known conditions.   In \cite{GillMM} it is studied the  legitimacy of the  Michaelis-Menten approximation of the  Enzyme-Substrate-Product {stochastic} reaction kinetics. Most important, it is shown that such a  stochastic approximation, i.e. the SQSSA previously discussed, obeys the same validity conditions for the  deterministic regime. This suggests the legitimacy of using - in case of low number of 
molecules -  the Gillespie algorithm not only for simulationg mass-action law kinetics, but more in general 
to simulate more complex rate laws, once  a simple conversion of deterministic Michaelis-Menten models is performed and provided - of course - that the SQSSA validity conditions are fulfilled. 
 
In this section we investigate numerically whether the Michaelis-Menten approximations and the stochastic results obtained in \cite{GillMM} still hold true in case that a bounded stochastic noise perturb the kinetic constants of the propensities of the exact mass-action law system Enzyme-Substrate-Product.
Let $E$ be an enzyme, $S$ a substrate and $P$ a product, the  {exact} mass-action model of enzymatic reactions comprises the following three reactions
\begin{align*}
E &+S \srewrites{c_1} ES && ES \srewrites{c_2} E + S 
&& ES \srewrites{c_3} E+P   
\end{align*}
where  $c_1$, $c_2$ and $c_3$ are the kinetic constants.  The network describes the transformation of substrate $S$ into product $P$, as driven by the formation of the enzyme-substrate complex $ES$, which is  reversible.

\begin{table}[t] 
\begin{center}
\[\small
\left(
\begin{matrix}
-1 & 1 & 1   \\
-1 & 1 & 0   \\
1 & -1 & -1   \\
0 & 0 & 1  
\end{matrix}
\right) 
\qquad 
\begin{aligned}
a_1&= c_1 E\cdot S   \\
a_2&= c_2 ES  \\
a_3&= c_3ES  
\end{aligned}
\qquad\qquad\qquad\qquad
\left(
\begin{matrix}
0    \\
-1  \\
0    \\
1  
\end{matrix}
\right) 
\qquad 
\begin{aligned}
a_1&= \dfrac{V_{max} S}{K_M + S}   
\end{aligned}
\]
\end{center}
\caption{Exact model (left) and Michaelis-Menten approximation (right) of enzymatic reactions:  the stoichiometry matrixes (rows in order $E$, $S$, $ES$, $P$) and  the  propensity functions.}
\label{table:enzyme}
\end{table}

\begin{figure}[!t]
\begin{center}$\scriptsize
\begin{array}{cc} 
(i) \; (E,S,ES,P)=(1,10,0,0) & (ii) \; (E,S,ES,P)=(100,10,0,0) \\
\includegraphics[width=7cm]{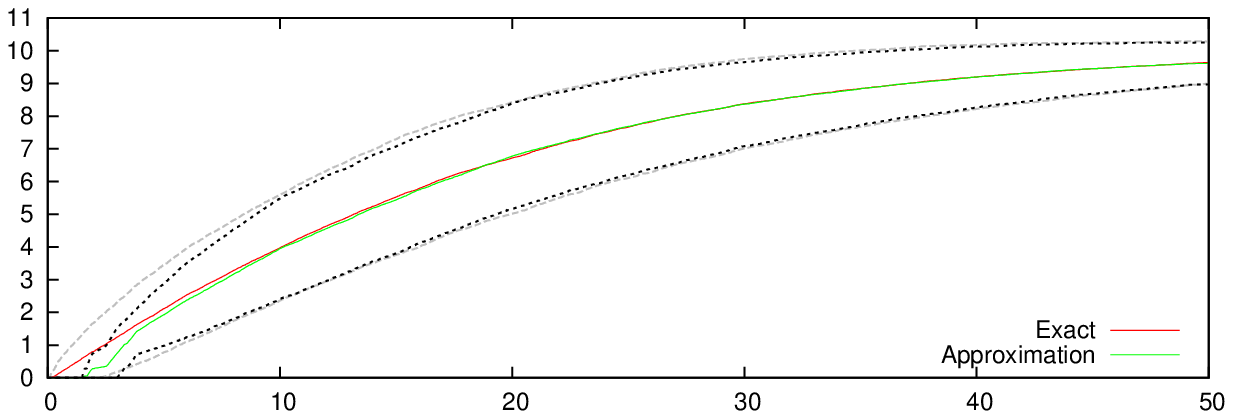} &
 \includegraphics[width=7cm]{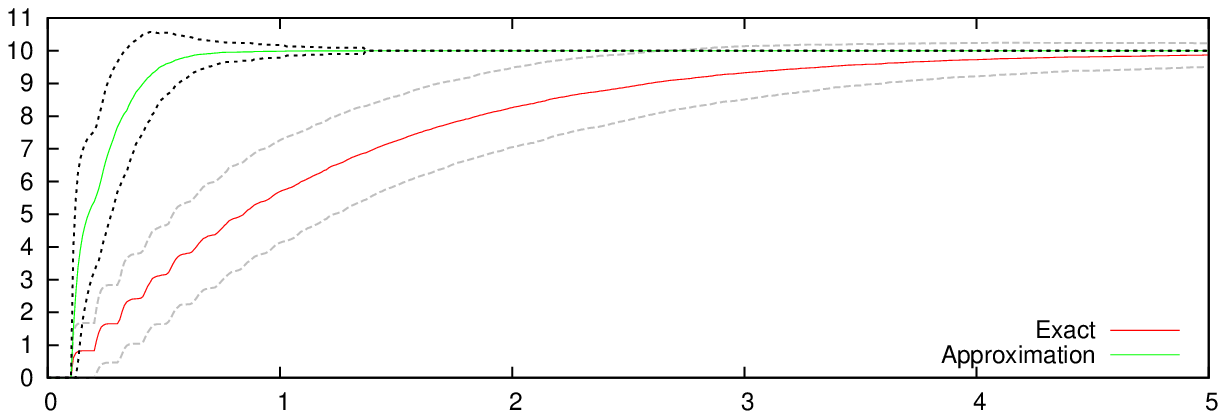}
  \end{array}$
\end{center}
 \caption{{\bf Noise-free Enzyme-Substrate-Product system.} Averages of $1000$ simulations for $P$, plotted with its standard deviation (dotted), for both exact and approximated Michaelis-Menten models. We have set  $c_1 = c_3 = 1$ and  $c_2 = 10$  and
   the initial configuration is $(i)$ in left $(E,S,ES,P)=(1,10,0,0)$ and  $(ii)$ in right 
   $(E,S,ES,P)=(100,10,0,0)$.}
\label{fig:enzyme-orig}
\end{figure}

\begin{figure}[t]
\begin{center}$\scriptsize
\begin{array}{cc} 
\text{(up) } \tau_{i}=1, \beta_i= 0.5  \quad\quad\quad  \text{(down) } \tau_{i}=1, \beta_i= 1 & 
\text{(up) } \tau_{i}=1, \beta_i= 0.5  \quad\quad\quad  \text{(down) } \tau_{i}=1, \beta_i= 1\\
 \includegraphics[width=7cm]{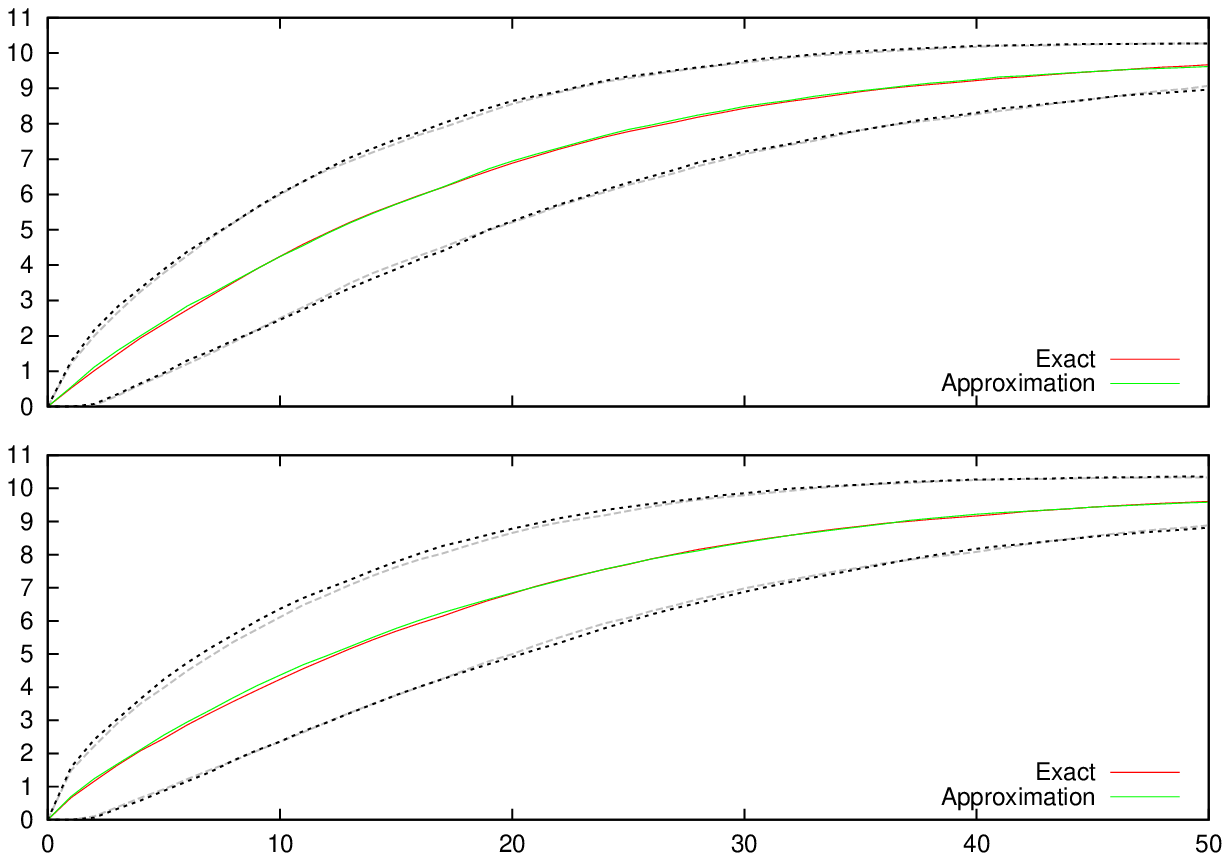} &
 \includegraphics[width=7cm]{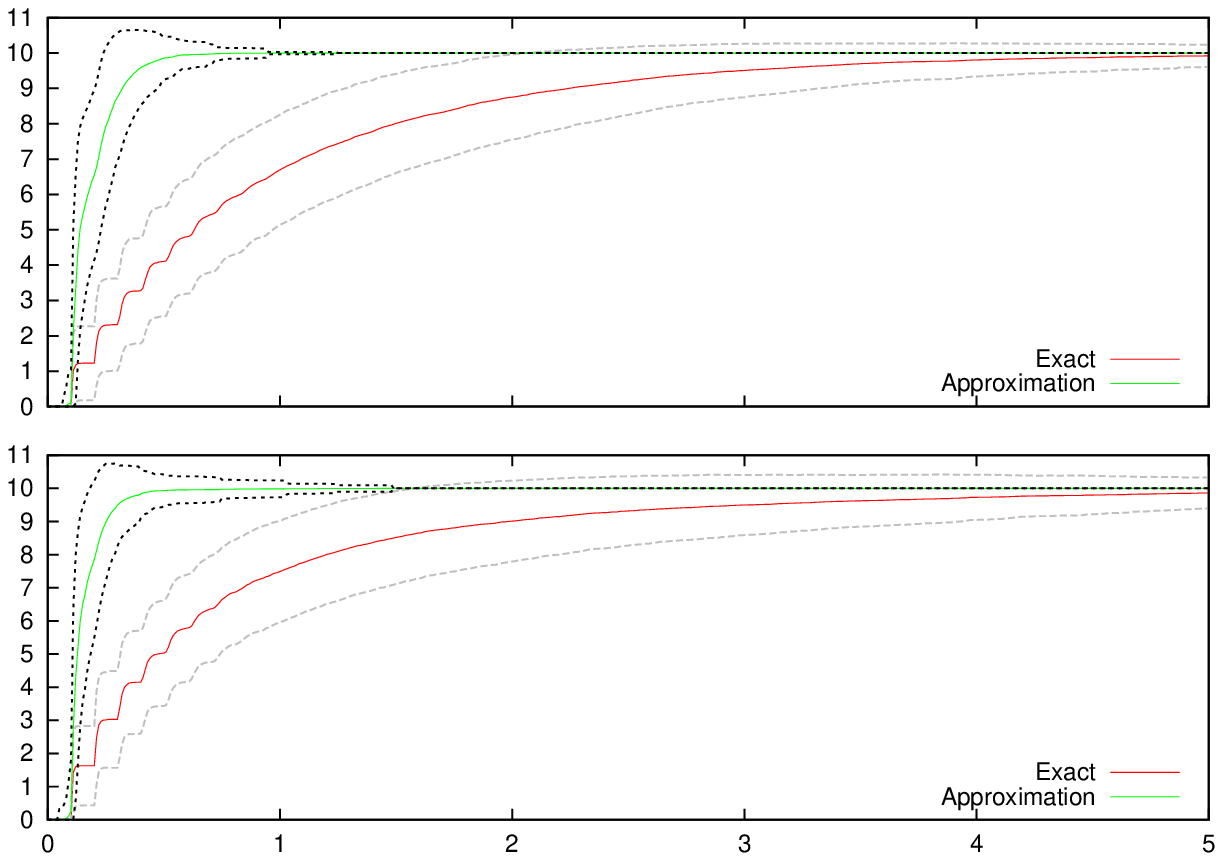} 
  \end{array}$
\end{center}
\begin{center}$\scriptsize
\begin{array}{cc} 
\text{(up) } \tau_{i}=5, \beta_i= 0.5  \quad\quad\quad  \text{(down) } \tau_{i}=5, \beta_i= 1  & 
\text{(up) } \tau_{i}=5, \beta_i= 0.5  \quad\quad\quad  \text{(down) } \tau_{i}=5, \beta_i= 1  \\
 \includegraphics[width=7cm]{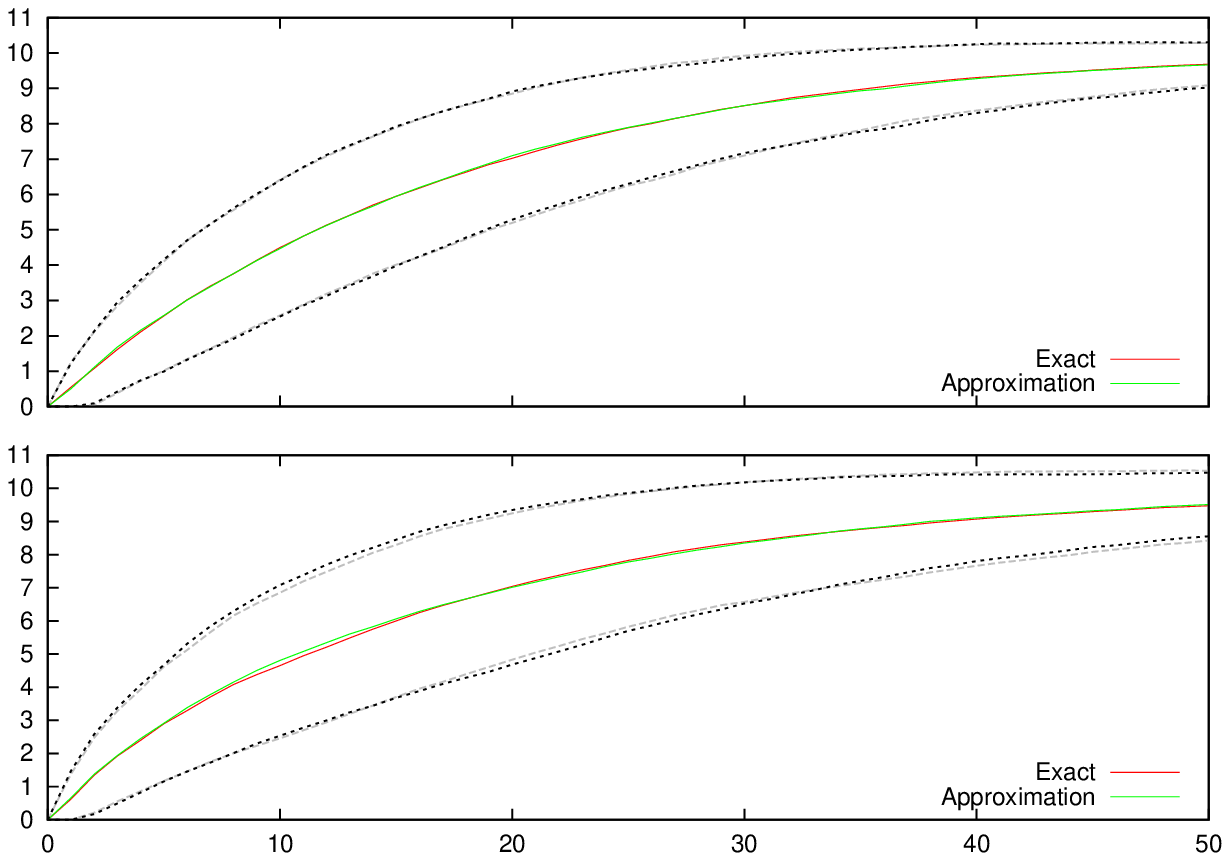}
 &
 \includegraphics[width=7cm]{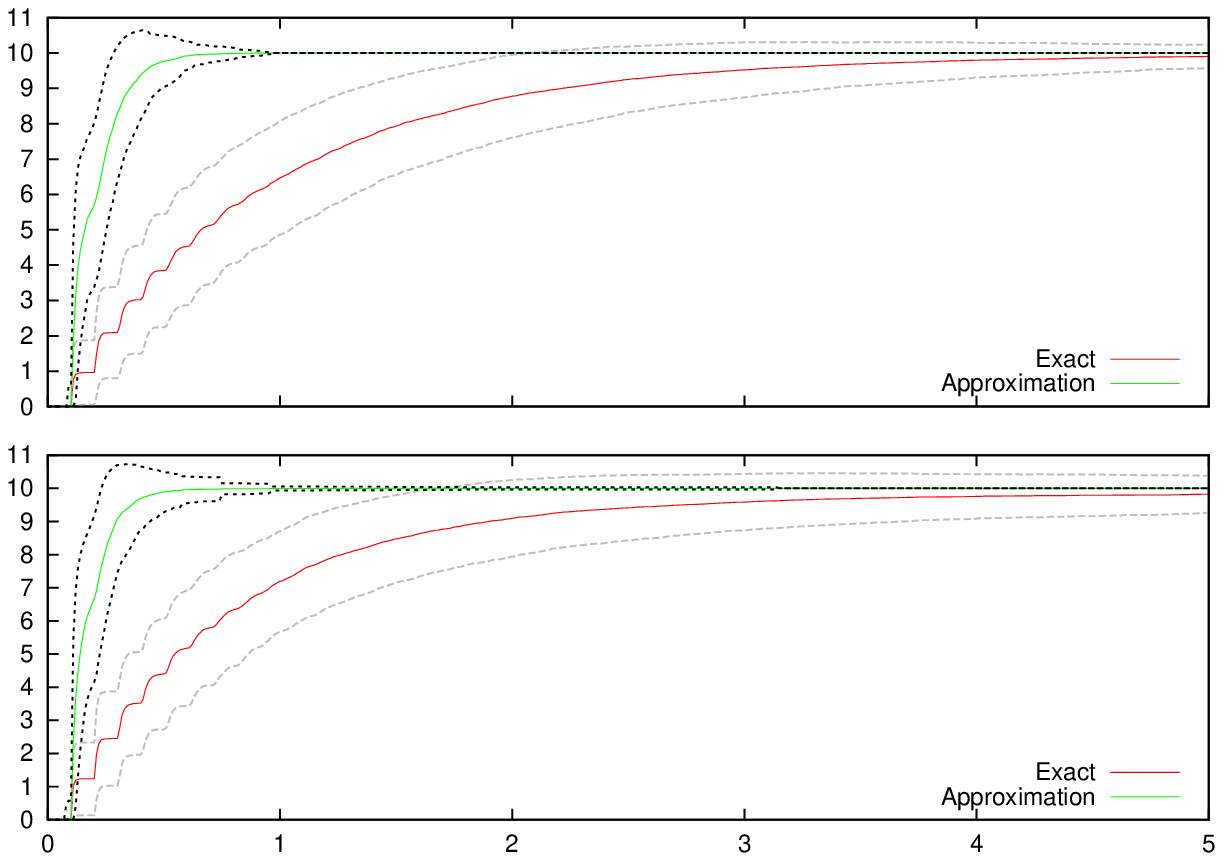}
 \end{array}$
\end{center}
 \caption{{\bf Stochastically perturbed Enzyme-Substrate-Product system.} Averages of $1000$ simulations for $P$, plotted with its standard deviation (dotted), for both exact and approximated Michaelis-Menten models. Parameters are as in Figure \ref{fig:enzyme-orig} $(i)$ in the top four panels, and as in  Figure \ref{fig:enzyme-orig} $(ii)$ in the bottom four. Independent Sine-Wiener noises  are present in all the reactions. For  $i=1,2,3$, in  the left column panels $\tau_{i}=1$,  in the right  $\tau_{i}=5$, in top  $\beta_i = 0.5$   and  in bottom $\beta_i = 1$.}
\label{fig:enzyme-T15NALL}
\end{figure}

The deterministic version of such reactions is
\begin{align}
S'&=-c_1S\cdot E+c_2ES && E'=-c_1S\cdot E+(c_2+c_3)ES  \\
ES'&=c_1S\cdot E-(c_2+c_3)ES&& P'=c_3ES    \, ,  \nonumber
\end{align}
where we write $S\cdot E$ to distinguish the multiplication of $E$ and $S$ from complex $ES$.
By the relations 
\begin{align}
E_T=E(t)+ES(t) && P(t) = P(t_0) + S(t_0) -(S(t) + ES(t))
\end{align}
a QSSA reduces to one the number of involved equations. Indeed, since $ES$ is in quasi-steady-state, i.e. $ES' = 0$, then 
\begin{align}
P'  \approx \dfrac{V_{max} S}{K_M + S} && \text{where}\quad  V_{max} = c_3E_T \qquad  \text{and}\quad  K_M = \dfrac{c_2+c_3}{c_1} \, .
\end{align}
Here $K_M$ is termed the {Michaelis-Menten} constant.
In practice, the QSSA permits to reduce the three-reactions model to the single-reaction model
\begin{align*}
S & \srewrites{} P 
\end{align*}
with non mass-action  non linear rate $(V_{max} S)/(K_M + S)$. In \cite{GillMM} the condition \begin{align} \label{eq:MM-cond}
E_T \ll S_0 + K_M
\end{align}
is used to determine a  region of the parameters space guaranteeing  the legitimacy of the Michaelis-Menten approximation. When condition (\ref{eq:MM-cond}) holds, a separation exists between the fast  pre-steady-state and the slower steady-state timescales \cite{GillMM9} and the solution of the  Michaelis-Menten 
approximation closely tracks the solution of the exact model on the slow 
timescale.

\begin{figure}[t]
\begin{center}$\scriptsize
\begin{array}{cc} 
\text{(up) } \tau_{1}=1, \beta_1= 0.5  \quad  \text{(down) } \tau_{1}=1, \beta_1= 1 & 
\text{(up) } \tau_{2}=1, \beta_2= 0.5  \quad  \text{(down) } \tau_{2}=1, \beta_2= 1 \\
 \includegraphics[width=5.0cm]{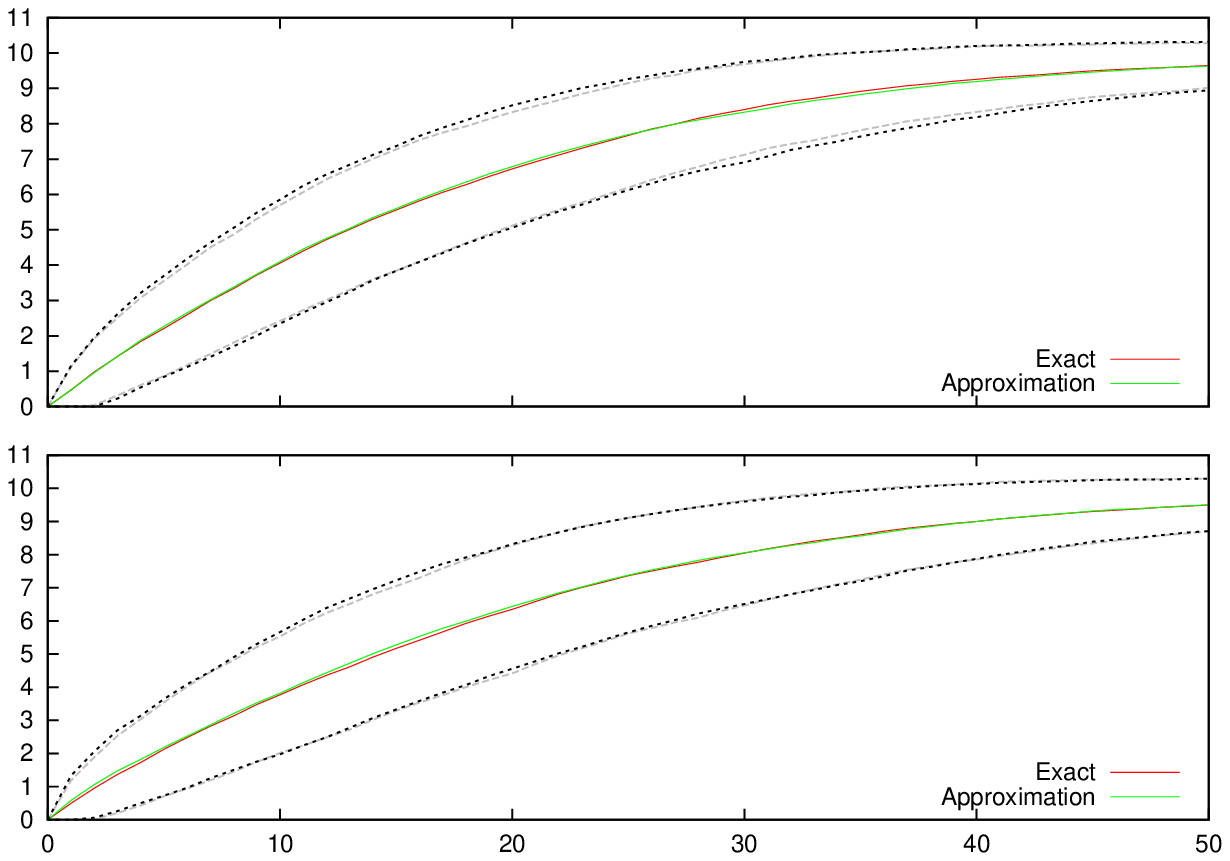} &
 \includegraphics[width=5.0cm]{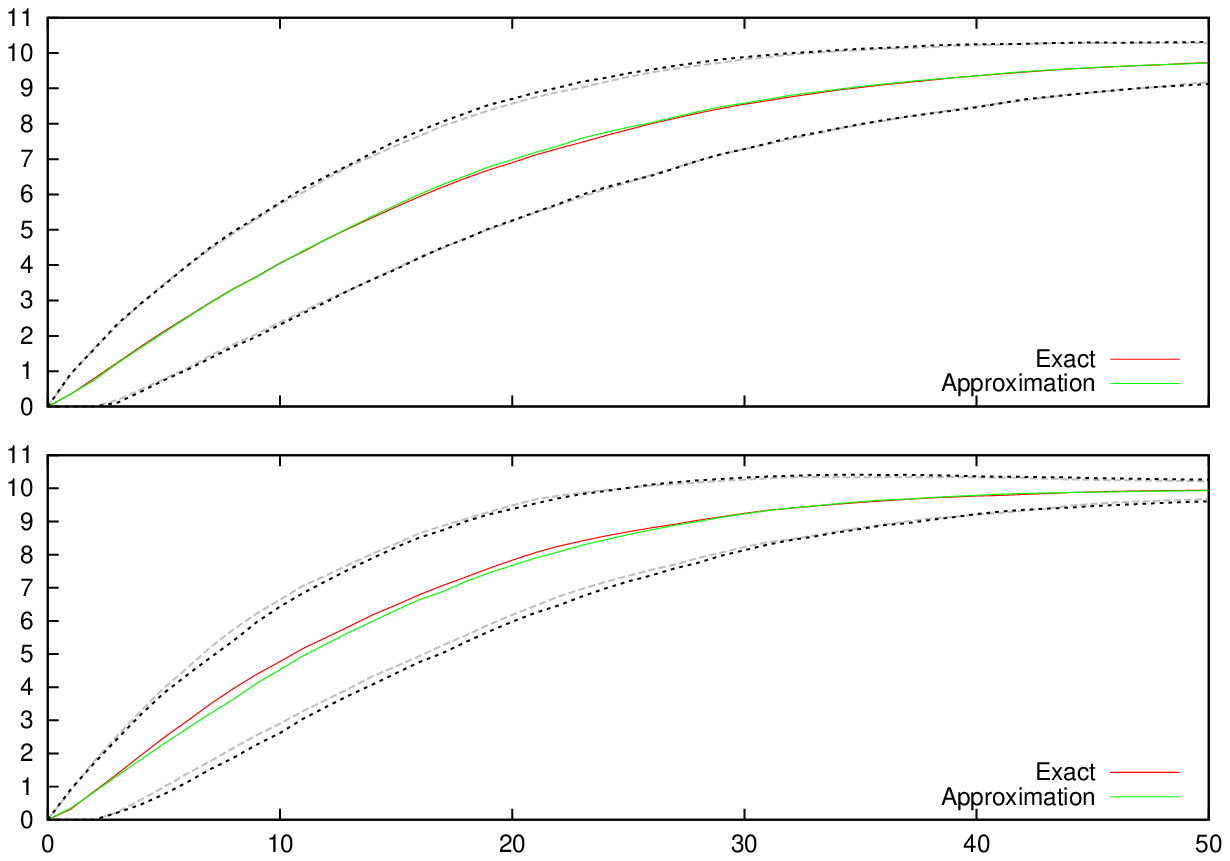} \\
\text{(up) } \tau_{3}=1, \beta_3= 0.5  \quad  \text{(down) } \tau_{3}=1, \beta_3= 1 & 
\text{(up) } \tau_{1}=5, \beta_1= 0.5  \quad  \text{(down) } \tau_{	1}=5, \beta_1= 1 \\
 \includegraphics[width=5.0cm]{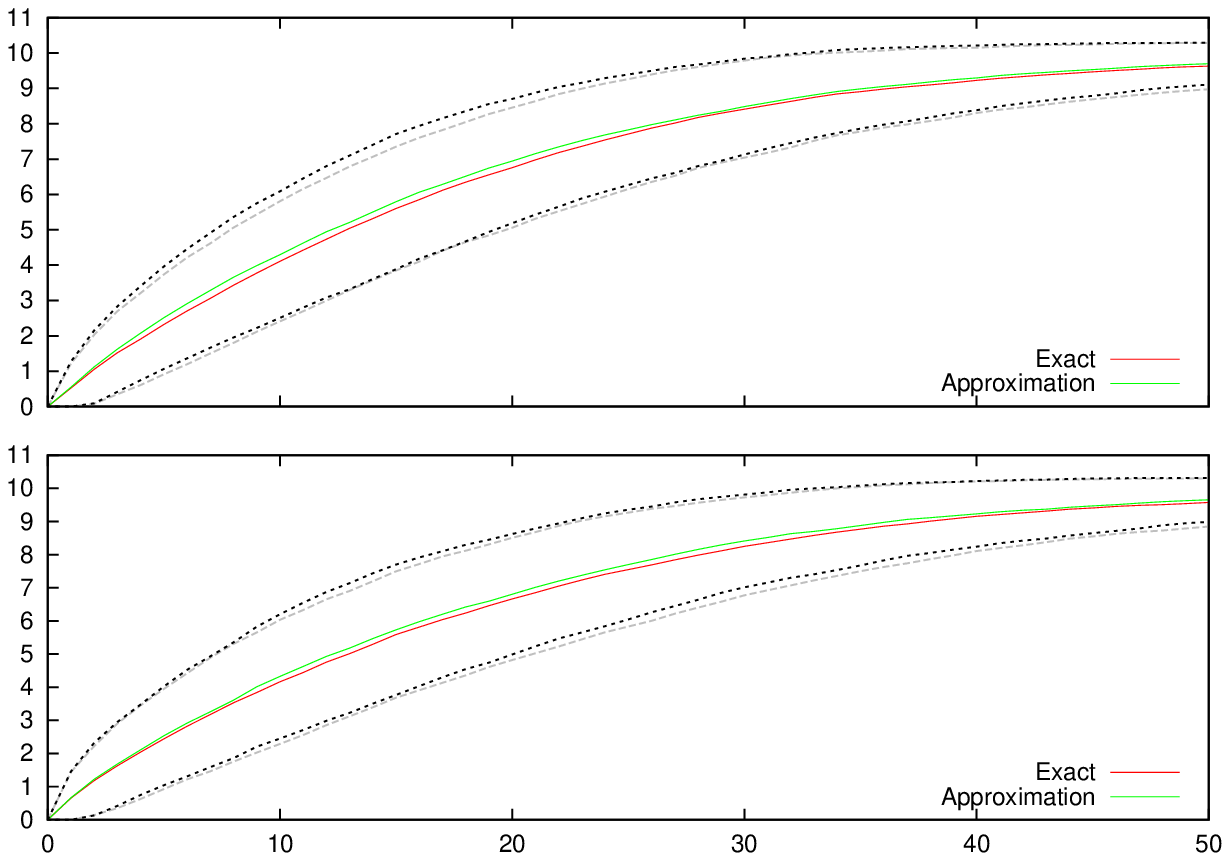} &
 \includegraphics[width=5.0cm]{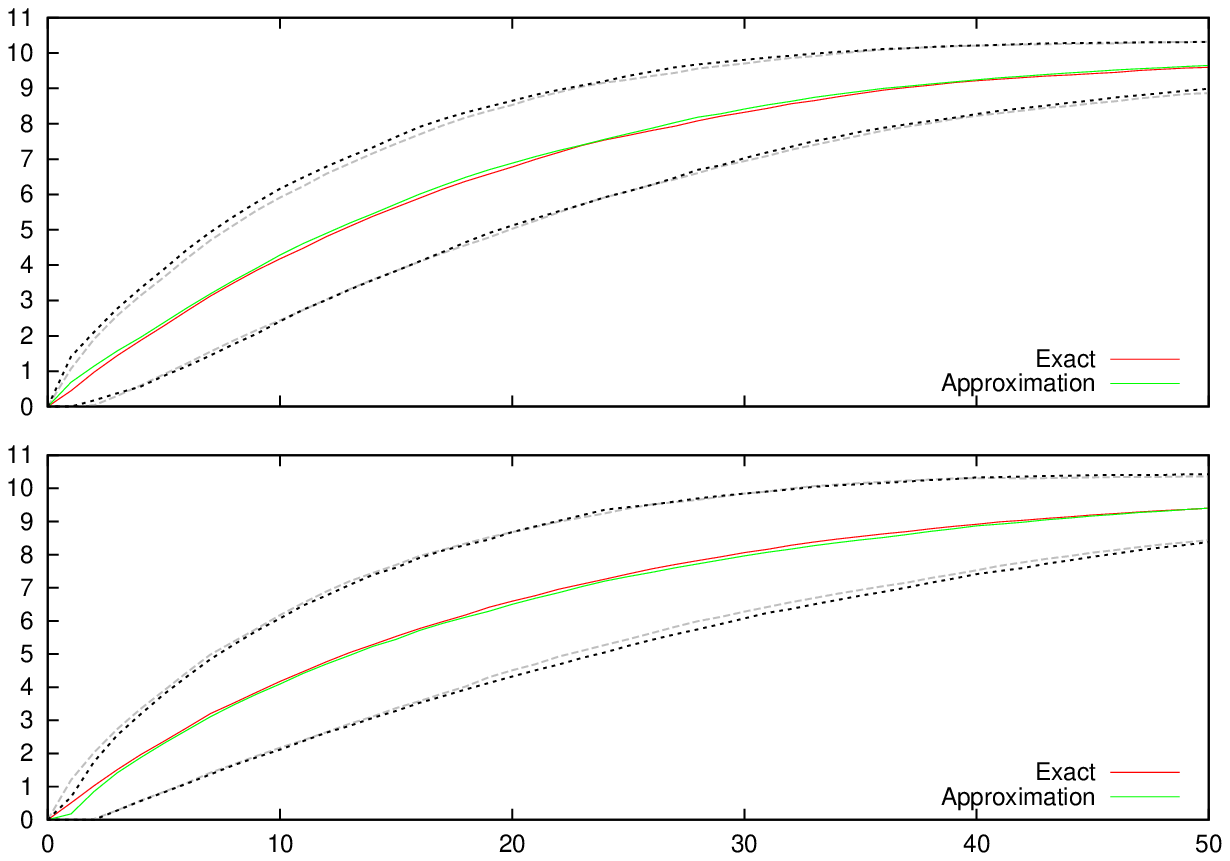}\\
 \text{(up) } \tau_{2}=5, \beta_2= 0.5  \quad  \text{(down) } \tau_{2}=5, \beta_2= 1 & 
\text{(up) } \tau_3=5, \beta_3= 0.5  \quad  \text{(down) } \tau_{3}=5, \beta_3= 1 \\
 \includegraphics[width=5.0cm]{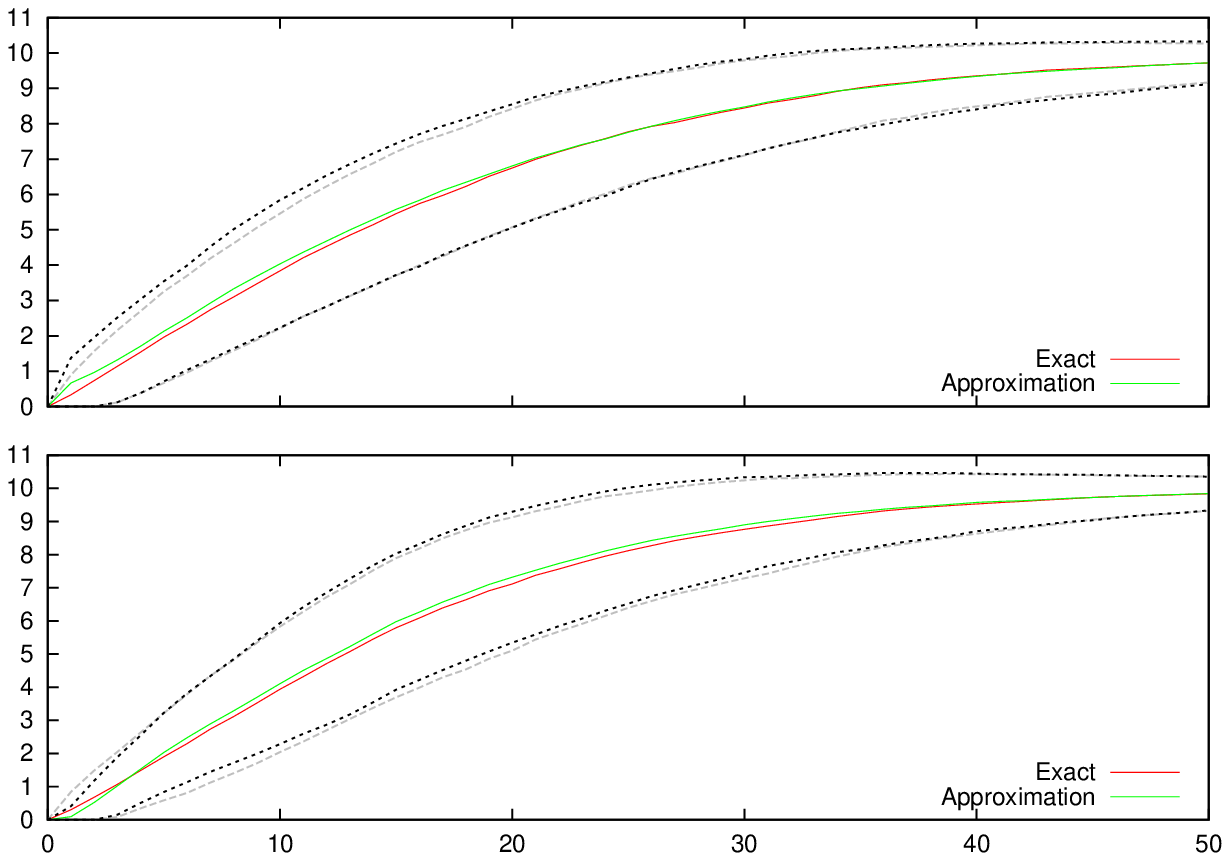} &
 \includegraphics[width=5.0cm]{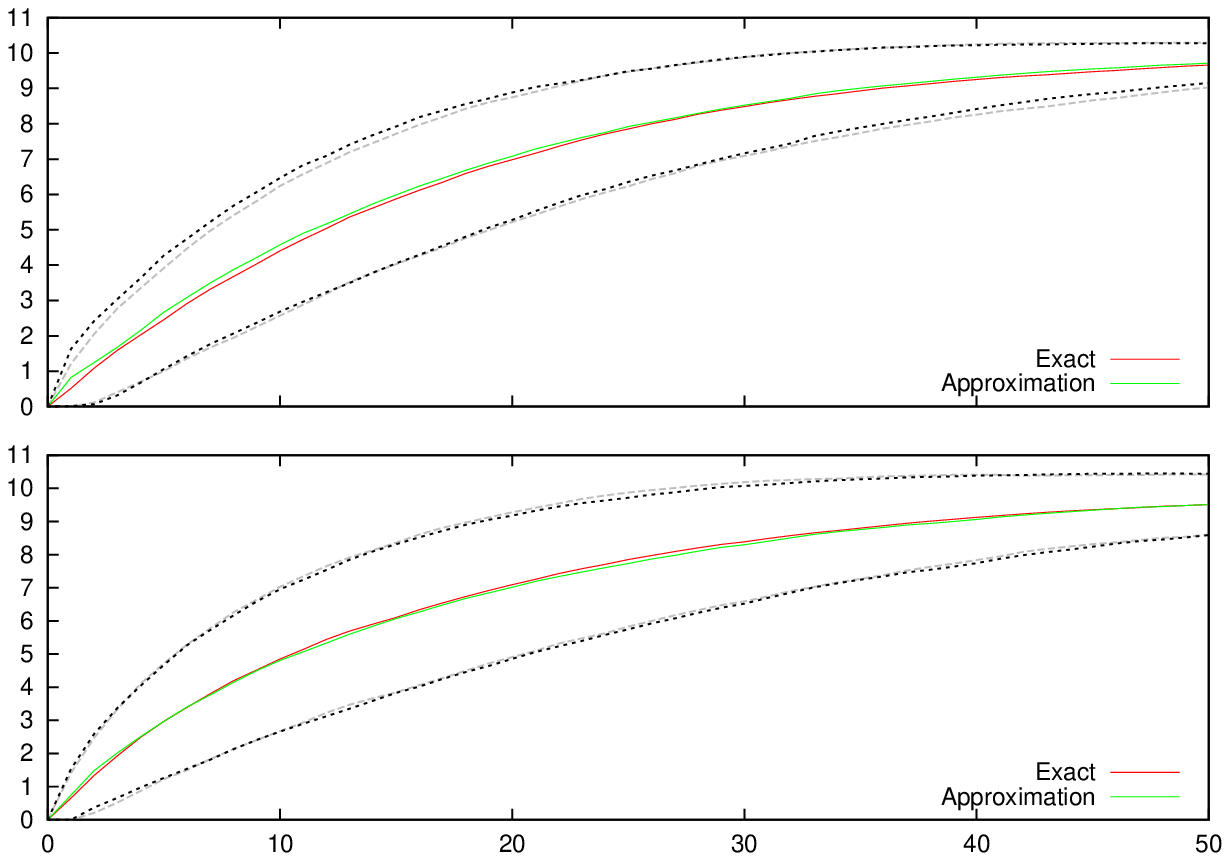}
\end{array}$
\end{center}
 \caption{{\bf Stochastically perturbed Enzyme-Substrate-Product system.} Averages of $1000$ simulations for $P$ with dotted standard deviation, for   both exact and approximated Michaelis-Menten models with parameters as in Figure \ref{fig:enzyme-orig} $(i)$. Here single noises  with two intensities  and different autocorrelations are used. The non-zero parameters are reported in top captions. }
\label{fig:enzyme-T15SingleNoise}
\end{figure}
\
Here we show that the same condition is sufficient to legitimate the  Michaelis-Menten approximation with bounded noises arbitrarily applied to any of the involved reactions. We start by recalling the result in \cite{GillMM} about the noise-free models  given in Table \ref{table:enzyme}. We considered two initial conditions: $(i)$ one with  $10$ copies of substrate, $1$ enzyme and $0$ complexes and products, and $(ii)$ one with  $10$ copies of substrate, $100$ enzyme and $0$ complexes and products. As in \cite{GillMM} we set   $c_1 = c_3 = 1$ and  $c_2 = 10$; notice that the parameters are  dimensionless and, more important, in $(i)$ they satisfy condition (\ref{eq:MM-cond}) since $E_T=1$ and $S_0+K_M=21$, in $(ii)$ no. In Figure \ref{fig:enzyme-orig} we reproduced the results in \cite{GillMM} for $(i)$ in right panel and $(ii)$ in left. As expected, in $(i)$ the approximation is valid on the slow time-scale, and not valid in the fast, i.e. for $t < 3$, in $(ii)$ it is not valid also in the slow time-scale.


If noises are considered the models in  Table \ref{table:enzyme} change accordingly. So, for instance when independent Sine-Wiener noises are applied to each reaction, the exact model becomes
\begin{gather*}
a_1= c_1\left[1 +  \beta_1 \sin\left( \sqrt{ {2}/{\tau_{1}}} W_1(t) \right) \right] E\cdot S    \qquad\qquad 
a_2= c_2 \left[1 +  \beta_2 \sin\left( \sqrt{{2}/{\tau_{2}}}W_2(t) \right) \right] ES  \\
a_3= c_3\left[1 +  \beta_3 \sin\left( \sqrt{{2}/{\tau_{3}}}W_3(t) \right) \right] ES  
\end{gather*}
and the Michaelis-Menten constant becomes the time-dependent function
\begin{align*}
K_M^*(t)= \dfrac{
c_2 \left[1 +  \beta_2 \sin\left( \sqrt{{2}/{\tau_{2}}}W_2(t) \right) \right]  +
c_3\left[1 +  \beta_3 \sin\left( \sqrt{{2}/{\tau_{3}}}W_3(t) \right) \right]
}
{
c_1\left[1 +  \beta_1 \sin\left( \sqrt{ {2}/{\tau_{1}}}W_1(t) \right) \right]
} \, .
\end{align*}
Notice that the nonlinear approximated propensity $a_1(t)$ is now time-dependent, and, moreover, it depends nonlinearly on the noises affecting the system. 

Thus condition (\ref{eq:MM-cond}) becomes time-dependent and we rephrase it to be
 \begin{align} \label{eq:MM-n-cond}
E_T \ll S_0 + \langle K_M^* (t) \rangle \, .
\end{align}
Note that if $\beta_1 >0$ then $\langle K_M^* (t) \rangle \neq K_M$, whereas if $\beta_1=0$ then $\langle K_M^* (t) \rangle = K_M$.

Each of the shown figures is the result of $1000$ simulations for model configuration  where the simulation times, which span from  few seconds to few minutes, depend on the noise correlation.
When the same system of Figure \ref{fig:enzyme-orig} $(i)$ is extended with these noises 
the approximation is still valid, as  shown in the top panels of Figure \ref{fig:enzyme-T15NALL}. In addition, the approximation is not valid when condition  (\ref{eq:MM-n-cond}) does not hold, as shown
in the bottom panels of Figure \ref{fig:enzyme-T15NALL}, as it was in Figure \ref{fig:enzyme-orig} $(ii)$. 
Notice that in  there we use two different noise correlations, i.e. $\tau_{i}=1$ in the left and $\tau_{i}=5$  for $i=1,2,3$ in the right column panels, thus mimicking noise sources with quite different charateristic kinetics. Also, we set two different noise intensities, i.e.
  $\beta_i = 0.5$ in top panels and $\beta_i = 1$ (maximum intensity)  in bottom panels, whereas all the other parameters are as in   Figure \ref{fig:enzyme-orig}. Summarizing, we get  a complete agreement between  enzymatic reactions with/without  noise, independently on the noise characteristics when it affects all of the reactions.

To strengthen this conclusion it becomes important to investigate whether  it still holds when  noises affects only a portion of the network and, also,  whether it holds  on the fast time-scale.  

As far as the number of noises is concerned, we  investigated various single-noise configurations in Figure \ref{fig:enzyme-T15SingleNoise}. In there we used  a single noise, i.e. two out of the three noises have $0$ intensity, with both low and high  intensities, i.e. $0.5$ and $1$. Also, in that figure we vary the noise correlation time as $\tau\in [1, 5]$. As hoped, the simulations show that the approximation is legitimate in the slow time-scale for all the various parameter configurations, thus independently on the presence of single or multiple noises. 

\begin{figure}[t]
\begin{center}$\scriptsize
\begin{array}{cc} 
\text{(up) } \tau_{1}=1, \beta_1= 0.5  \quad\quad\quad  \text{(down) } \tau_{1}=1, \beta_1= 1 & 
\text{(up) } \tau_{1}=5, \beta_1= 0.5  \quad\quad\quad  \text{(down) } \tau_{1}=5, \beta_1= 1 \\
 \includegraphics[width=7cm]{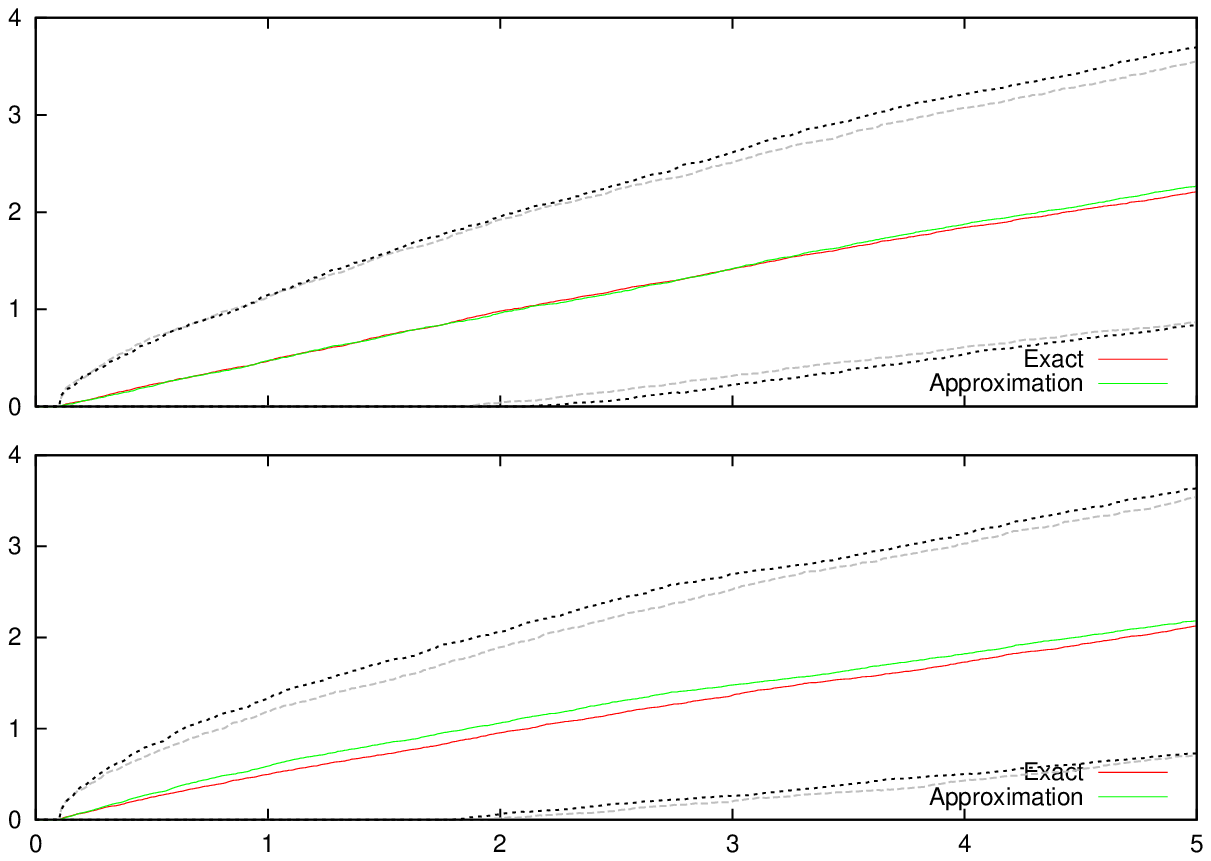} &
 \includegraphics[width=7cm]{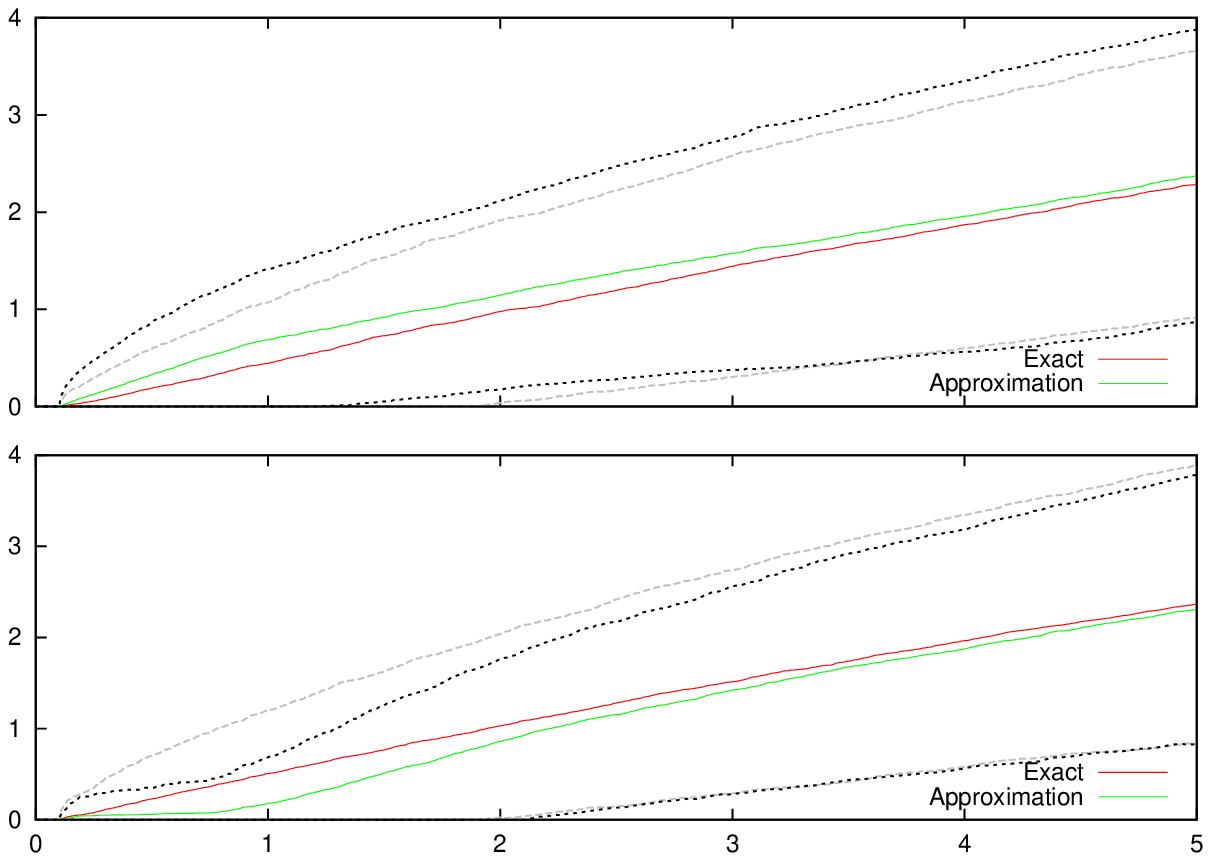} 
\end{array}$
\end{center}
 \caption{{\bf Stochastically perturbed Enzyme-Substrate-Product system.} Averages of $1000$ simulations for $P$, plotted with its standard deviation (dotted) in the fast time-scale $0\leq t \leq 5$  for both exact and approximated Michaelis-Menten models with parameters as in Figure \ref{fig:enzyme-orig} $(i)$.
 Here we use independent Sine-Wiener noises with $\tau_{1}=1$ in left and $\tau_{1}=5$ in right panels. 
 For every parameters configuration both low, i.e. $0.5$, and high, i.e. $1$, noise level intensities are used. }
\label{fig:enzyme-T15SingleNoise-fast-zoom}
\end{figure}

Finally, as far as the legitimacy of the approximation in the fast time-scale is concerned, i.e. $t \in [0,  5]$, our simulations show a result of interest: if the noise correlation is small compared to the reference fast time-scale and if single noises are considered  the noisy Michaelis-Menten approximation performs well also on the fast time-scale. We remark that this was not the case for the analogous noise-free scenario in Figure \ref{fig:enzyme-orig} $(i)$. In support of this  we plot in Figure \ref{fig:enzyme-T15SingleNoise-fast-zoom} the fast time-scale for $\tau_{i}=1$ and  $\tau_{i}=5$ for the single noise model with a noise in the enzyme-substrate complex formation, i.e. $\beta_2=\beta_3=0$. Similar evidences were found  in the configurations plotted in Figure \ref{fig:enzyme-T15SingleNoise} (not shown). 

\subsection*{Futile cycles}

\begin{table}[!t] 
\begin{center}
\[\small
\left(
\begin{matrix}
-1 & 1 & 0 & 0 & 0 &1  \\
1 & -1 & 1 & 0 & 0 &0  \\
1 & -1 & -1 & 0 & 0 & -1  \\
0 & 0 & 1 & -1 & 1 &0  \\
0 & 0 & 0 & -1 & 1 &1\\
0 & 0 & 0 & 1 & -1 &-1\\  
\end{matrix}
\right) 
\qquad \qquad
\begin{matrix}
a_1=  k_1 E\cdot S_0   & a_2= k_{-1} ES_0  \\
a_3= k_2 ES_0  & a_4= k_3 F\, S_1  \\
a_5= k_{-3} FS_1 & a_6= k_4 FS_1    
\end{matrix}
\]
\end{center}
\caption{{\bf Futile cycle model.} The noise-free enzymatic  futile cycle  \cite{Arkin}:  the stoichiometry matrix  (rows in order $S_0$, $E$, $ES_0$, $S_1$, $F$, $FS_1$) and  the  propensity functions.}
\label{table:arkin}
\end{table}

\begin{figure}[t]
\begin{center}$\scriptsize
\begin{array}{cc} 
\text{noise-free} & \text{futile cycle with $N$  } \\
 \includegraphics[width=7cm]{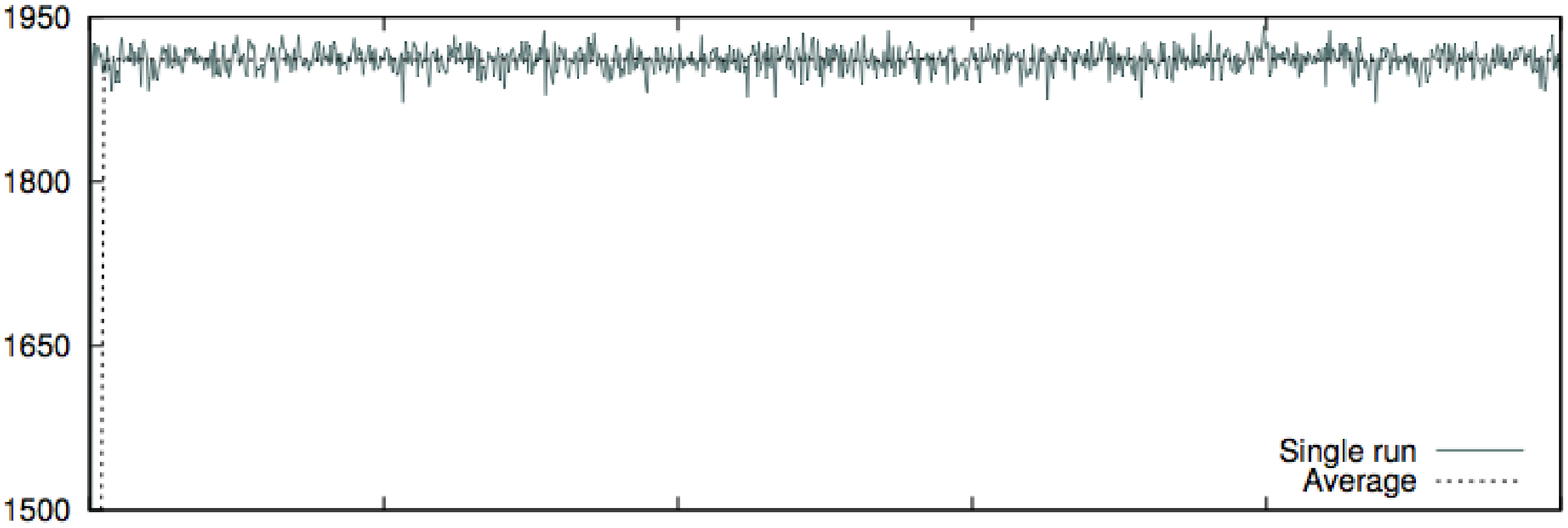}   &
  \includegraphics[width=7cm]{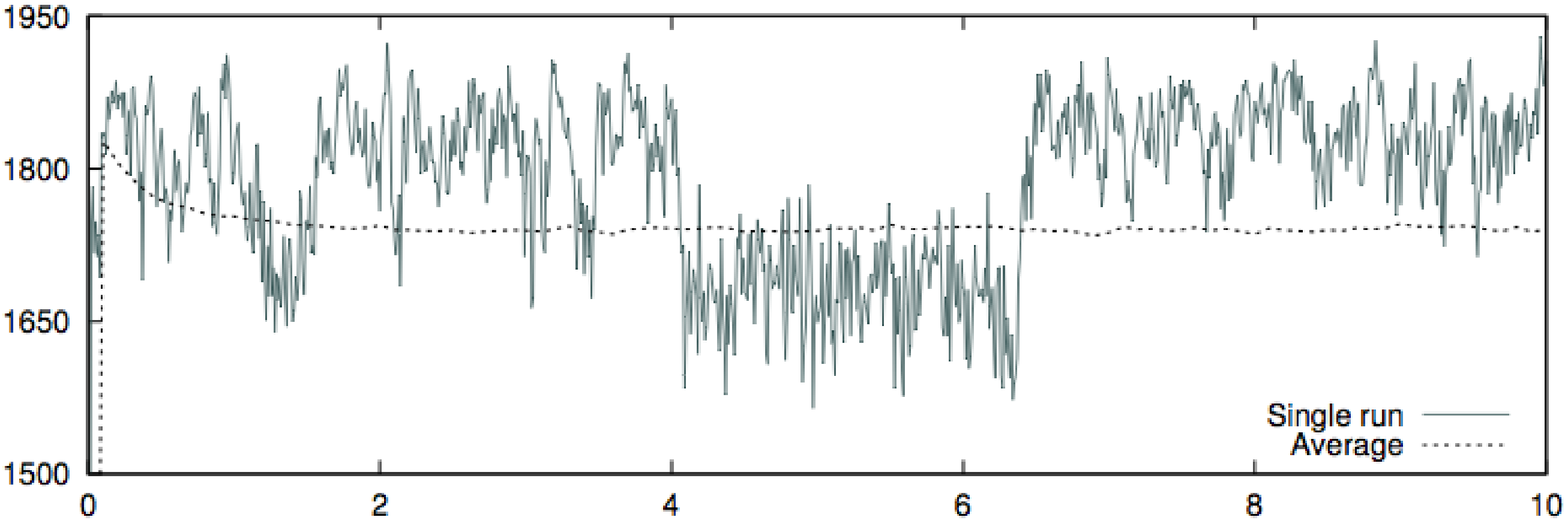}   
 \end{array}$
\end{center}
\begin{center}$\scriptsize
\begin{array}{cc} 
\text{cycle with bounded noise with $\tau=0.1$} & \text{cycle with bounded noise with $\tau=1$}\\
\text{(top) $\beta=0.5$  $\quad\qquad$ (bottom) $\beta=1$} &\text{(top) $\beta=0.5$  $\quad\quad$ (bottom) $\beta=1$}\\
 \includegraphics[width=7cm]{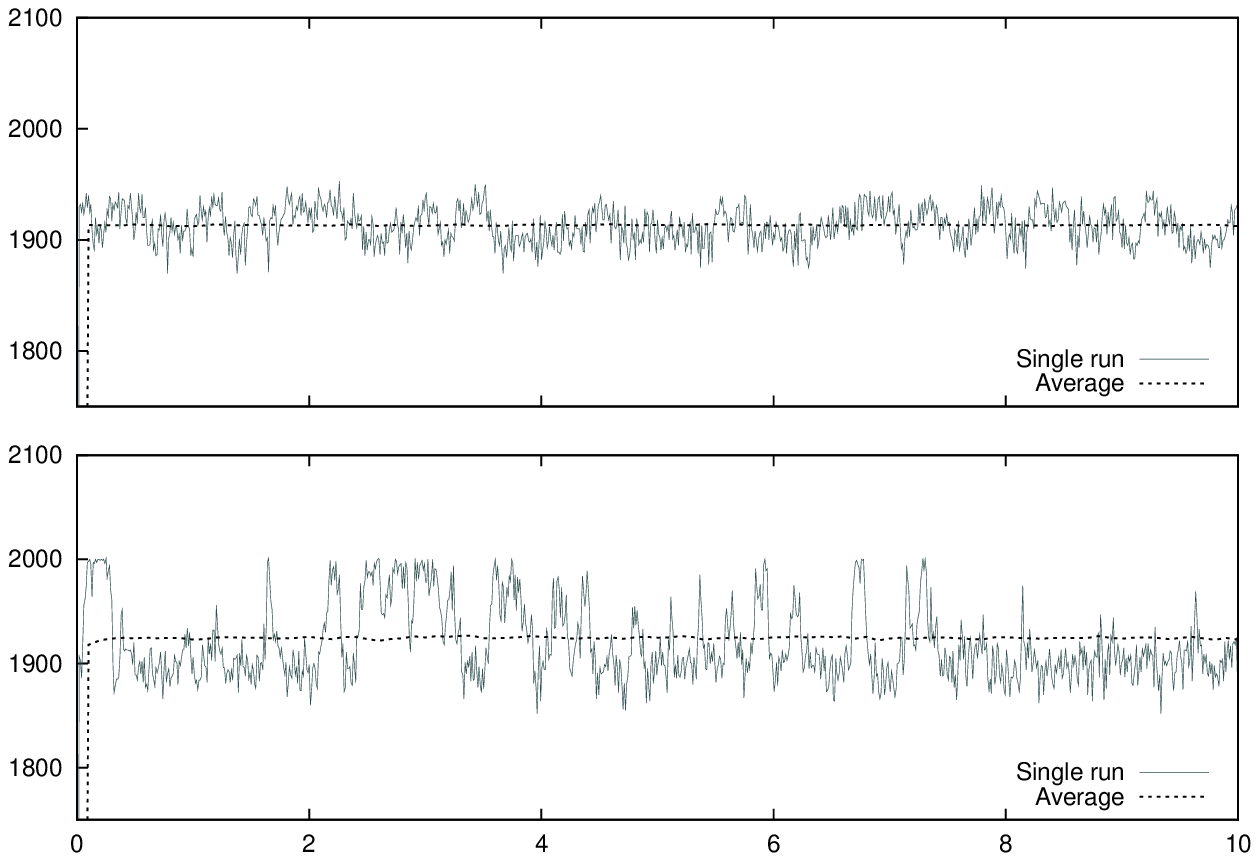} &
 \includegraphics[width=7cm]{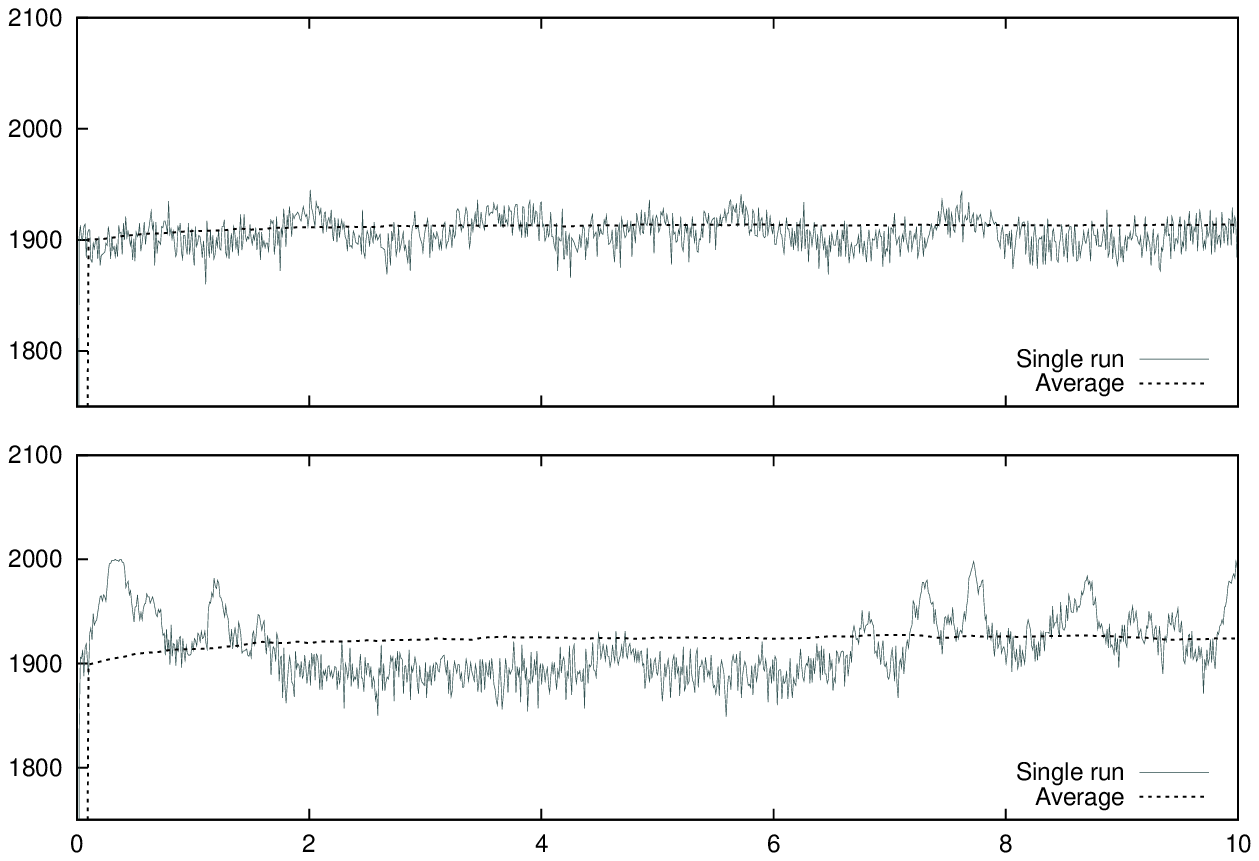}
   \end{array}$
\end{center}
 \caption{{\bf Stochastic models of futile cycles.} Single run and averages of $1000$ simulations for substrate $S_0$ of the futile cycle models. In top panels the noise-free (left) and the cycle unimodal noise as $N$ (right). In bottom panels the cycle with bounded noise and $\tau=0.1$ in (left) and $\tau=1$ in (right). In both cases in the top panel the noise intensity is $\beta=0.5$ (top) and $\beta=1$ (bottom). The initial configuration is always
   $(S_0, E, ES_0, S_1, F, FS_1,N) = (0, 20,0,2000,50,0,10)$ and the kinetic parameters are $k_1=40$,  $k_{-1}=k_2=10000$, $k_3=200$, $k_{-3}=100$, $k_4=5000$ for the noise-free and the bounded noise case, and $k_5=k_6=10$, $k_{-5}=5$ and $k_{-6}=0.2$ \cite{Arkin}.}
\label{fig:arkin-srun-averages}
\end{figure}

In  this section we consider a model of {\em futile cycle}, as the one computationally studied in \cite{Arkin}. The model consists of the following mass-action reactions
\begin{align*}
E + S_0 &\srewrites{k_1} ES_0 &&  ES_0 \srewrites{k_{-1}} E + S_0 && ES_0 \srewrites{k_{2}}  E + S_1 \\
F + S_1 &\srewrites{k_3} FS_1 &&  FS_1 \srewrites{k_{-3}} F + S_1 && FS_1 \srewrites{k_{4}}  F + S_0
\end{align*}
where $E$ and $F$ are enzymes, $S_0$ and $S_1$ substrate molecules, and $ES_0$ and $FS_1$ the complexes enzyme-substrate. 
Futile cycles are an unbiquitous class of biochemical reactions, acing as a motif in many signal transduction pathways \cite{Voet}. 

Experimental evidences related the presence of enzymatic cycles with bimodalities in stochastic chemical activities \cite{Ferrell}. As already seen in the previous section, Michaelis-Menten kinetics is not sufficient to describe such complex behaviors, and further enzymatic processes are often introduced to induce more complex behaviors. For instance, in deterministic models of enzymatic reactions  feedbacks are necessary to induce  bifurcations and oscillations.  Instead, in \cite{Arkin} it is shown that, although the determinstic version of the model has a unique and attractive equilibrium state, stochastic fluctuations in the total number  of $E$ molecules may induce a transition from a unimodal to a bimodal behaviour of the chemicals. This phenomenon was shown both by the analytical study of a continuous SDE model where the random fluctuations in the total number of enzyme $E$ (both free and as a complex with $S$) is modeled by means of a white gaussian noise on the one hand, and  in a totally stochastic setting on the other hand. In the latter case it was assumed the presence of a third molecule $N$ interacting with enzyme $E$ according to the following reactions
\begin{align*}
N + N &\srewrites{k_5} E + N &&  E + N \srewrites{k_{-5}} N + N \\
N &\srewrites{k_6} E &&  E \srewrites{k_{-6}}  N \, .
\end{align*}
By using $N$ the stochastic model results to be both quantitatively and qualitatively different from the deterministic equivalent.
These differences serve to confer additional functional modalities on the enzymatic futile cycle mechanism that include stochastic amplification and signaling, the characteristics of which depend on the noise.

Our aim here is to investigate whether bounded noises affecting the kinetic constant, and thus not modifying the topology of the futile cycle network, may as well induce transition to bimodality in the system behavior. To this aim, here we analyze three model configurations: $(i)$ the noise-free futile cycle, namely only the first six reactions, $(ii)$ the futile cycle with the external noise as given by $N$ and $(iii)$ the futile cycle with a bounded noise on the binding of $E$ and $S_0$, i.e. the formation of $ES_0$, and $N$ is absent.

In Table \ref{table:arkin} the noise-free futile cycle is given as a stoichiometry matrix and $6$ mass-action reactions. 
The model simulated in \cite{Arkin} is obtained by extending the model in the table with a stoichiometry matrix containing
$N$ and four more mass-action reactions. For the sake of shortening the presentation we omit to show them here. The model with a bounded noise in $a_1$ is obtained by defining
\[
a_1(t) = k_1 E\cdot S_0  \left[1 + \beta \sin\left( \sqrt{ \dfrac{2}{\tau}}W(t) \right) \right] \, .
\]

We simulated the above three models according to the initial condition used in \cite{Arkin} $(S_0, E, ES_0, S_1, F, FS_1) = (0, 20,0,2000,50,0)$ which is extended to account for $10$ initial molecules of $N$, when necessary. The kinetic parameters are dimensionless and defined as $k_1=40$,  $k_{-1}=k_2=10000$, $k_3=200$, $k_{-3}=100$, $k_4=5000$ for the noise-free and the bounded noise case, and $k_5=k_6=10$, $k_{-5}=5$ and $k_{-6}=0.2$ when the unimodal noise is considered \cite{Arkin}. Furthermore, when the bounded noise is considered the autocorrelation is chosen as $\tau\in[ k_5^{-1}, 1]=[0.1, 1]$ according to the highest rate of the reactions generating the unimodal noise.

\begin{figure}[!t]
\begin{center}$\scriptsize
\begin{array}{c} 
\text{noise-free  cycle} \hspace{1.50cm} \text{cycle with species $N$} \hspace{0.5cm} 
\text{noisy cycle $\tau=0.1$, $\beta=0.5$}\\
\includegraphics[width=10cm]{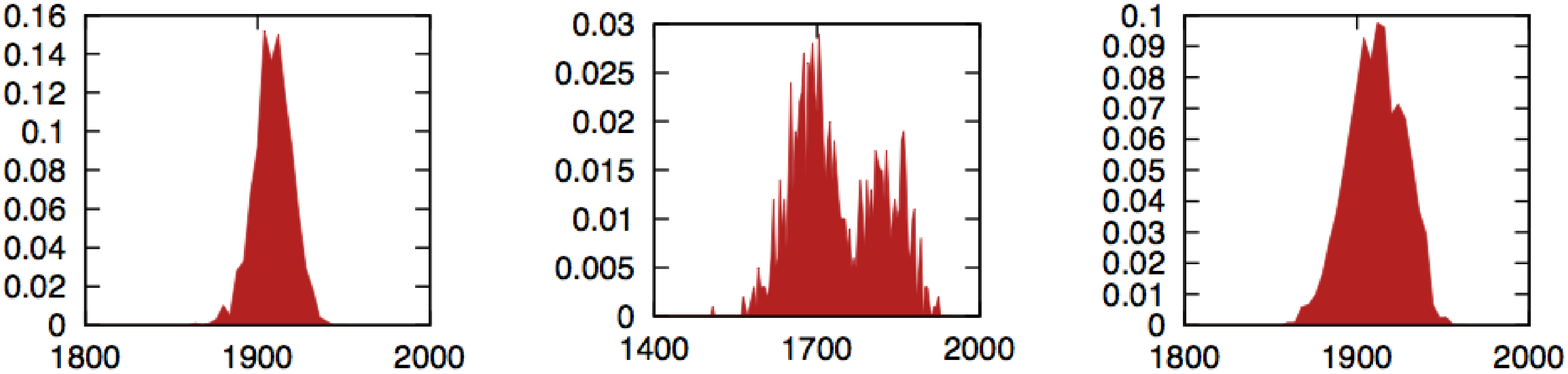} \\
\text{noisy cycle $\tau=0.1$, $\beta=1$ $\quad$  noisy cycle $\tau=1$, $\beta=0.5$ $\quad$  noisy cycle $\tau=1$, $\beta=1$}\\
  \includegraphics[width=10cm]{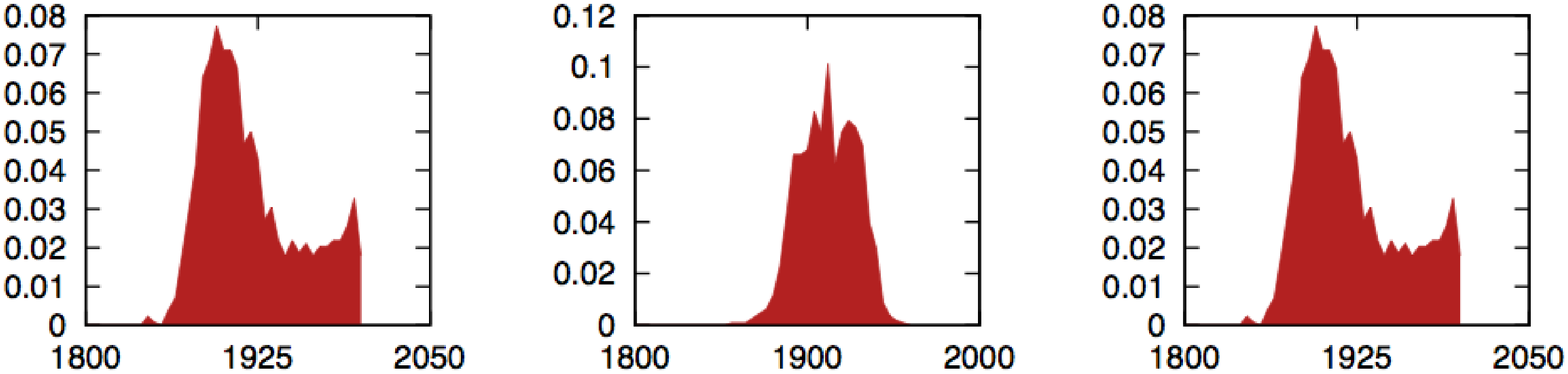} \\
\end{array}$
\end{center}
 \caption{{\bf Stochastic models of futile cycles.} Empirical probability density function for $S_0$ at $t = 10$ after  $1000$ simulations for the futile cycle models with  the  parameter configurations considered in Figure \ref{fig:arkin-srun-averages}.}
\label{fig:arkin-pdfs}
\end{figure}

In Figure \ref{fig:arkin-srun-averages}  a single run and averages of $1000$ simulations for the futile cycle models are shown. In this case  the simulation times span from $20 \div 30 \, min$ to $60 \div 80\, min$, thus making the choice of good parameters more crucial than in the other cases.
In Figure \ref{fig:arkin-srun-averages}   the substrate $S_0$ is plotted, and  $S_1$ behaves complementarily. In top panels the noise-free (top) and the cycle unimodal noise as $N$ (bottom). In bottom panels the cycle with bounded noise and  autocorrelation $\tau=0.1$ in (left) and $\tau=1$ in (right). In both cases in the top panel the noise intensity is $\beta=0.5$ (top) and $\beta=1$ (bottom). The initial configuration is always
   $(S_0, E, ES_0, S_1, F, FS_1,N) = (0, 20,0,2000,50,0,10)$ and the kinetic parameters are $k_1=40$,  $k_{-1}=k_2=10000$, $k_3=200$, $k_{-3}=100$, $k_4=5000$ for the noise-free and the bounded noise case, and $k_5=k_6=10$, $k_{-5}=5$ and $k_{-6}=0.2$ \cite{Arkin}. We also show in Figure \ref{fig:arkin-pdfs} the empirical probability density function for the concentration of $S_0$, i.e. $\Probab(\XX(t)=\xx)$ given the considered initial configuration, at $t\in\{2, 5, 7, 10\}$ after  $1000$ simulations for the futile cycle models with  the  parameter configurations considered in Figure \ref{fig:arkin-srun-averages}.
The analysis of such distributions outline that for the noise-free system the distributions are clearly unimodal, whereas for noisy futile cycle, in both cases, they are bi-modal. Moreover, it is  important to notice that the smallest peak of the distribution, i.e. the rightmost, has a bigger variance when $N$ is considered, rather than when a bounded noise is considered.

\subsection*{Bistable kinetics of gene expression}

\begin{table}[t] 
\begin{center}
\[\small
\left(
\begin{matrix}
1 & -1 & 0 & 0 & 0 &0 & 0 & 0  \\
0 & 0 & 1 & -1 & 0 &0 & 0 & 0  \\
0 & 0 & 0 & 0 & 1 & -1 & 0 & 0  \\
0 & 0 & 0 & 0 & 0 &0 & 1 & -1  
\end{matrix}
\right) 
\qquad \qquad
\begin{matrix}
a_1(t)= \xi(t) [{K}/({K+P_{2}})]^2  & a_2= \delta_R R_1  \\
a_3(t)= \xi(t) [{K}/({K+P_{1}})]^2  & a_4= \delta_R R_2  \\
a_5= \alpha_P R_1 & a_6= \delta_P P_1  \\
a_7=\alpha_P R_2  & a_8= \delta_P P_2  
\end{matrix}
\]
\end{center}
\caption{The bistable model of gene expression in \cite{Zhdanov}:  the stoichiometry matrix  (rows in order $R_1$, $R_2$, $P_1$, $P_2$) and  the  propensity functions.}
\label{table:zhdanov}
\end{table}

\begin{figure}[t]
\begin{center}$\scriptsize
\begin{array}{cc} 
\text{(single run) }\alpha=0.5  &   \text{(single run) } \alpha=1 \\
 \includegraphics[width=7cm]{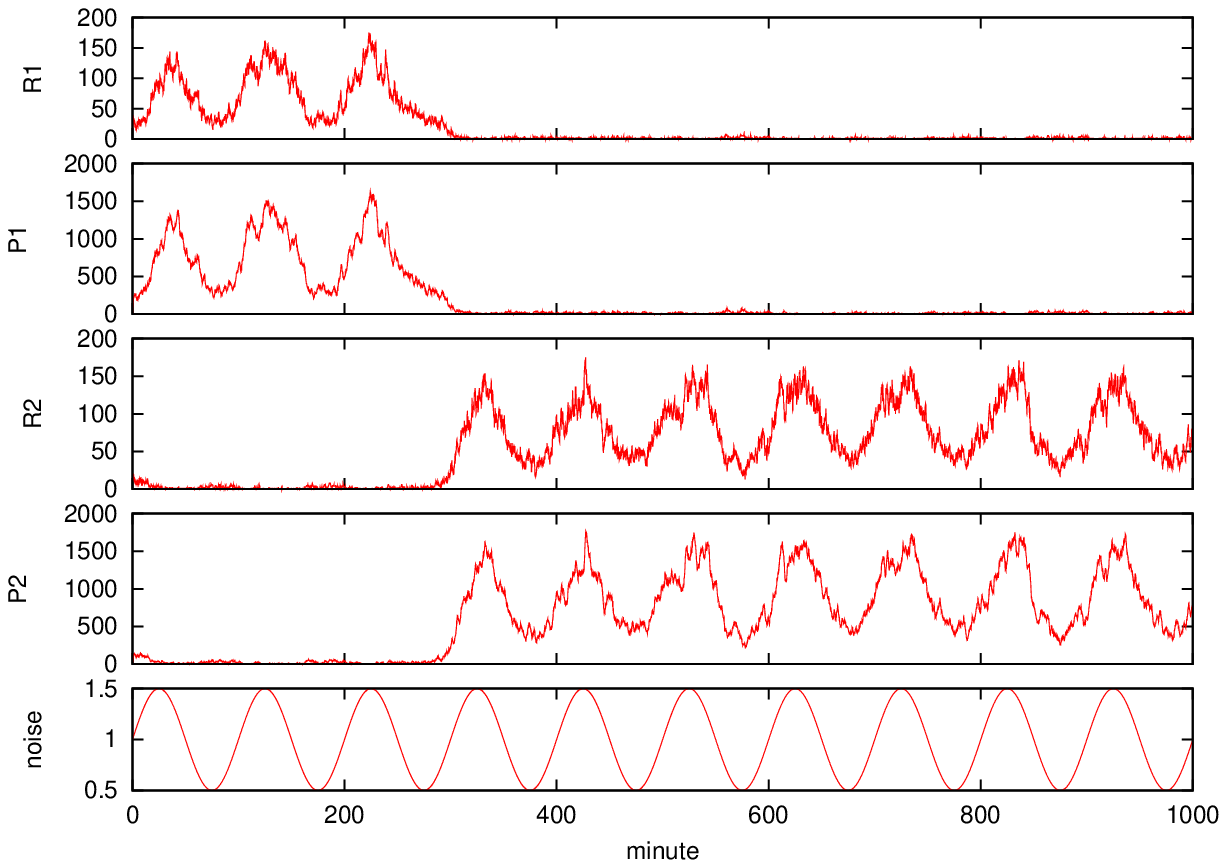} &
 \includegraphics[width=7cm]{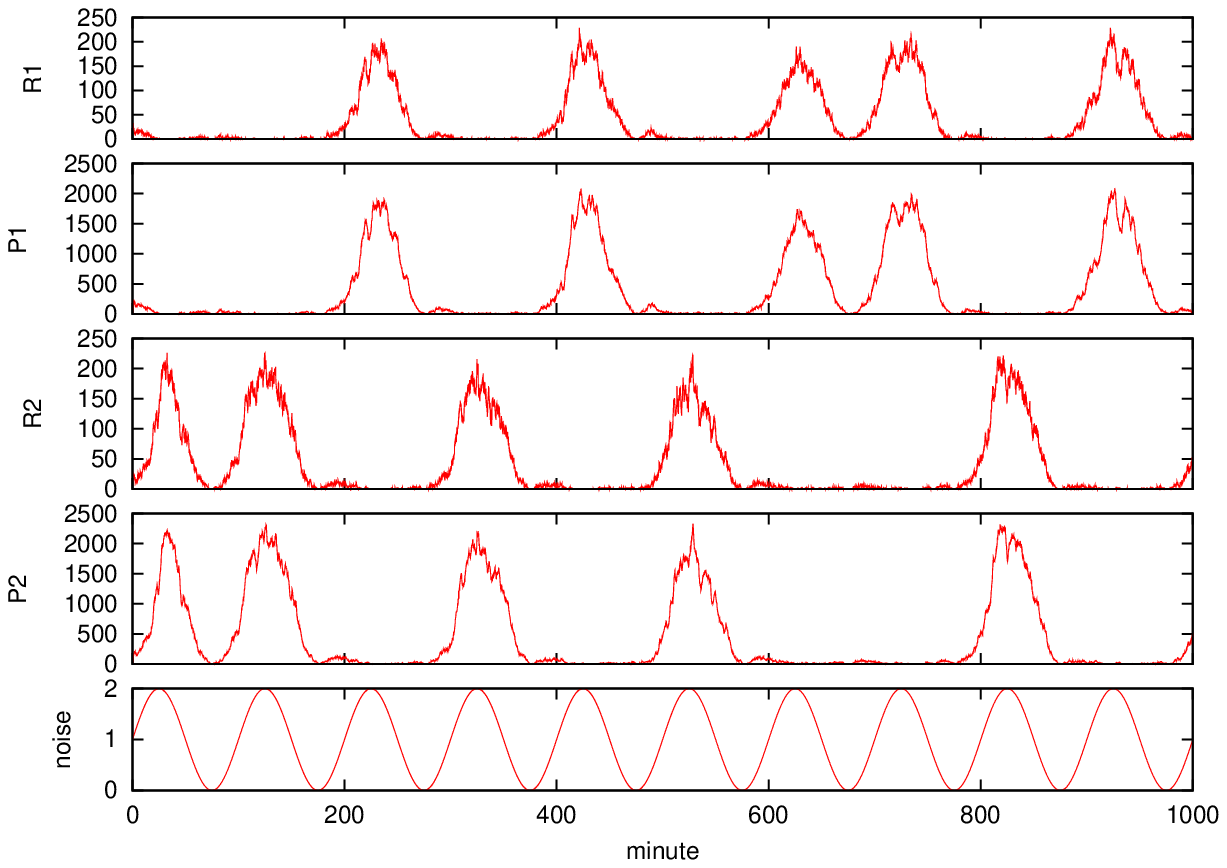}
   \end{array}$
\end{center}
\begin{center}$\scriptsize
\begin{array}{cc} 
\text{(averages) }\alpha=0.5  &   \text{(averages) } \alpha=1 \\
 \includegraphics[width=7cm]{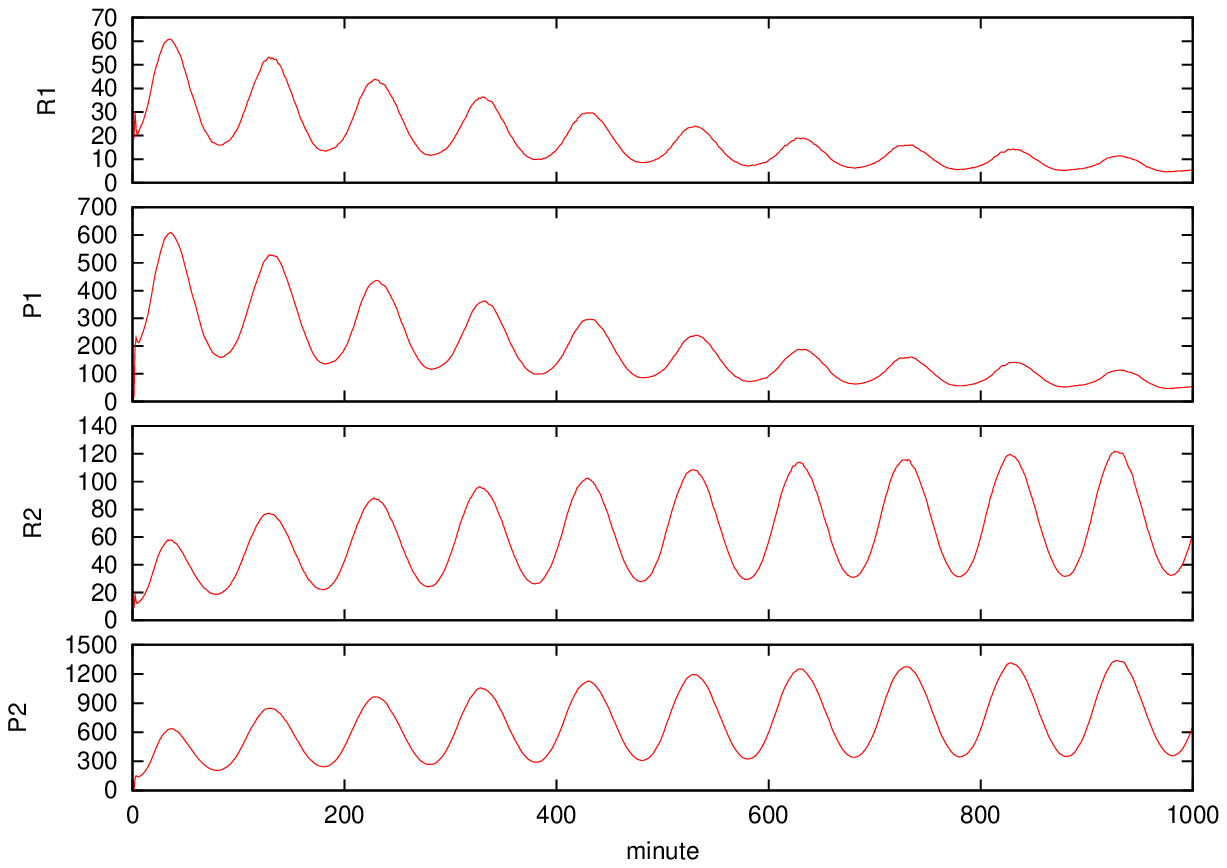}&
 \includegraphics[width=7cm]{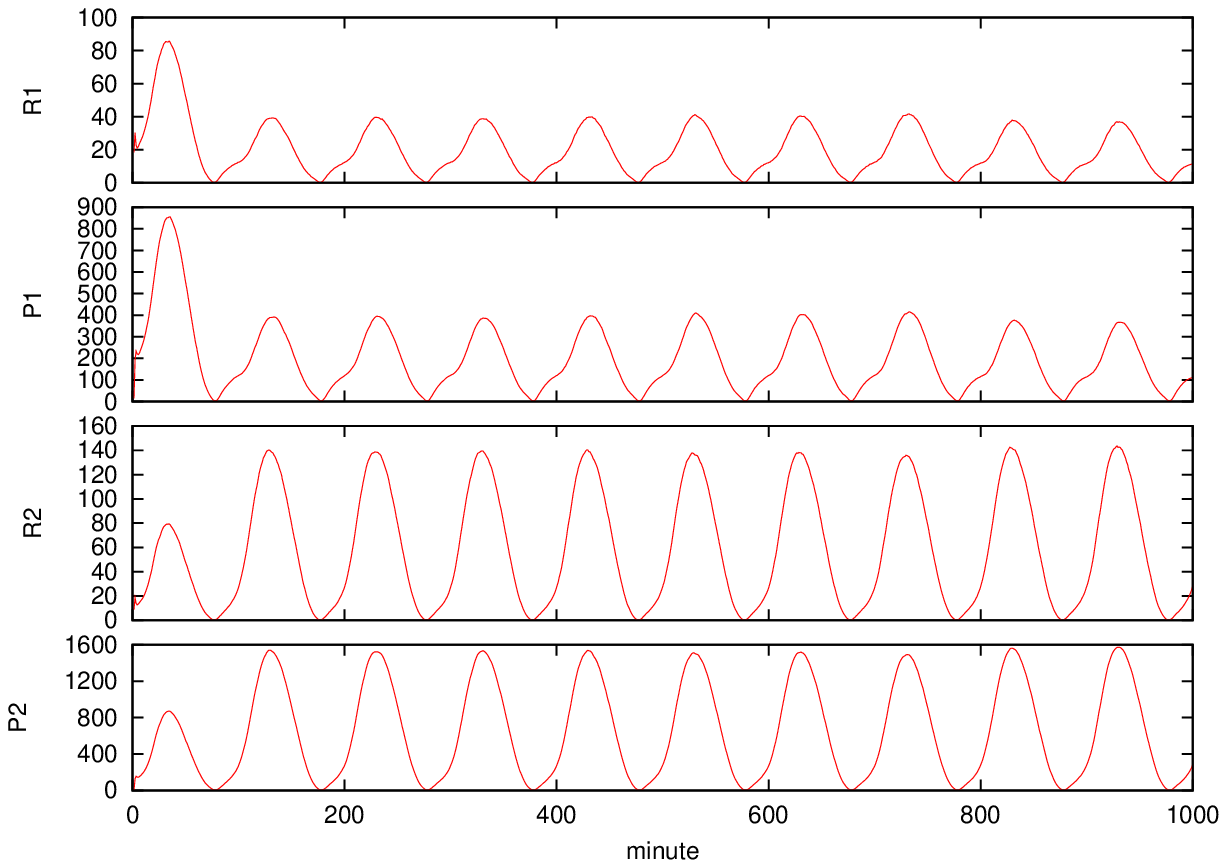}
 \end{array}$
\end{center}
 \caption{{\bf Periodically perturbed  toggle switch.} In  the top panels a single run for Zhdanov model (\ref{eq:zhmodel}) with $\alpha=0.5$ (left) and $\alpha=1$ (right). In bottom panels averages of $1000$ simulations.  In all cases $\alpha_R = 100\, min^{-1}$, $\alpha_P = 10\, min^{-1}$,  $\delta_R=\delta_P=1\, min^{-1}$, $K=100$ and $\tau=100\, min^{-1}$ and the initial configuration is $(R_1,P_1,R_2,P_2)=(10,0,0,0)$. The populations and the noise are plotted for the single run.}
\label{fig:zhdanov-orig}
\end{figure}

\begin{figure}[t]
\begin{center}$\scriptsize
\begin{array}{cc} 
t=900, \alpha=0.5 & t=900,  \alpha=1 \\
 \includegraphics[width=5.0cm]{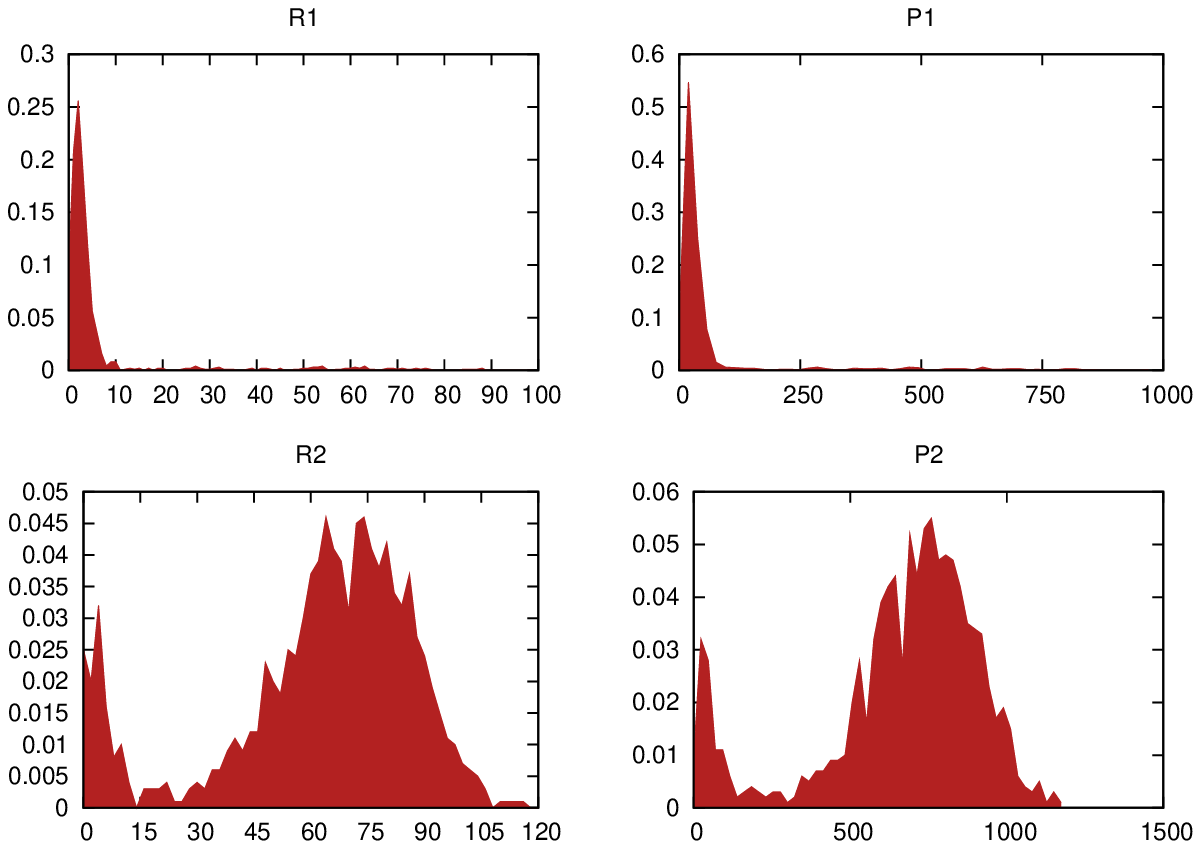}&
 \includegraphics[width=5.0cm]{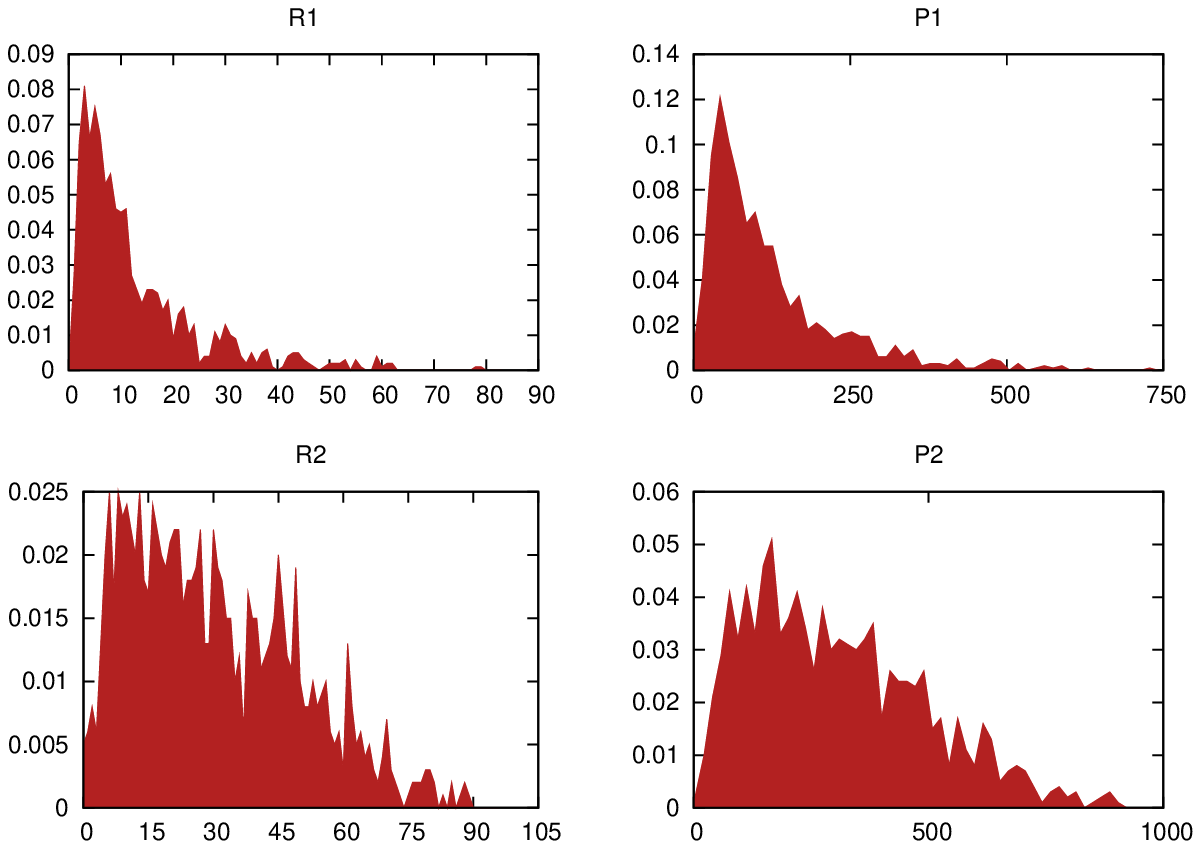}
 \end{array}$
\end{center}
\begin{center}$\scriptsize
\begin{array}{cc} 
t=950,  \alpha=0.5 & t=950,  \alpha=1 \\
 \includegraphics[width=5.0cm]{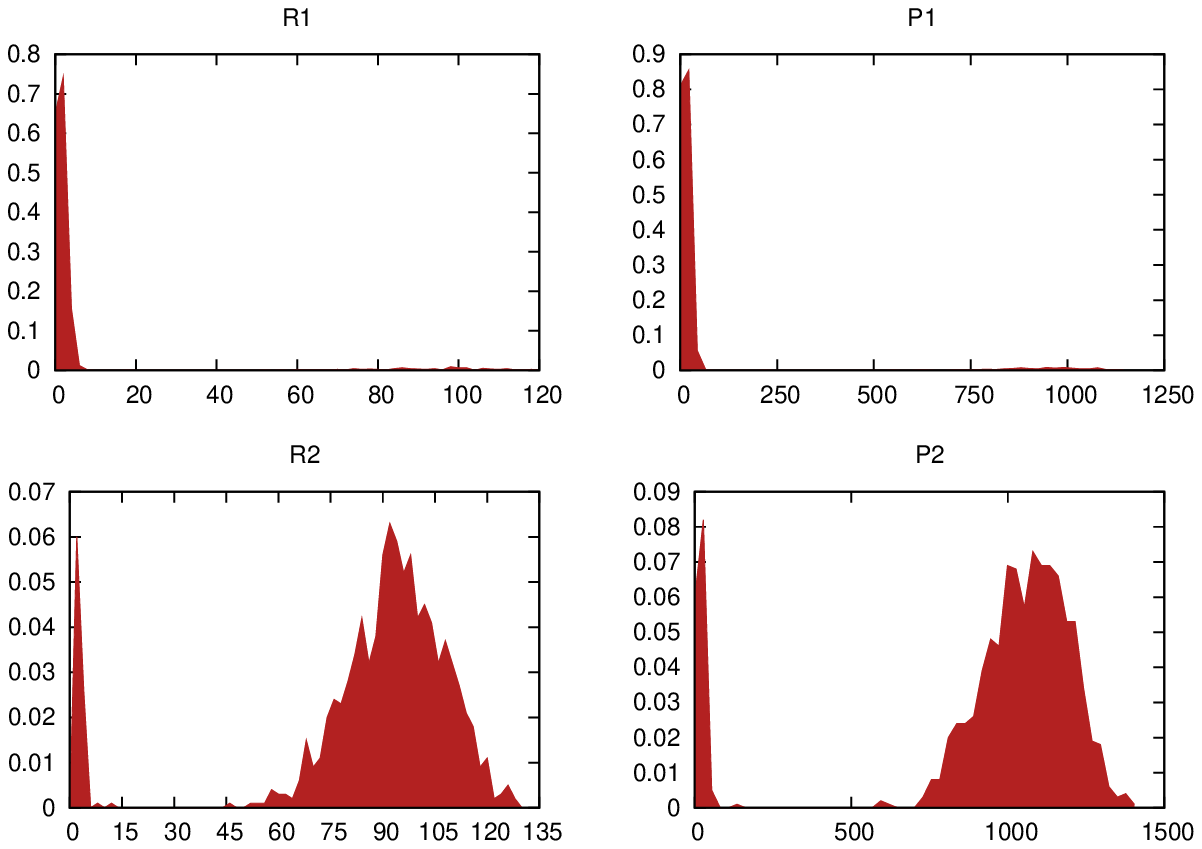}&
 \includegraphics[width=5.0cm]{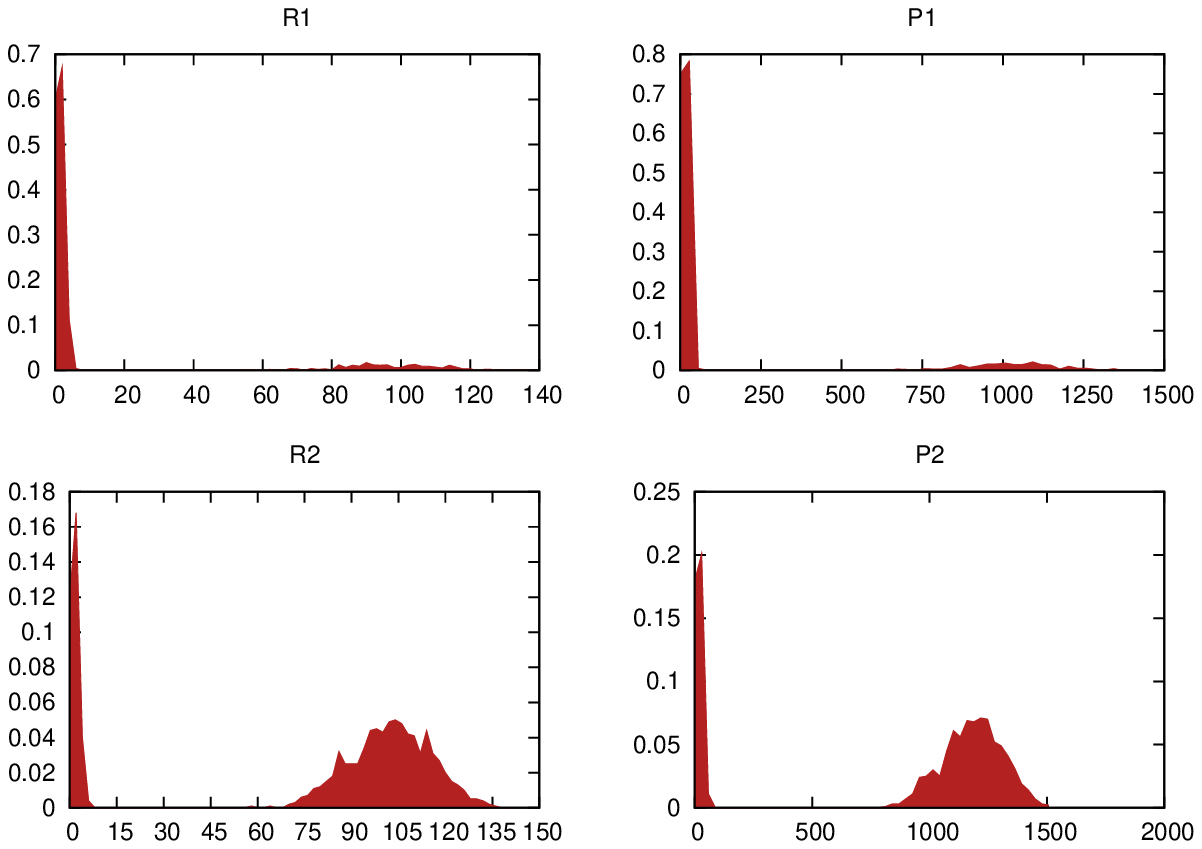}
 \end{array}$
\end{center}
\begin{center}$\scriptsize
\begin{array}{cc} 
t=1000,  \alpha=0.5 & t=1000,  \alpha=1 \\
 \includegraphics[width=5.0cm]{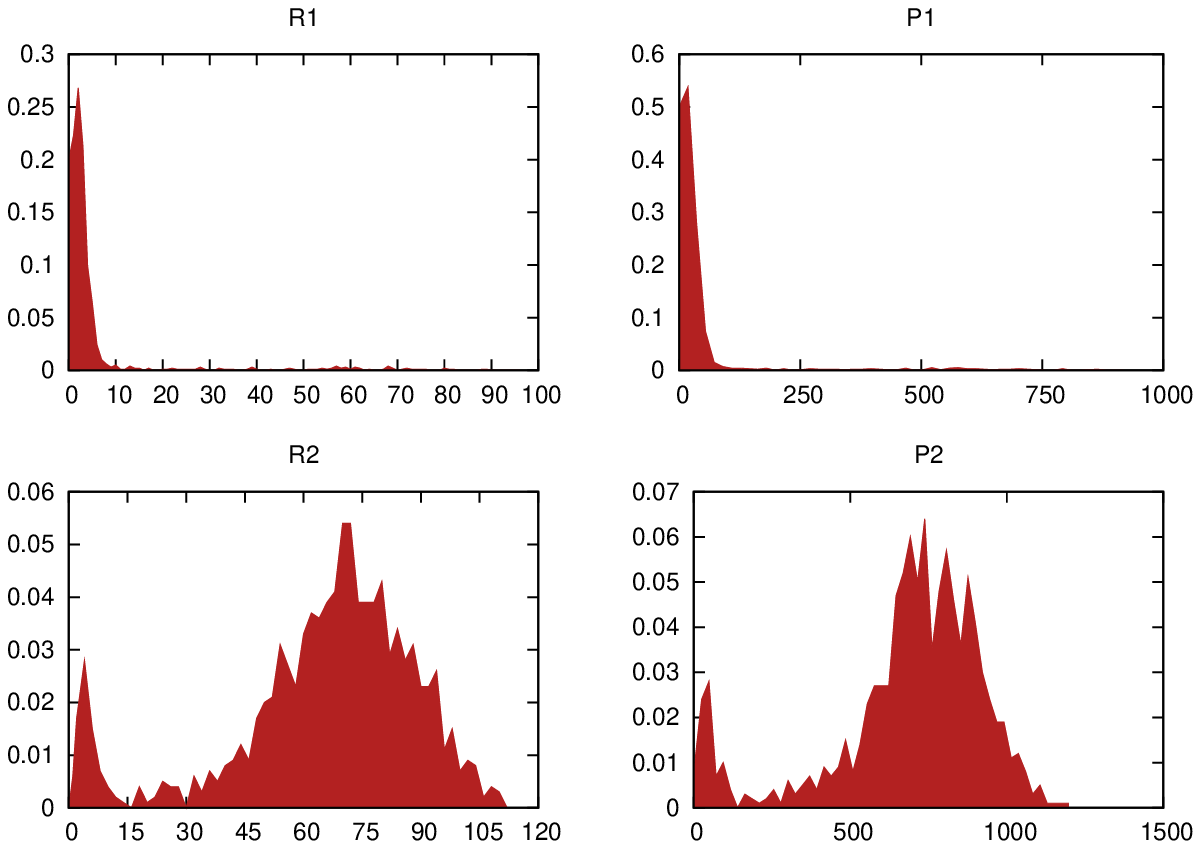}&
 \includegraphics[width=5.0cm]{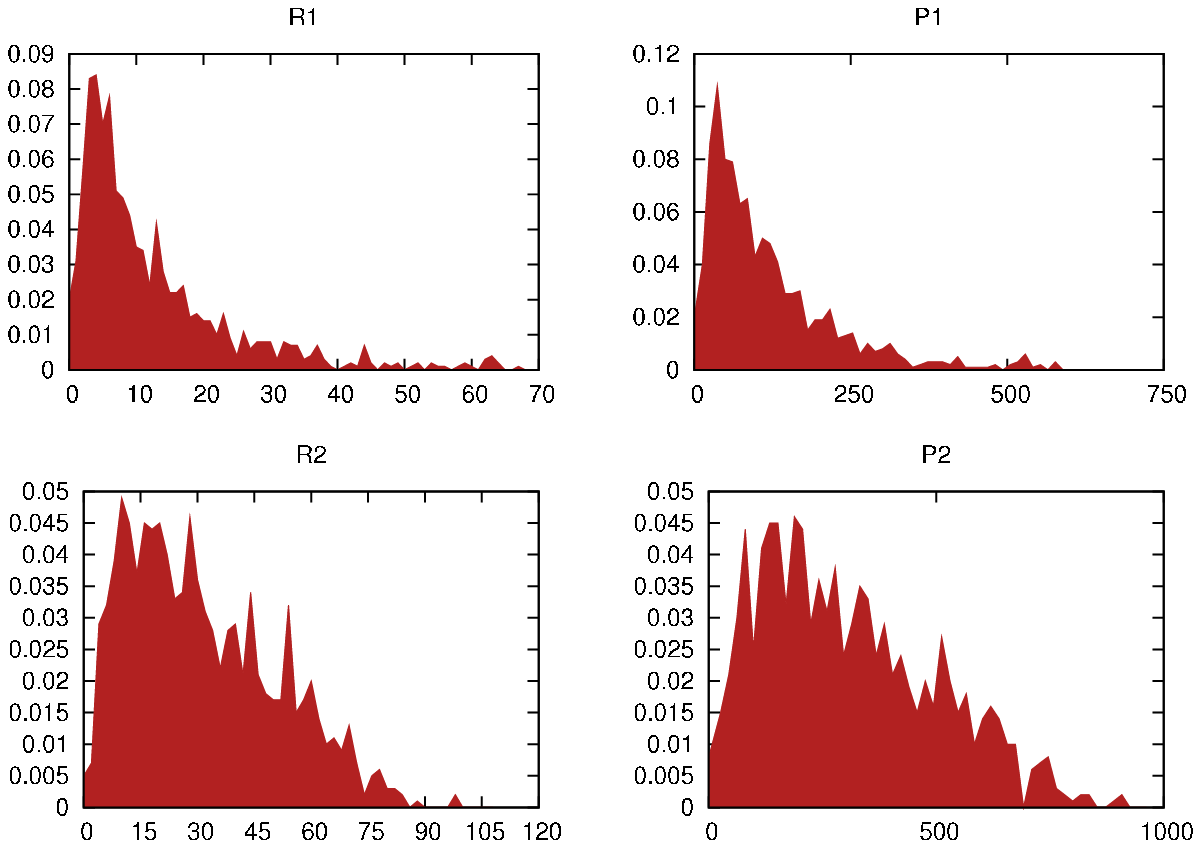}
 \end{array}$
\end{center}
\caption{{\bf Periodically perturbed  toggle switch.} Empirical probability density function at $t\in\{900, 950, 1000\}$, after  $1000$ simulations for Zhdanov model with  the  parameter configurations considered in Figure 
\ref{fig:zhdanov-orig}. In left  $ \alpha=0.5$ and in right $ \alpha=1$.}
\label{fig:zhdanov-orig-finalPDF}
\end{figure}

Let us consider a model by Zhdanov \cite{Zhdanov, Zhdanov2} where two genes $G_1$ and $G_2$, two RNAs $R_1$ and $R_2$ and two proteins $P_1$ and $P_2$ are considered. In such a model synthesis and degradation correspond to
\begin{align*}
G_1 &\srewrites{} G_1 + R_1 && R_1 \srewrites{} R_1 + P_1 
&& R_1 \srewrites{} \ast && P_1 \srewrites{} \ast \\
G_2 &\srewrites{} G_2 + R_2 && R_2 \srewrites{} R_2 + P_2 
&& R_2 \srewrites{} \ast && P_2 \srewrites{} \ast  \, .
\end{align*}
 Such a reaction scheme is a genetic toggle switch if the formation of $R_1$ and $R_2$ is suppressed by  $P_2$ and $P_1$, respectively \cite{ZH4,ZH2,ZH3,ZH1,Zhdanov3}. Zhdanov further simplifies the schema by considering kinetically equivalent  genes, and by assuming that the mRNA synthesis occurs only if $2$ regulatory sites  of either $P_1$ or $P_2$ are free. The deterministic model of the simplified switch when synthesis is  perturbed   is 
\begin{align} \label{eq:zhmodel}
\deriv{R_1}  &=  \xi(t) \left(  \dfrac{K}{K+P_{2}} \right)^2 - \delta_R R_1
&& 
\deriv{R_2}  =  \xi(t) \left(  \dfrac{K}{K+P_{1}} \right)^2 - \delta_R R_2
\\
\deriv{P_1}  &=  \alpha_P R_1 - \delta_P P_1 \nonumber
&& 
\deriv{P_2}  =  \alpha_P R_2 - \delta_P P_2
\end{align}
where the perturbation is
\[
\xi(t) = \alpha_R \left[1 + \alpha \sin\left(\dfrac{2\pi t}{\tau}\right) \right] \, .
\]
Here $\alpha_R$, $\delta_R$, $\alpha_P$ and $\delta_P$ are the rate constants of the reactions involved, term $[{K}/(K+P_{i})]^2$ is the probability that $2$ regulatory sites are free and $K$ is the  association constant for protein $P$. Notice that here perturbations are given in terms of a time-dependent kinetic function for synthesis, rather than a stochastic differential equation. Before introducing a realistic noise in spite of a perturbation we perform some analysis of this model. As in \cite{Zhdanov} we re-setted model (\ref{eq:zhmodel}) in a stochastic framework
by defining the  reactions described in Table \ref{table:zhdanov}. Notice that in there two reactions have a time-dependent propensity function, i.e. $a_1(t)$ and $a_3(t)$ modeling synthesis.

In the top panels of Figure \ref{fig:zhdanov-orig} we show single runs for Zhdanov model where simulations are performed with the exact SSA with time-dependent propensity function\footnote{In  \cite{Zhdanov} an exact SSA \cite{G77}  is used to simulated the model under the assumption  that  variations in the propensity functions are slow between two stochastic jumps. This is true  for $\tau=100$ as in \cite{Zhdanov}, but not true in general for small values of $\tau$.}. We considered an initial configuration with only $10$ RNAs $R_1$. As in \cite{Zhdanov} we set   $\alpha_R = 100\, min^{-1}$, $\alpha_P = 10\, min^{-1}$,  $\delta_R=\delta_P=1\, min^{-1}$, $K=100$ and $\tau=100\, min^{-1}$; notice that this parameters are realistic since, for instance, protein and mRNA degradation usually occur on the minute time-scale \cite{ZHparams}. We considered two possible noise intensities, i.e.    $\alpha=0.5$ in left and   $\alpha=1$ in right and, as expected, when $\alpha$ increases the number of switches increases. To investigate more in-depth this model we performed $1000$ simulations for both the configurations.  In the bottom panels of Figure \ref{fig:zhdanov-orig} the averages of the simulations are shown. The average  of our simulations evidences a major expression of  protein $P_2$ against $P_1$, for both values of $\alpha$, with dumped oscillations for $\alpha=0.5$ and almost persistent oscillations for $\alpha=1$. 

 \begin{figure}[!t]
\begin{center}$\scriptsize
\begin{array}{c} 
\alpha=0.5  \qquad \qquad\qquad \qquad \qquad\qquad \qquad \alpha=1 \\
 \includegraphics[width=10cm]{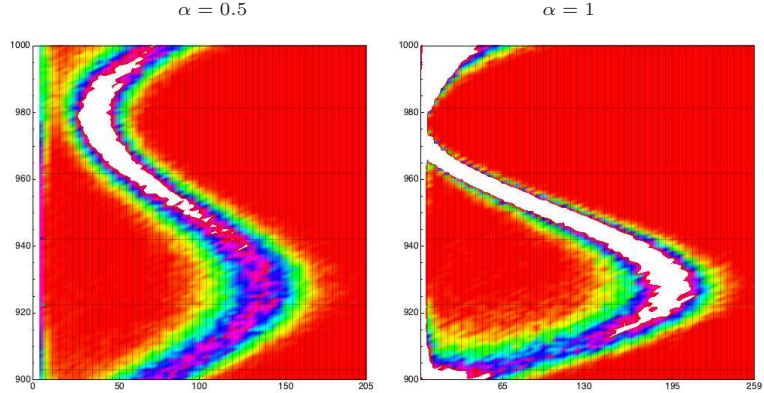}
 \end{array}$
\end{center}
\caption{{\bf Periodically perturbed toggle switch.} Empirical probability density function for $R_2$ plotted against time, i.e. the probability of being in any reachable state $\xx$ for $900 \leq t \leq 1000$. Lighter gradient denotes higher probability values. We used data collected with  $1000$ simulations of    model (\ref{eq:zhmodel}) where $\tau=100$  and two perturbation intensities are used, i.e. $\alpha \in \{0.5, 1\}$ as reported in the top captions. In the $x$-axis the  species concentration is represented, in the $y$-axis minutes are given. }
\label{fig:zhdanov-orig-pdf}
\end{figure}

\begin{figure}[t]
\begin{center}$\scriptsize
\begin{array}{cc} 
\text{(single run) }\alpha=0.5  &   \text{(single run) } \alpha=1 \\
 \includegraphics[width=7cm]{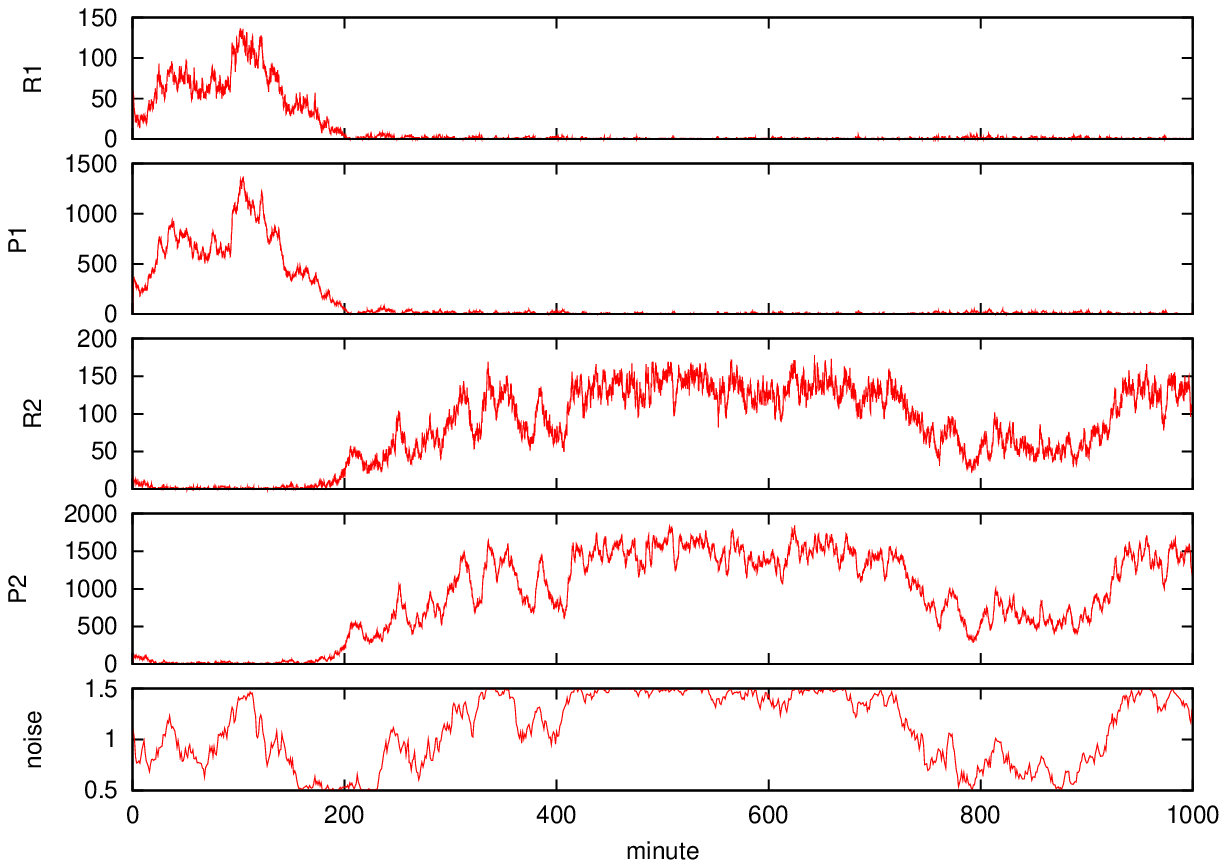} &
  \includegraphics[width=7cm]{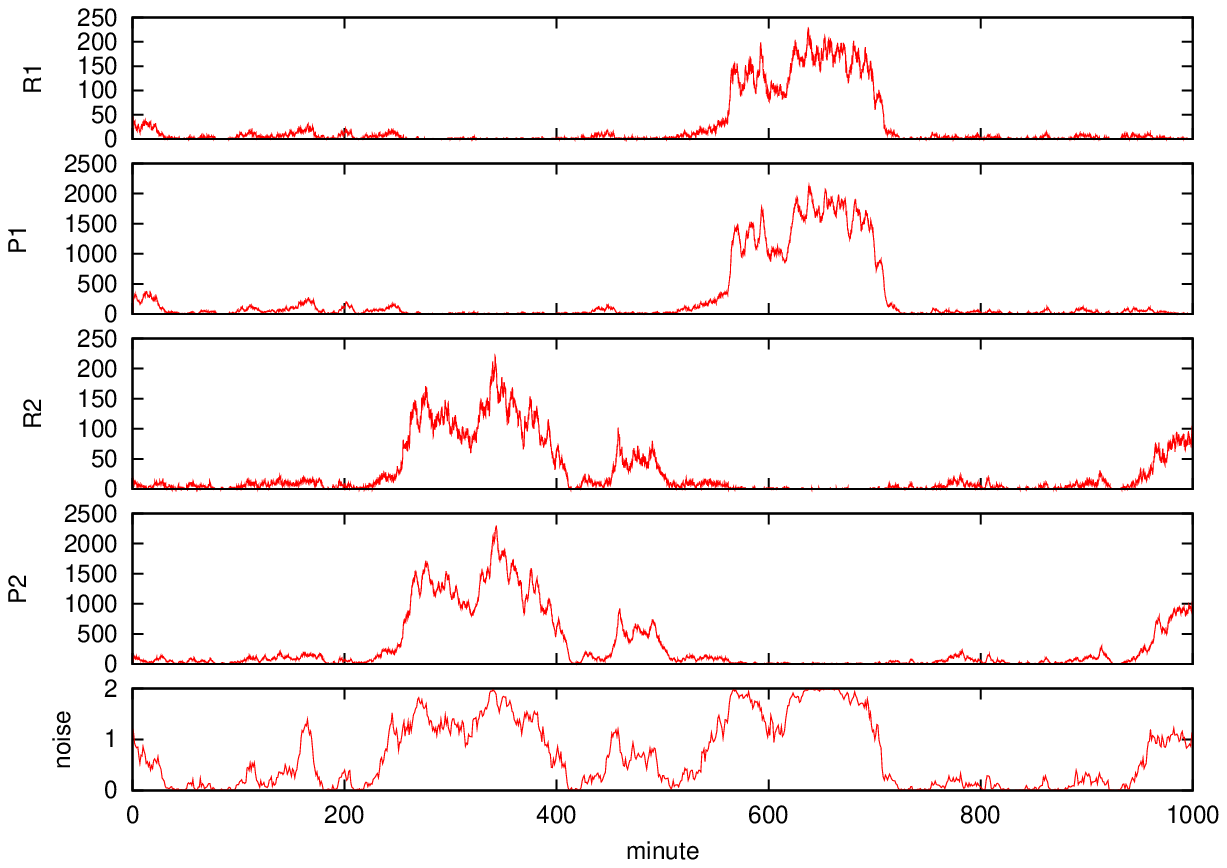}
 \end{array}$
\end{center}
\begin{center}$\scriptsize
\begin{array}{cc} 
\text{(averages) } \alpha=0.5 &   \text{(averages) } \alpha=1  \\
 \includegraphics[width=7cm]{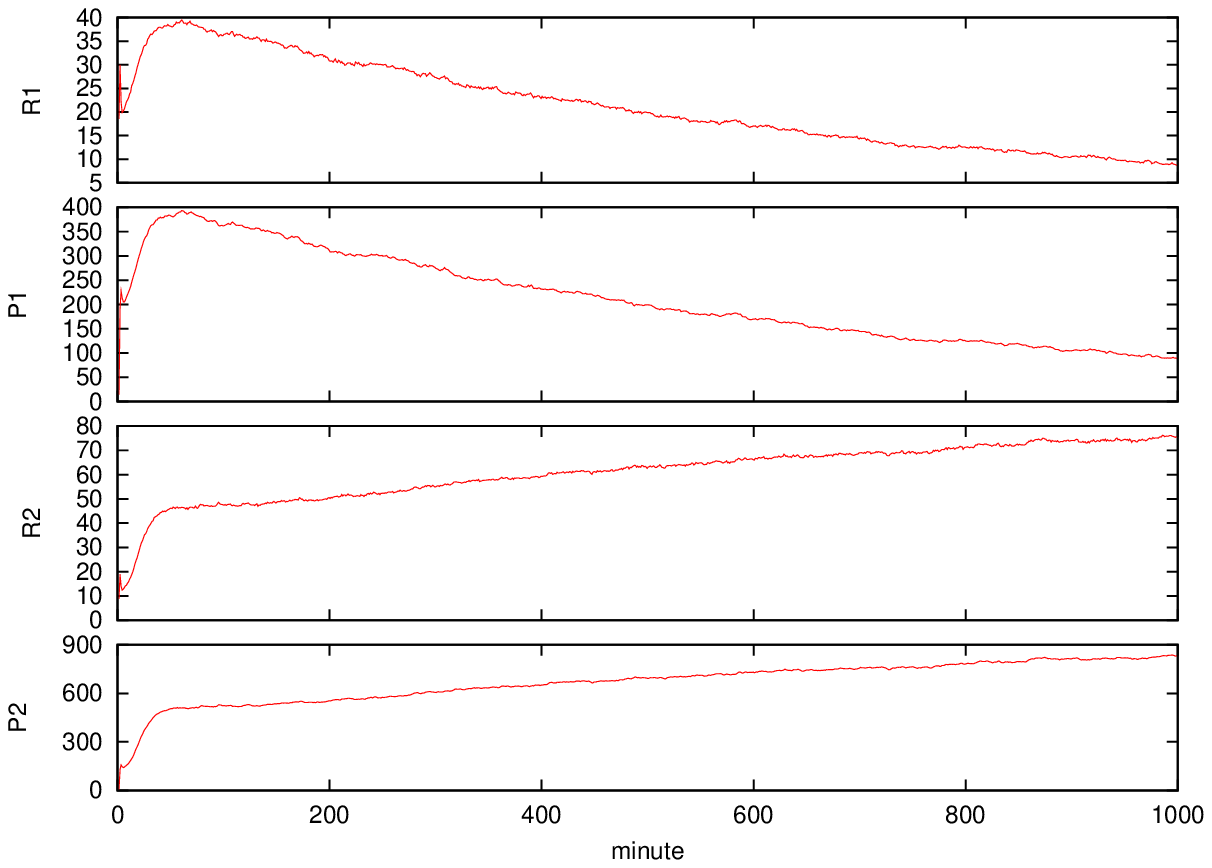} &
 \includegraphics[width=7cm]{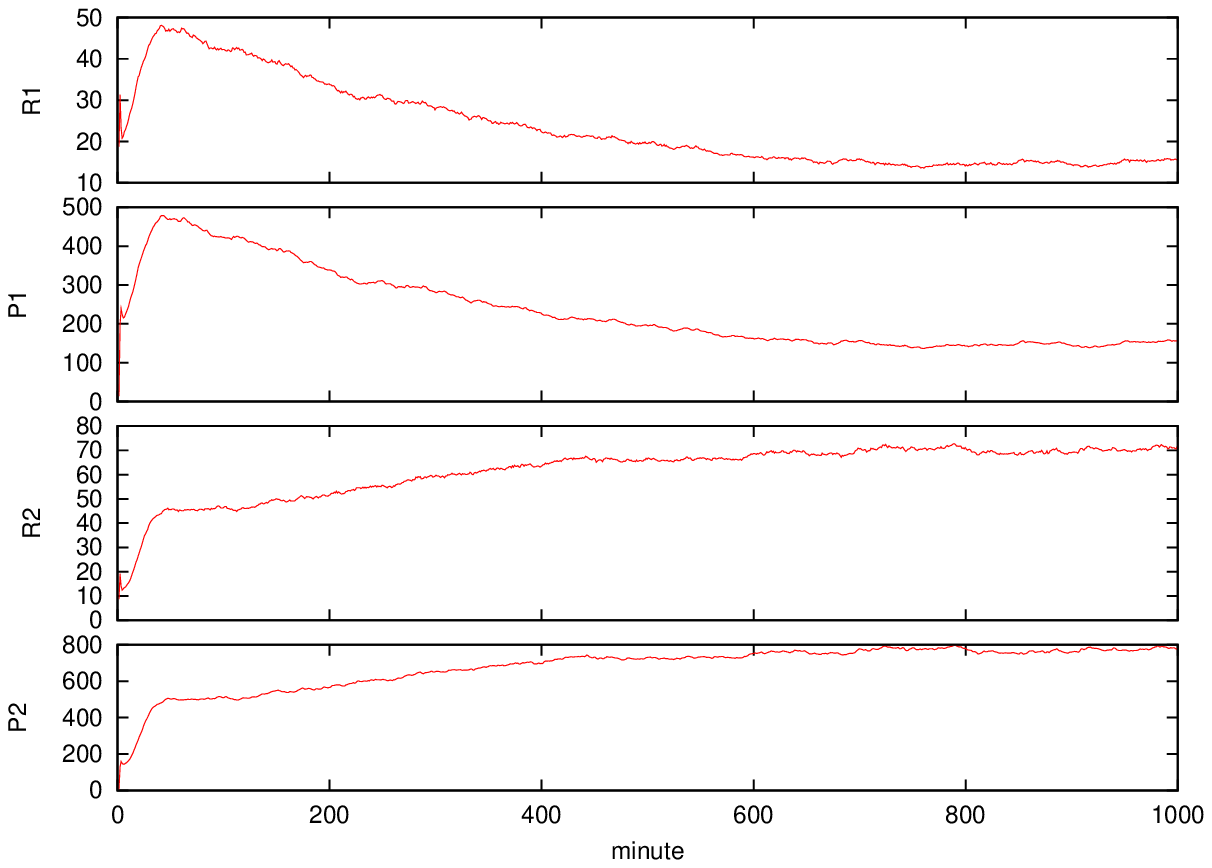}
 \end{array}$
\end{center}
\caption{{\bf Stochastically perturbed toggle switch.} In top panels, single run for Zhdanov model with Sine-Wiener  bounded noise: in left  $\alpha=0.5$, in  right $\alpha=1$. In bottom panels the averages of $1000$ simulations. In all cases  $\alpha_R = 100\, min^{-1}$, $\alpha_P = 10\, min^{-1}$,  $\delta_R=\delta_P=1\, min^{-1}$, $K=100$ and $\tau=100\, min^{-1}$  and the initial configuration is $(R_1,P_2,R_2,P_2)=(10,0,0,0)$.  The populations and the noise are plotted for the single runs.}
\label{fig:zhdanov-sw-1}
\end{figure}

 In Figure  \ref{fig:zhdanov-orig-finalPDF} we plot the empirical   probability density function of the species concentrations, i.e. $\Probab[\XX(t)=\xx]$ given the considered initial configuration, at $t \in \{ 900, 950, 1000\} \, min$ as obtained by   $1000$ simulations. Interestingly, these bi-modal probability distributions  immediately evidence the presence of  stochastic bifurcations  in the more expressed populations  $R_2$ and $P_2$. In addition, the distributions for the protein seem to  oscillate with period around $100$, i.e. for $\alpha=1$ they are unimodal at $t \in \{900, 1000\}$ and bi-modal at $t=950$.

For the sake of confirming this hypothesis in Figure \ref{fig:zhdanov-orig-pdf}  the probability density function of $P_2$ is plotted against time, i.e. the probability of being in state $\xx$ at time $t$, for any reachable state $\xx$ and time $900\leq t \leq 1000$. In there we plot  a heatmap with  time on the $y$-axis and protein concentration on the $x$-axis; in the figure the lighter gradient denotes higher probability values. Clearly, this figure shows the oscillatory behavior of the probability distributions for both value of $\alpha$ and, more important, explains the uni-modality of the distribution   at $t=900$ and $t=1000$ with $\alpha=1$, i.e. the higher variance of the rightmost peak at $\alpha=1$ makes the two modes collapse. Finally, we omit to show but, as one should expect, the oscillations of the probability distribution, which are caused by the presence of a sinusoidal perturbation in the parameters, are present and periodic over all the time window $0 \leq t \leq 1000$.

\paragraph{Bounded noises.}

We investigated the effect of  a  Sine-Wiener noise  affecting protein synthesis rather than a perturbation, i.e. a new $\xi(t)$ is considered
\[
\xi_{\text{sw}}(t) = \alpha_R \left[1 +  \alpha \sin\left( \sqrt{ \dfrac{2}{\tau}}W(t) \right) 
 \right] 
\]
with $W$ a Wiener process.  Here simulations are performed by using the  \SSAL{}
where the reactions in Table \ref{table:zhdanov} are left unchanged, and the propensity functions  $a_1(t)$ and $a_3(t)$ are modified to
\begin{align*}
a_1(t)= \xi_{\text{sw}}(t) [{K}/({K+P_{2}})]^2    &&
a_3(t)= \xi_{\text{sw}}(t) [{K}/({K+P_{1}})]^2   \, .
\end{align*}

For the sake of comparing the simulations with those in Figures \ref{fig:zhdanov-orig}--\ref{fig:zhdanov-orig-pdf}, we used the same initial condition  and the same values for  $\alpha_R$, $\alpha_P$,  $\delta_R$, $\delta_P$ and $K$.  To compare the effect of a realistic noise against the original  perturbation we  simulated the system with the same values i.e. the noise intensity $\alpha =0.5$ in left  and $\alpha=1$ in right of  the top panels in Figure  \ref{fig:zhdanov-sw-1}, and in both cases  $\tau=100$. {As expected, in this case the trajectories are more scattered than those in Figure \ref{fig:zhdanov-orig}, and the switches are still present. However, for maximum noise intensity $\alpha=1$ time-slots emerge where the stochastic systems predicts a more complex outcome of the interaction. In fact,  for $t \in [0, 200] \cup [800, 900]$  neither protein $P_1$ nor $P_2$ seem to be as expressed as in the other portions of the simulation,  thus suggesting the presence of noise-induced equilibria absent when periodic perturbations are present.}

\begin{figure}[!t]
\begin{center}$\scriptsize
\begin{array}{cc} 
 \alpha=0.5 &  \alpha=1 \\
 \includegraphics[width=5.5cm]{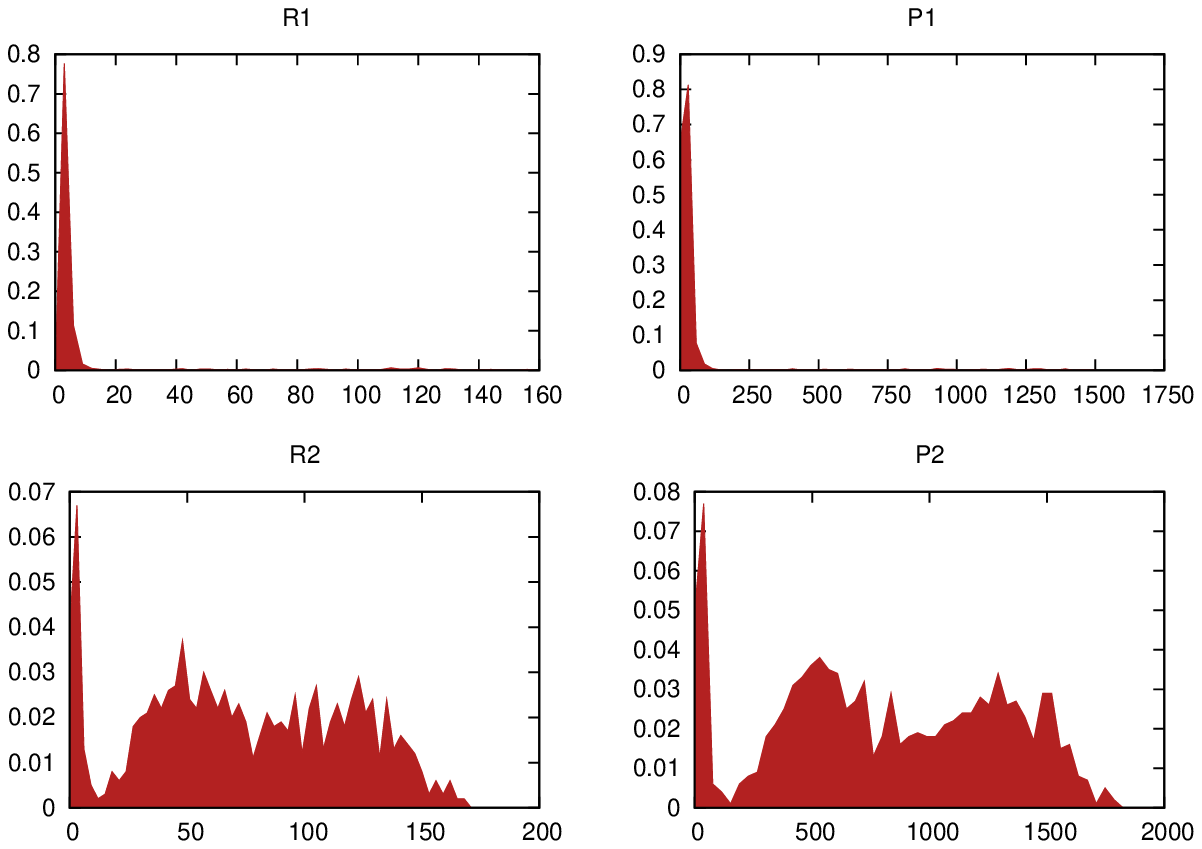}&
 \includegraphics[width=5.5cm]{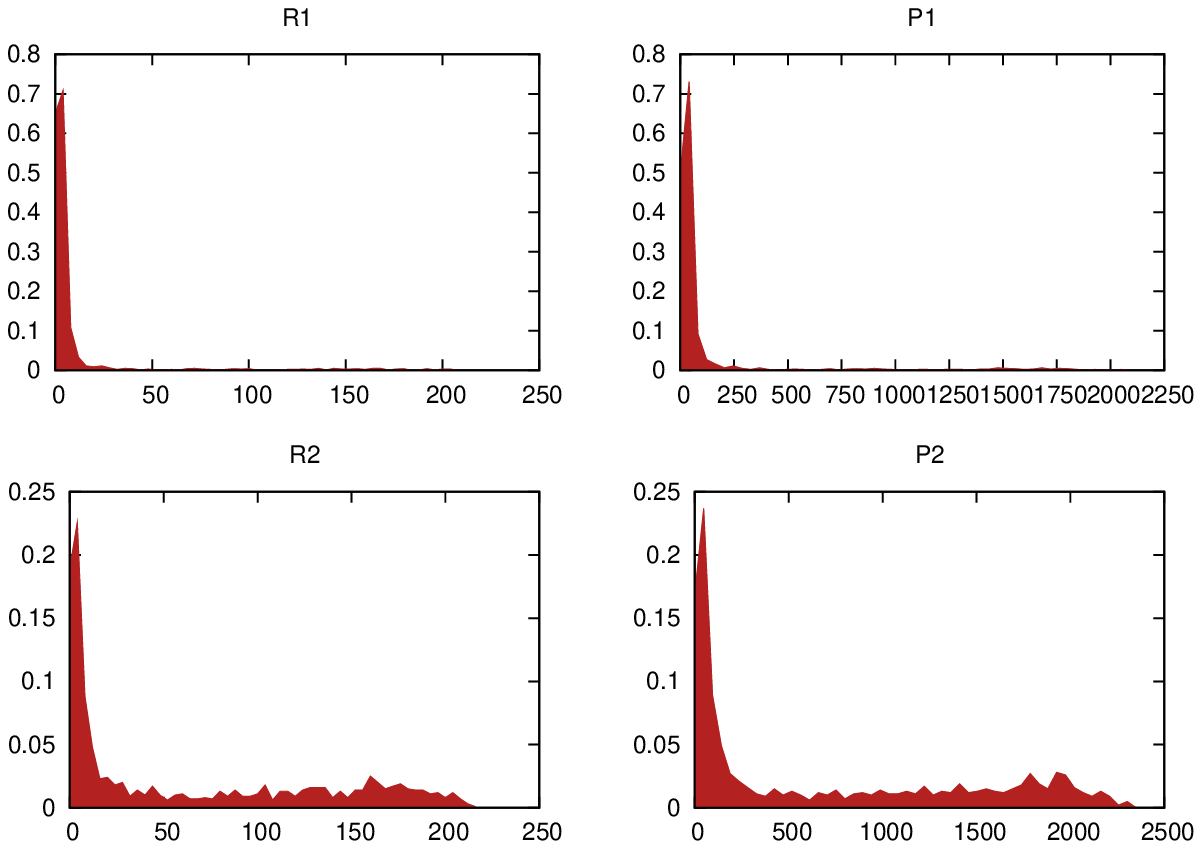}
 \end{array}$
\end{center}
\caption{{\bf Stochastically perturbed toggle switch.} Empirical probability density function at $t=1000$, after  $1000$ simulations for Zhdanov model  with Sine-Wiener noise. Parameters  are as in Figure 
\ref{fig:zhdanov-sw-1} and two perturbation intensities are used, i.e. $\alpha \in \{0.5, 1\}$ as reported in the top captions.}
\label{fig:zhdanov-sw-finalPDF}
\end{figure}

\begin{figure}[!t]
\begin{center}$\scriptsize
\begin{array}{c}
\alpha=0.5  \quad\qquad\qquad\qquad\qquad\qquad\qquad\qquad  \alpha=1 \\
 \includegraphics[width=10cm]{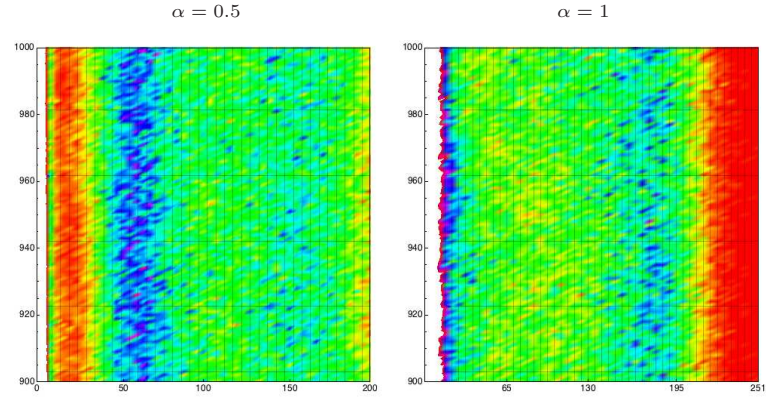}
 \end{array}$
\end{center}
\caption{{\bf Stochastically perturbed toggle switch.}  Empirical probability density function for $R_2$ plotted against time, i.e. the DCKE solution  for  $R_2$ in $900 \leq t \leq 1000$. Lighter gradient denotes higher probability values. We used data collected with  $1000$ simulations of  Zhdanov model  with Sine-Wiener noise  where $\tau=100$ and two perturbation intensities are used, i.e. $\alpha \in \{0.5, 1\}$ as reported in the top captions. In the $x$-axis the  species concentration is represented, in the $y$-axis minutes are given.}
\label{fig:zhdanov-sw-pdf-1}
\end{figure}

To investigate more in-depth this hypothesis we again performed $1000$ simulations for both the configurations, the averages of which are shown in the bottom panels  of Figure \ref{fig:zhdanov-sw-1}. 
In this case,  the simulation times, which again depend on the noise correlation, span from $3 \div 5 \, min$ to $30 \div 40\, min$, thus making the choice of good parameters crucial.
Differently from the case in which a sinusoidal perturbation is considered, i.e. Figure \ref{fig:zhdanov-orig},  in this case the averages are not oscillatory, but instead show a stable convergency. Also, the final outcome seems again to predict the expression of $P_2$ inhibiting $P_1$. To understand better this point we  plotted in Figure   \ref{fig:zhdanov-sw-finalPDF} the probability density of reachable states at $t= 1000\, min$, i.e. $\Probab[\XX(t)=\xx]$ given the considered initial configuration, and in Figure  \ref{fig:zhdanov-sw-pdf-1} we plotted that distribution against time for  $R_2$. It is worth noting that we also ranged $t$ over $[900, 1000]$ but since  $\Probab[\XX(t)=\xx]$ did not change we omitted to plot it here.
Again, Figure  \ref{fig:zhdanov-sw-pdf-1} is  a heatmap where on the $y$-axis time in minutes is  given, on the $x$-axis the possible concentration for  $R_2$ and the lighter gradient denotes higher probability values. Notice that in this case Figure  \ref{fig:zhdanov-sw-pdf-1} represents an empirical evaluation of the solution the DCKE for this system, i.e. equation (\ref{eq:DCKE}).
Both graphics are obtained by  $1000$ simulations with $\alpha=0.5$ (left panels) and $\alpha=1$ (right  panels). These figures show that a low-intensity  noise makes  the probability distribution become three-modal, i.e. notice the two rightmost peaks in 
Figure  \ref{fig:zhdanov-sw-finalPDF} and the white/light-blue gradients in Figure \ref{fig:zhdanov-sw-pdf-1}. Differently, when the noise intensity is higher,  the two rightmost peaks almost merge, thus forming a bi-modal distribution where the smaller peak  almost spreads uniformly on the state space for the variables. Notice that, in this case, the amplitude of such a peak is higher than for $\alpha=0.5$, i.e. notice the intensity of the blue gradient in Figure  \ref{fig:zhdanov-sw-pdf-1}.  For $\alpha=0.5$ it is possible to notice two red gradients: one approximatively for $\xx \to 200$ and one for $\xx \in (10, 30)$. The   major peaks in the distribution for  $R_2$ are for  $\xx  < 10$, for $\xx \in (50, 100)$ and for  $\xx \in [130, 180]$. The probability of each of these peaks is decreasing as $\xx$ increases, thus confirming the intuition of Figure \ref{fig:zhdanov-sw-finalPDF}. Similar considerations can be done when $\alpha=1$ where, as shown by Figure \ref{fig:zhdanov-sw-finalPDF}, the first dark-red area separating the 
first two peaks in $\alpha=0.5$ is vanished, thus forming a bi-modal instead that a three-modal probability distribution.


\begin{figure}[t]
\begin{center}$\scriptsize
\begin{array}{cccc} 
\tau=1 & \tau=10 & \tau=25 & \tau=100 \\
 \includegraphics[width=3.3cm]{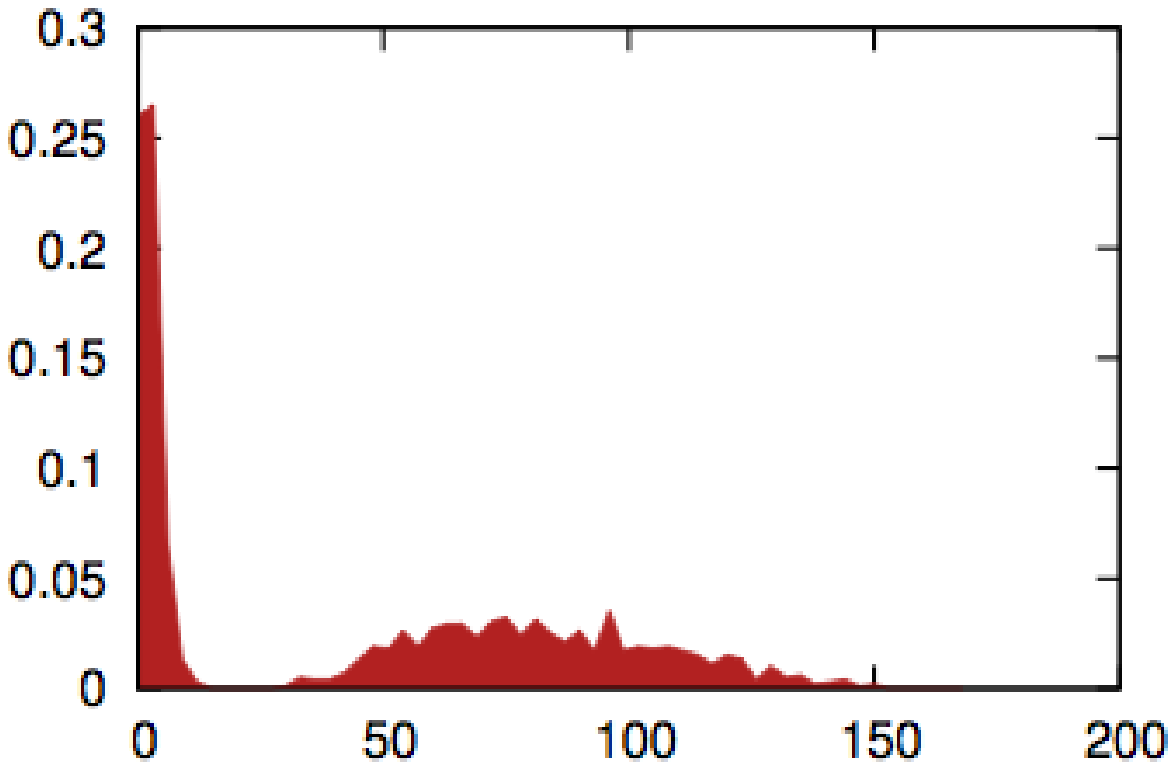} &
 \includegraphics[width=3.3cm]{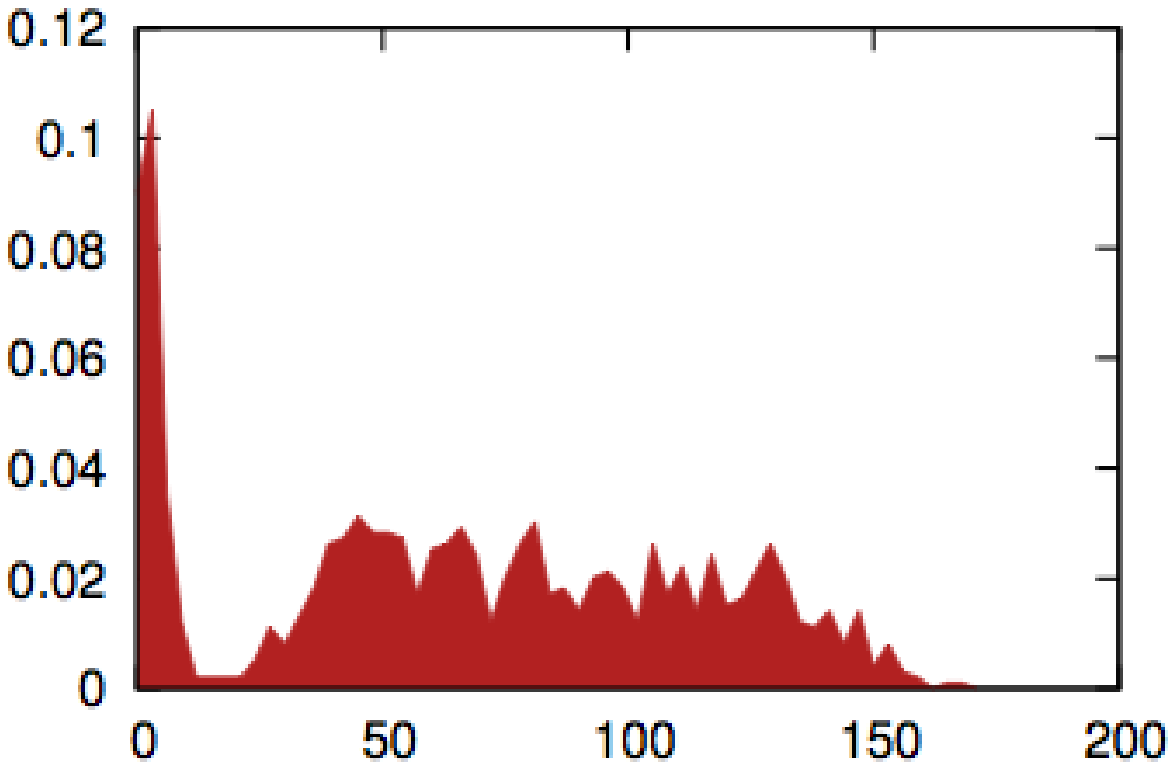}&
 \includegraphics[width=3.3cm]{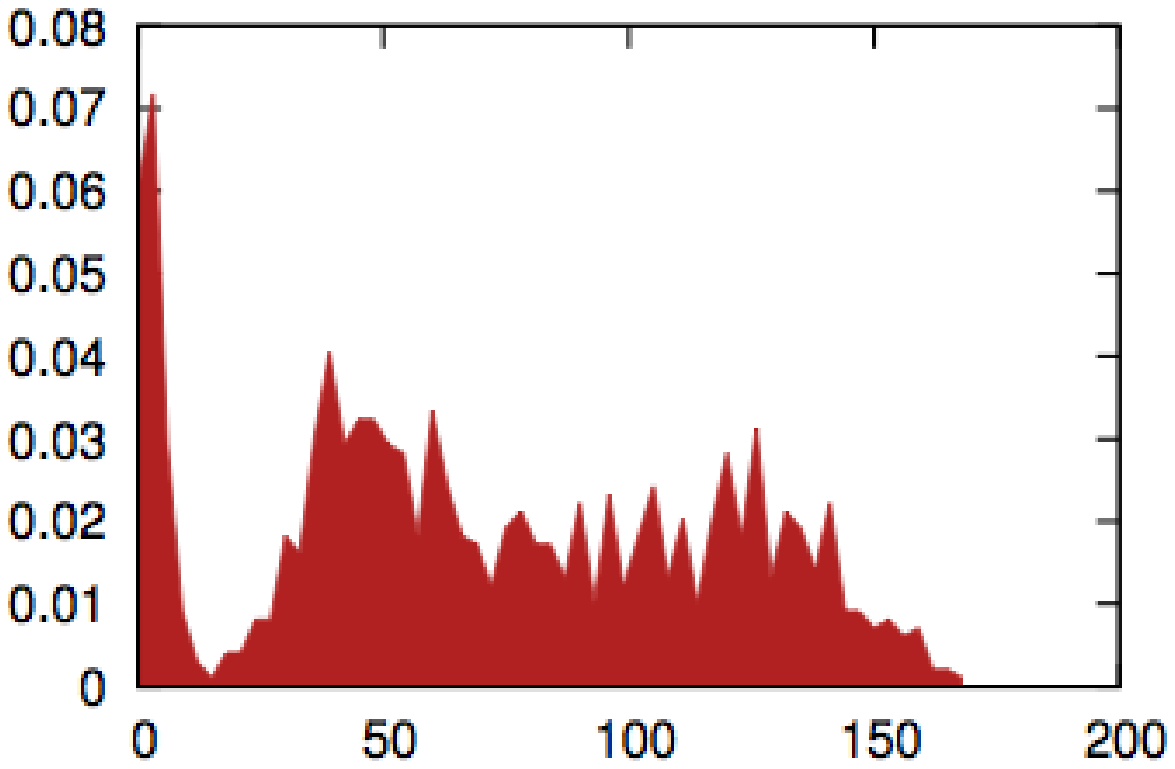}&
 \includegraphics[width=3.3cm]{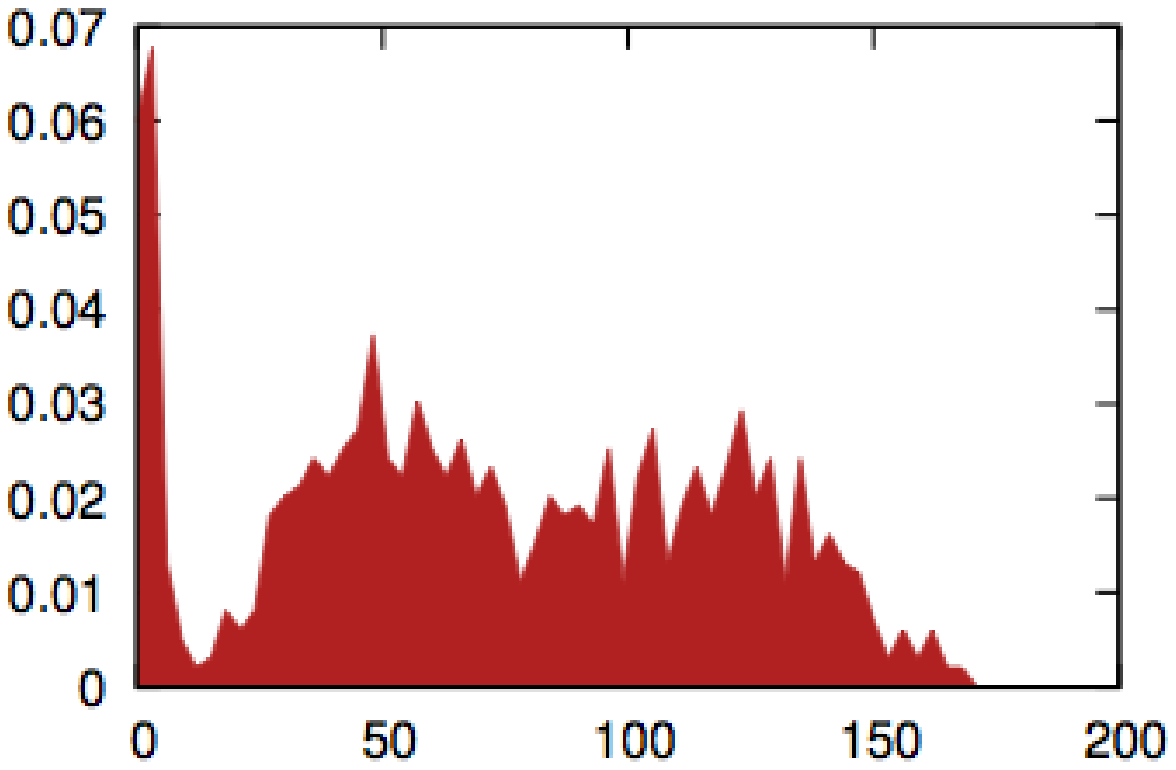}
  \end{array}$
\end{center}
\caption{{\bf Stochastically perturbed toggle switch.} Empirical probability density function at $t =1000$ for $R_2$, after  $1000$ simulations for Zhdanov model  with Sine-Wiener noise. From left to right  $\tau=1$,  $\tau=10$,  $\tau=25$ and $\tau=100$. In all cases $\alpha=0.5$ and other parameters  are as in Figure 
\ref{fig:zhdanov-sw-1}.}
\label{fig:zhdanov-sw-finalPDF-various}
\end{figure}

\begin{figure}[!t]
\begin{center}$ \scriptsize
\begin{array}{cccc}
\tau=1 & \tau=10 & \tau=25 & \tau=100 \\
 \includegraphics[width=3.5cm]{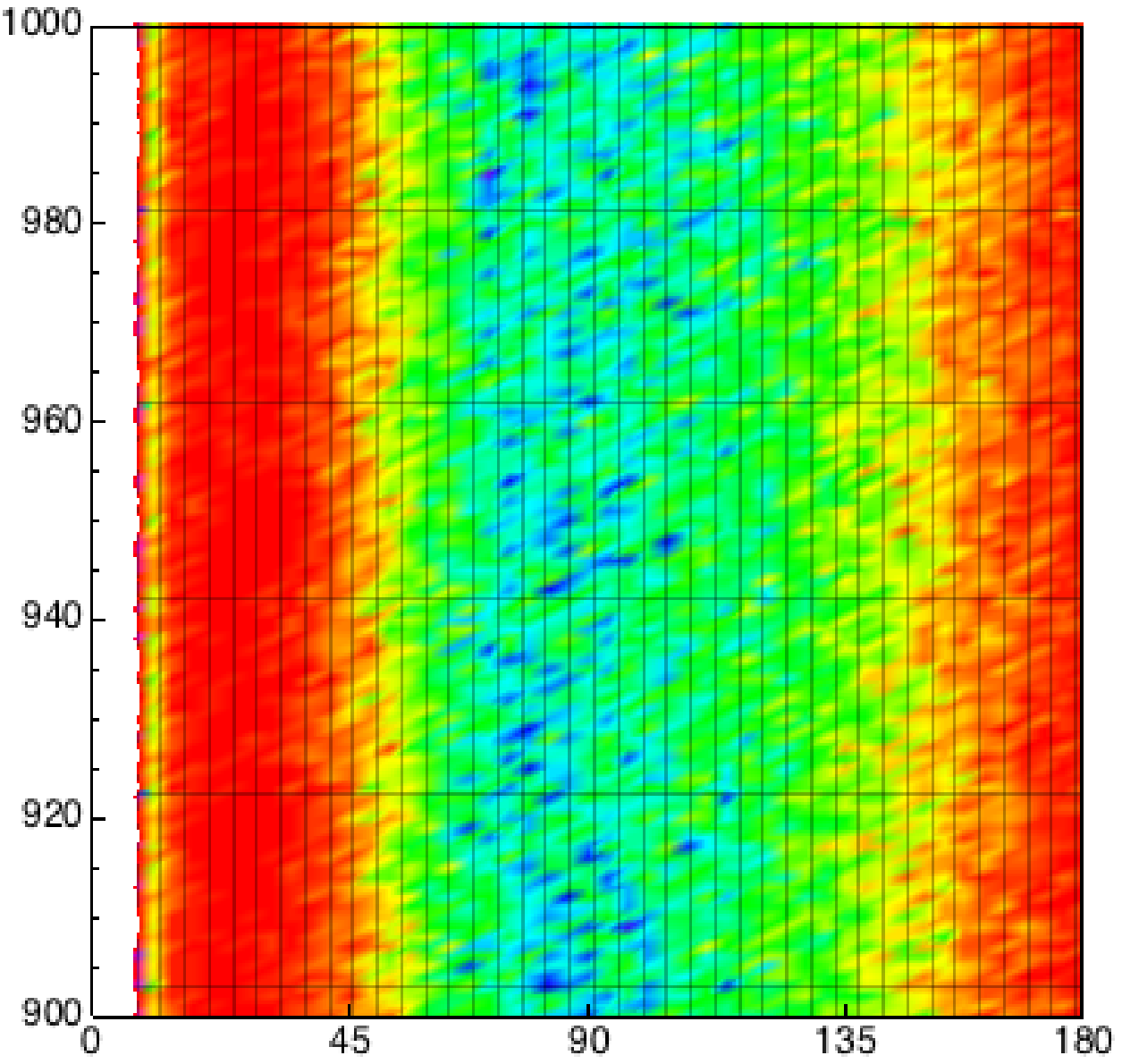}&
 \includegraphics[width=3.5cm]{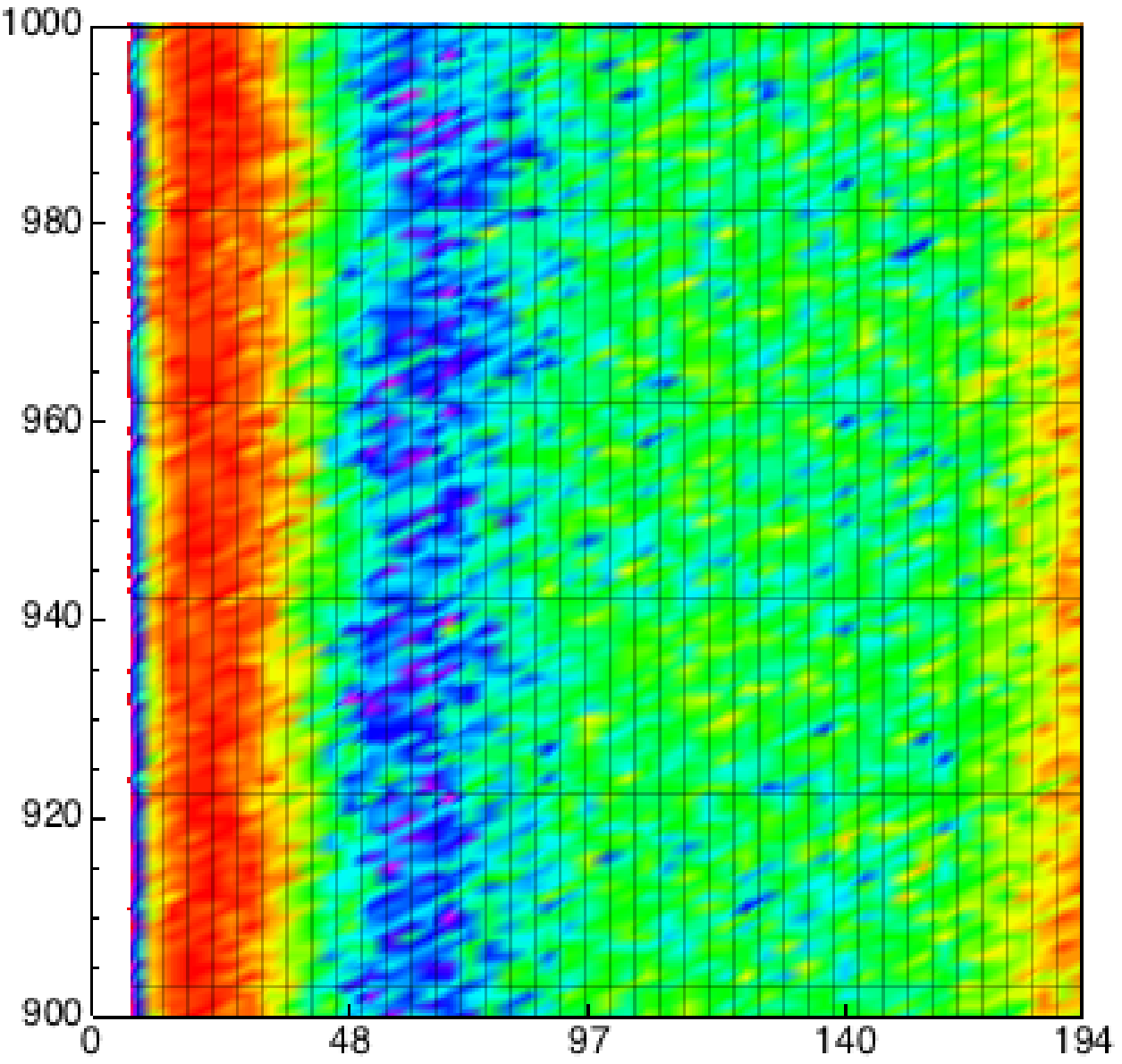} &
 \includegraphics[width=3.5cm]{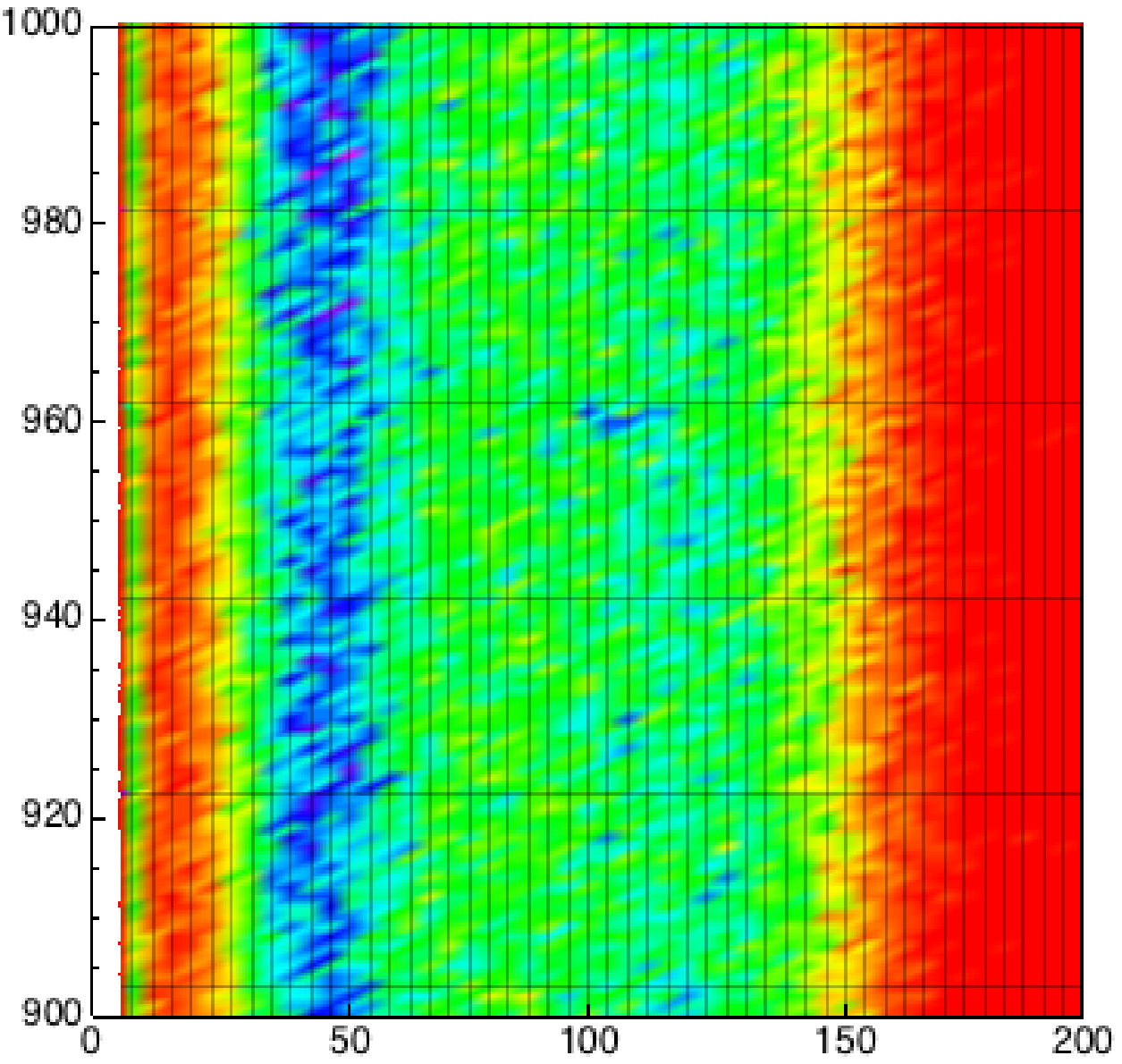} &
 \includegraphics[width=3.5cm]{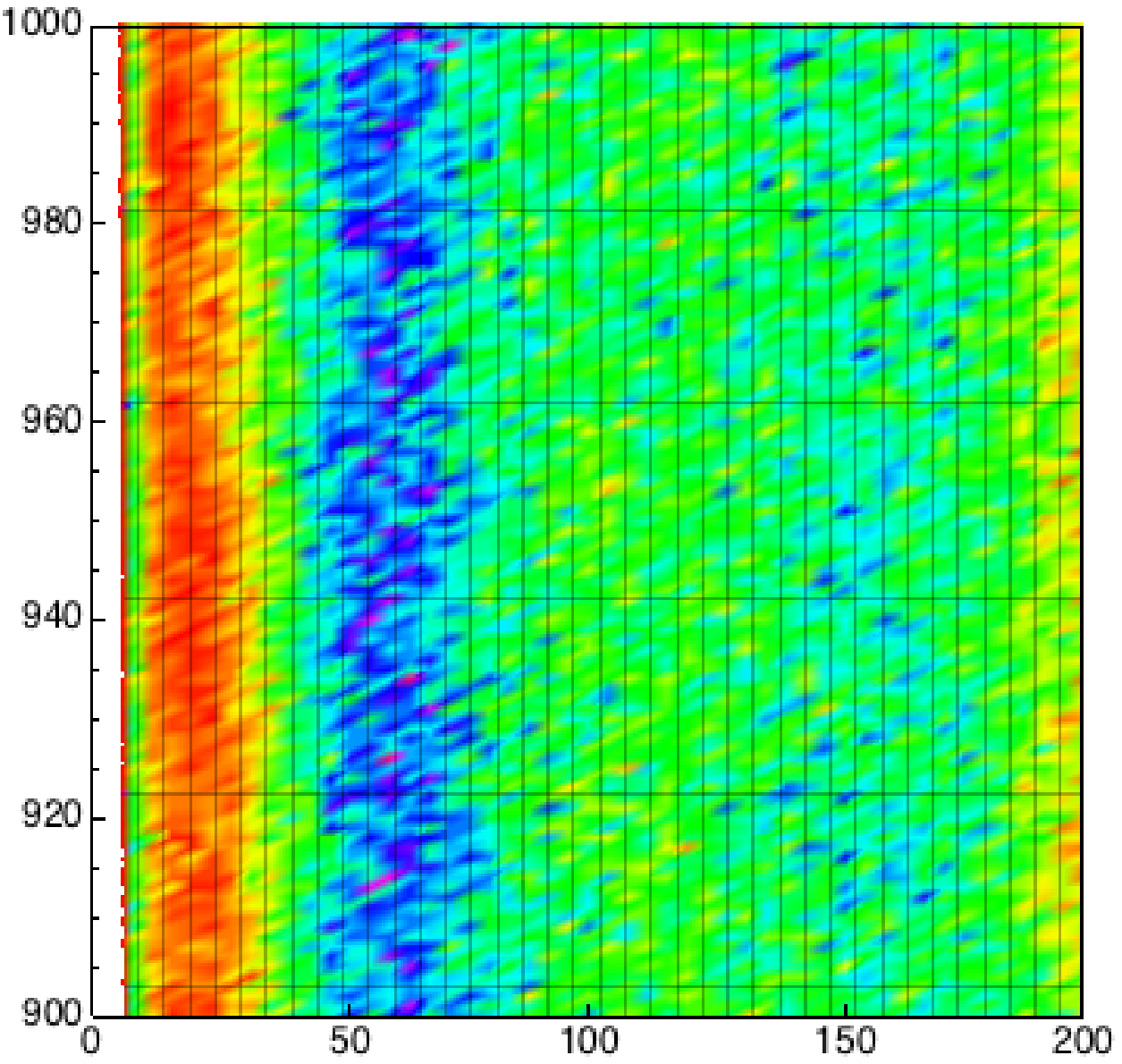} \\
 \end{array}$
\end{center}\caption{{\bf Stochastically perturbed toggle switch.}
Empirical probability density function for $R_2$ plotted against time, i.e. the DCKE solution  for  $R_2$ in  $900 \leq t \leq 1000$. Lighter gradient denotes higher probability values. We used data collected with  $1000$ simulations of  Zhdanov model  with Sine-Wiener noise. From left to right  $\tau=1$,  $\tau=10$,  $\tau=25$ and $\tau=100$. In all cases  the noise intensity is $\alpha=0.5$. In the $x$-axis the species concentration  is represented, in the $y$-axis minutes are given.}
\label{fig:zhdanov-sw-pdf-all}
\end{figure}

Finally, for the sake of considering a wide range of biologically meaningful values for $\tau$, which we recall it represents a measure of the speed of noise variation, we evaluated the solution of the DCKE  for $R_2$  for the same configuration used in Figure \ref{fig:zhdanov-sw-pdf-1} and  $\tau \in \{1, 10, 25, 100\} \, min$. 
We performed $1000$ simulations  of the model for each value of $\tau$ with $\alpha=0.5$,  the value showing a more interesting behavior. In Figure \ref{fig:zhdanov-sw-finalPDF-various} the probability of the reachable states at $t=1000 \, min$ is plotted. If is immediate to notice that the height of the first peak increases as $\tau$ decreases, and more precisely the distribution seems to switch from a three-modal one to a bi-modal when $\tau \leq 25$. In  each panel of  Figure  \ref{fig:zhdanov-sw-pdf-all} we plot the variation of such probability distribution for $900 \leq t \leq 1000$. By that figure it is possible to observe that  by ranging $\tau$  the dark-red gradient increases in size as far as $\tau$ decreases. This means that the amplitude between the peaks of the density strictly depends on the value of $\tau$, thus suggesting a strong role for extrinsic noise in determining the network functionalities.

%% file: sections/conclusions.tex
\section*{Discussions}

In this paper we investigated the effects of joint extrinsic and intrinsic randomness in nonlinear genetic networks, under the assumption of non-gaussian bounded external perturbations. Our applications have shown that the combination of both intrinsic and extrinsic noise-related phenomena may have a constructive functional role also when the extrinsic noise is bounded. This is in line with other researches - only focusing on either intrinsic or extrinsic noise - recasting the classical interpretation of noise as a disturbance more or less obfuscating the real behavior of a network.

This work required the combination of two well-known frameworks, often used to separately describe biological systems. We combined the theory of stochastic chemically reacting systems developed by Gillespie with Langevin systems describing the bounded variations of kinetic parameters. The former shall allow considering the inherent stochastic fluctuations of small numbers of interacting entities, often called intrinsic noise, and clearly opposed to classical deterministic models based on differential equations. The latter permits to consider the influence of bounded extrinsic noises. These noises are modeled as stochastic differential equations. For these kind of systems, although an analytical characterization is unlikely to be feasible, we were able to derive a {differential Chapman-Kolgomorov equation} (DCKE) describing the probability of the system to occupy each one of a set of states. Then, in order to analyze these models by sampling from this equation we defined an extension of the Gillespie's Stochastic Simulation Algorithm (SSA) with a state-dependent Langevin system affecting the model jump rates. This algorithm, despite being more costly than the classical Gillespie's SSA, allows for the exact simulation of these doubly stochastic systems.

We outlined the role of extrinsic noise for some biological networks of interest. In particular, we were able to extend classical results on the validity of the Michaelis-Menten approximation to the prototypical Enzyme-Substrate-Product enzymatic reaction by drawing a Stochastic Quasi Steady State Assumption (SQSSA) for noisy reactions. Along the line of the classical deterministic or stochastic uses of the Michaelis-Menten approximation, this should permit to reduce the size of more general enzymatic networks even in presence of extrinsic bounded noises. 

Moreover, we showed that in a recurrent pattern of genetic and enzymatic networks, i.e. the futile cycle, the presence of extrinsic noises induces the switching from a monomodal probability density (in absence of external perturbations) to a multimodal density. 

Similarly, in the case of the toggle switch, which is inherently multistable, the presence of extrinsic noise significantly modulates the probability density of the genes concentration. In this  important network motif  we also investigated the role of periodic perturbations against a realistic noise.

Thus in general the co-presence of both intrinsic stochasticity and  bounded extrinsic random perturbations might suggest the presence of possibly unknown functional roles for noise for these and other networks. 
The described noise-induced phenomena are shown to be strongly related to physical characteristics of the extrinsic noise such as the noise amplitude and its autocorrelation time.

A relevant issue that we are going to investigate in the next future is the role of the specific extrinsic bounded perturbations. Indeed, in non-genetic systems affected by bounded noises it has been shown that the effects of the perturbations depend not only on the above general characteristics of the noise, but also on its whole model \cite{onofr1,onofr4,onofr2,onofr3}. In other words the transitions of a system perturbed by a sine-Wiener noise might be quite different from those induced by another bounded perturbation, for example the Cai-Lin noise \cite{CaiLin} or the Tsallis noise \cite{wio}, also when their amplitude and autocorrelation times are equal. 
In other words, a single network in two different environments might show two different behaviors depending of fine details of the kind of perturbations that are present. This might also suggest that a same network might exhibit many different functions depending on its ``locations".

Finally, concerning these points, we stress that these peculiar properties of bounded extrinsic perturbations make it even more important the investigations, such as those of \cite{You}, aimed at inferring by deconvolution the external noise from the experimental data, in order to infer which kind of noise affect a given network in a well determined environment.